\documentclass[aps,prd,superscriptaddress,floats,floatfix,preprint]{revtex4}
\usepackage{amsmath}
\usepackage{amsfonts}
\usepackage{graphicx}
\usepackage{bbm}
\usepackage{multirow}
\usepackage{pifont}

\newcommand{\tbox}[1]{\mbox{\tiny #1}}
\newcommand{\tmbx}[1]{\mbox{\tiny{$#1$}}}

\newcommand{\tto}{G(221)}

\newcommand{\uL}{u_{\tmbx{L}}}
\newcommand{\dL}{d_{\tmbx{L}}}
\newcommand{\nL}{\nu_{\tmbx{L}}}
\newcommand{\eL}{e_{\tmbx{L}}}
\newcommand{\uR}{u_{\tmbx{R}}}
\newcommand{\dR}{d_{\tmbx{R}}}
\newcommand{\nR}{\nu_{\tmbx{R}}}
\newcommand{\eR}{e_{\tmbx{R}}}

\newcommand{\Zp}{Z^{\prime}}

\newcommand{\tgo}{\tilde{g}_{1}}
\newcommand{\tgt}{\tilde{g}_{2}}
\newcommand{\tgL}{\tilde{g}_{\tmbx{L}}}

\newcommand{\tgY}{\tilde{g}_{\tmbx{Y}}}
\newcommand{\tgX}{\tilde{g}_{\tmbx{X}}}

\newcommand{\gZLt}{\left[g_{\tmbx{L}}^{\tmbx{Z}}(f)\right]^2}

\newcommand{\gZRt}{\left[g_{\tmbx{R}}^{\tmbx{Z}}(f)\right]^2}

\newcommand{\gNLt}{\left(g_{\tmbx{L}}^{\tmbx{\nu N}}\right)^2}
\newcommand{\gNLSMt}{\left(g_{\tmbx{L},\tbox{SM}}^{\tmbx{\nu N}}\right)^2}

\newcommand{\geV}{g_{\tmbx{V}}^{\tmbx{\nu e}}}
\newcommand{\geA}{g_{\tmbx{A}}^{\tmbx{\nu e}}}

\newcommand{\tphi}{\tilde{\phi}}
\newcommand{\tth}{\tilde{\theta}}
\newcommand{\tbet}{\tilde{\beta}}

\newcommand{\stphi}{s_{\tphi}}
\newcommand{\ctphi}{c_{\tphi}}

\newcommand{\VEV}[1]{\langle  #1 \rangle}

\newcommand{\tu}{\tilde{u}}

\newcommand{\tx}{\tilde{x}}

\newcommand{\Lag}{\mathcal{L}}

\newcommand{\GZ}[1]{\Gamma_Z\left(#1\bar{#1}\right)}
\newcommand{\GZhad}{\Gamma_Z\left(\mbox{had.}\right)}
\newcommand{\GZpm}[1]{\Gamma_Z\left(#1^-#1^+\right)}

\newcommand{\epsL}[1]{\varepsilon_{\tmbx{L}}\left(#1\right)}
\newcommand{\epsLt}[1]{\varepsilon_{\tmbx{L}}^2\left(#1\right)}
\newcommand{\epsR}[1]{\varepsilon_{\tmbx{R}}\left(#1\right)}
\newcommand{\epsRt}[1]{\varepsilon_{\tmbx{R}}^2\left(#1\right)}

\newcommand{\ffLR}[3]{\big(\bar{#1}#2\big)_{\tmbx{#3}}}

\newcommand{\ffLRm}[3]{\big(\bar{#1}#2\big)_{\tmbx{#3},\mu}}

\newcommand{\picwidth}{3.00in}
\newcommand{\smbox}[1]{\mbox{\scriptsize #1}}
\newcommand{\mfrac}[2]{\frac{ \mbox{$#1$} }{ \mbox{$#2$} }}

\newcommand{\ttheta}{\tilde{\theta}}
\newcommand{\tbeta}{\tilde{\beta}}

\newcommand{\PL}{P_{\tbox{L}}}
\newcommand{\PR}{P_{\tbox{R}}}

\newcommand{\uT}{\tilde{u}_{\tbox{T}}}
\newcommand{\uD}{\tilde{u}_{\tbox{D}}}

\newcommand{\tM}{\widetilde{M}}

\newcommand{\cphi}{c_{\tphi}}
\newcommand{\sphi}{s_{\tphi}}

\newcommand{\Eref}[1]{(\ref{#1})}

\begin{document}
\title{Global Analysis of General $SU(2)\times SU(2)\times U(1)$ Models with Precision Data}

\author{Ken Hsieh}
\email[]{kenhsieh@pa.msu.edu}
\affiliation{Department of Physics and Astronomy,
Michigan State University, East Lansing, Michigan 48824, USA}
\author{Kai Schmitz}
\email[]{kai.schmitz@desy.de}
\affiliation{DESY, Theory Group, Notkestrasse 85, D-22607 Hamburg, Germany}
\affiliation{Department of Physics and Astronomy,
Michigan State University, East Lansing, Michigan 48824, USA}

\author{Jiang-Hao Yu}
\email[]{yujiangh@msu.edu}
\affiliation{Department of Physics and Astronomy,
Michigan State University, East Lansing, Michigan 48824, USA}

\author{C.--P. Yuan}
\email[]{yuan@pa.msu.edu}
\affiliation{Department of Physics and Astronomy,
Michigan State University, East Lansing, Michigan 48824, USA}

\preprint{DESY 09-205}
\preprint{MSUHEP-091123}

\begin{abstract}
We present the results of a global analysis of a class of models with
an extended electroweak gauge group of the form $SU(2)\times SU(2)\times U(1)$,
often denoted as $\tto$ models,
which include as examples the left-right, the lepto-phobic, the hadro-phobic, the fermio-phobic, the un-unified,
and the non-universal models.
Using an effective Lagrangian approach, we compute the shifts to
the coefficients in the electroweak Lagrangian due to the new heavy gauge
bosons, and obtain the lower bounds on the masses of the $Z^\prime$ and $W^\prime$
bosons.
The analysis of the electroweak parameter bounds reveals a consistent pattern of
several key observables that are especially sensitive to the effects of new physics and
thus dominate the overall shape of the respective parameter contours.
\end{abstract}

\date{\today}
\maketitle
\tableofcontents

\section{Introduction}
Despite the tremendous success of the Standard Model, there are still
open questions that are unanswered and motivate further model-building.
One of the most common model-building tools is to extend the
gauge structure of the Standard Model.
The simplest extension involves an additional $U(1)_X$ gauge symmetry
(and thus an extra gauge boson $Z^{\prime}$).
One of the next-simplest extensions involves an additional $SU(2)$,
with the left-right model \cite{Mohapatra:1974gc}\cite{Mohapatra:1974hk}\cite{Mohapatra:1980yp}
being perhaps the most widely-studied case of such models.
On the other hand, given the extended gauge
group $SU(2)_1\times SU(2)_2\times U(1)_X$ in the electroweak sector,
there are many other models
besides the left-right model that can be constructed,
and these models, despite having a common fundamental gauge group,
may have very different low-energy phenomenology.
In this paper we present a unified, systematic study of
many such models, which are commonly called $\tto$ models in the literature.

The most important feature of $G(221)$ models is the existence of new heavy gauge bosons, $W'$ and $Z'$.
The existence of the gauge boson $Z'$ has influences on the
low-energy neutral-current processes, the $Z$-pole data at LEP-I and
high energy LEP-II data~\cite{Rizzo:2006nw}\cite{Langacker:2008yv}.
The existence of the $W^\prime$ boson has implications to the search of new physics beyond the Standard Model (SM)
via studying charged-current processes.
In low energy experiments, the most sensitive probes of charged currents come from flavor physics,
such as the $K\bar{K}$, $b\bar{b}$ mixing processes
and semileptonic decays of the $b$ quark~\cite{Langacker:1989xa}\cite{Barenboim:1996nd}.
However, the low energy impact depends sensitively on the details of
the flavor sectors,
for which there is little experimental input~\cite{Rizzo:1997ts}.
There is thus a large uncertainty on the constraints on $W'$ and its
interactions.

In this paper, we classify the $G(221)$ models by the patterns of symmetry breaking
summarized in Table~\ref{tb:P-summary} (see section~\ref{sec:Models}).
Our main goals are to obtain the bounds on the masses of the $W'$
and $Z'$ bosons for these various models, and, through the results
of the global-fit analysis, to identify the key observables that are
most sensitive to the new physics in these models.
Our key results are that,  at the 95\% confidence level, the lower bounds on the masses
of new heavy gauge bosons can be very light  for breaking pattern I,
which includes left-right, lepto-phobic, hadro-phobic and fermio-phobic models, for example,
$M_{Z'}\sim 1.6~\mbox{TeV}$ and $M_{W'}\sim 0.3~\mbox{TeV}$ in the left-right model and hadro-phobic model; $M_{Z'}\sim 1.7~\mbox{TeV}$ and $M_{W'}\sim 0.7~\mbox{TeV}$ in the lepto-phobic and  fermio-phobic models. In breaking pattern II,
which includes un-unified and non-universial models,
because of the degeneracy of the masses of the $W'$ and $Z'$, the lower bounds on their masses are quite heavy, for example,
$M_{Z'}=M_{W'}\sim 2.5~\mbox{TeV}$ in the un-unified model.

We organize this paper as follows.
In Section \ref{sec:Models}, we lay out the various $\tto$ models and
discuss the results of the relevant literature.
In Section \ref{sec:EffLang}, we give the effective Lagrangians, both
at the electroweak scale (obtained by integrating out $W'$ and $Z'$)
and below the electroweak scale (by integrating out the $W$ and $Z$).
In Section \ref{sec:fit}, we discuss the global-fit procedure and
present our results obtained using the code \textit{Global Analysis
for Particle Properties} (GAPP) \cite{Erler:1999ug}, a software that
utilizes the CERN library MINUIT \cite{James:1975dr} and was used
for the Particle Data Group global analysis~\cite{Amsler:2008zzb}.
We also discuss which observables are especially sensitive to the
new physics contributions in these various models.
We conclude in Section \ref{sec:conc} with a summary and outlook of our key
findings.
The Appendix contains the explicit effective Lagrangians for
the $\tto$ models.

\section{The $\tto$ Models}
\label{sec:Models}
We focus on the so-called $\tto$ models having a
$SU(2)_1\times SU(2)_2\times U(1)_{X}$ gauge structure
that ultimately breaks to $U(1)_{\tbox{em}}$.
Relative to the Standard Model, these models have three additional
massive gauge bosons, and their phenomenology depends
on the specific patterns of symmetry breaking as well as
the charge assignments of the SM fermions.
For our studies, we consider the following different
$\tto$ models:
left-right (LR) \cite{Mohapatra:1974gc}\cite{Mohapatra:1974hk}\cite{Mohapatra:1980yp},
lepto-phobic (LP),
hadro-phobic (HP),
fermio-phobic (FP) \cite{Chivukula:2006cg}\cite{Barger:1980ix}\cite{Barger:1980ti},
un-unified (UU) \cite{Georgi:1989ic}\cite{Georgi:1989xz}, and non-universal (NU) \cite{Malkawi:1996fs}\cite{Li:1981nk}\cite{He:2002ha}.
The charge assignments of the SM fermions in these models
are given in Table~\ref{tb:models},
and these models
can be categorized by
two patterns of symmetry breaking
(summarized in Table \ref{tb:P-summary}):
\begin{itemize}
\item
Breaking pattern I (the LR, LP, HP, and FP models):\\
We identify $SU(2)_1$ as
$SU(2)_L$ of the SM.
The first stage of symmetry breaking then is
$SU(2)_2\times U(1)_X\rightarrow U(1)_Y$, giving
rise to three heavy gauge bosons $W^{\prime\pm}$ and $Z^{\prime}$
at the TeV-scale.
The second stage is $SU(2)_L\times  U(1)_Y\rightarrow U(1)_{\tbox{em}}$
at the electroweak scale.

\item
Breaking pattern II (the UU and NU models):\\
We identify $U(1)_X$ as $U(1)_Y$ of the SM.
The first stage of symmetry breaking is
$SU(2)_1\times SU(2)_2 \rightarrow SU(2)_L$.
The second stage is $SU(2)_L\times  U(1)_Y\rightarrow U(1)_{\tbox{em}}$
at the electroweak scale.
\end{itemize}

\begin{table}[h]
\begin{center}
\caption{
The charge assignments of the SM fermions under
the $\tto$ gauge groups.
Unless otherwise specified, the charge assignments
apply to all three generations.
}
\label{tb:models}
\vspace{0.125in}
\begin{tabular}{|c|c|c|c|}
\hline Model & $SU(2)_1$ & $SU(2)_2$ & $U(1)_{X}$ \\
\hline
Left-right (LR) &
$\begin{pmatrix} \uL \\ \dL \end{pmatrix}, \begin{pmatrix} \nL \\ \eL \end{pmatrix}$ &
$\begin{pmatrix} \uR \\ \dR \end{pmatrix}, \begin{pmatrix} \nR \\ \eR \end{pmatrix}$ &
$\begin{matrix} \tfrac{1}{6}\ \mbox{for quarks,} \\ -\tfrac{1}{2}\ \mbox{for leptons.} \end{matrix}$
\\
\hline
Lepto-phobic (LP) &
$\begin{pmatrix} \uL \\ \dL \end{pmatrix}, \begin{pmatrix} \nL \\ \eL \end{pmatrix}$ &
$\begin{pmatrix} \uR \\ \dR \end{pmatrix}$ &
$\begin{matrix} \tfrac{1}{6}\ \mbox{for quarks,} \\ Y_{\tbox{SM}}\ \mbox{for leptons.} \end{matrix}$
\\
\hline
Hadro-phobic (HP) &
$\begin{pmatrix} \uL \\ \dL \end{pmatrix}, \begin{pmatrix} \nL \\ \eL \end{pmatrix}$ &
$\begin{pmatrix} \nR \\ \eR \end{pmatrix}$ &
$\begin{matrix} Y_{\tbox{SM}}\ \mbox{for quarks,} \\ -\tfrac{1}{2}\ \mbox{for leptons.} \end{matrix}$
\\
\hline
Fermio-phobic (FP) &
$\begin{pmatrix} \uL \\ \dL \end{pmatrix}, \begin{pmatrix} \nL \\ \eL \end{pmatrix}$ &
 &
$ \begin{matrix} Y_{\tbox{SM}}\ \mbox{for all fermions.} \end{matrix}$
\\
\hline
Un-unified (UU) &
$\begin{pmatrix} \uL \\ \dL \end{pmatrix}$ &
$\begin{pmatrix} \nL \\ \eL \end{pmatrix}$ &
$ \begin{matrix} Y_{\tbox{SM}}\ \mbox{for all fermions.} \end{matrix}$
\\
\hline
Non-universal (NU) &
$\begin{pmatrix} \uL \\ \dL \end{pmatrix}_{1^{\tbox{st}},2^{\tbox{nd}}},
 \begin{pmatrix} \nL \\ \eL \end{pmatrix}_{1^{\tbox{st}},2^{\tbox{nd}}}$ &
$\begin{pmatrix} \uL \\ \dL \end{pmatrix}_{3^{\tbox{rd}}},
 \begin{pmatrix} \nL \\ \eL \end{pmatrix}_{3^{\tbox{rd}}}$ &
$ \begin{matrix} Y_{\tbox{SM}}\ \mbox{for all fermions.} \end{matrix}$
\\
\hline
\end{tabular}
\end{center}
\end{table}
\begin{table}[h]
\begin{center}
\caption{
Summary of the two different breaking patterns
and the two different stages of symmetry breaking
in $\tto$ models.
}
\label{tb:P-summary}
\vspace{0.125in}
\begin{tabular}{|l|l|l|l|}
\hline  Pattern &  Starting Point & First stage breaking & Second stage breaking \\
\hline  I & Identify $SU(2)_1$ as $SU(2)_L$ & $SU(2)_2\times
U(1)_X\rightarrow U(1)_Y$ & $SU(2)_L\times U(1)_Y\rightarrow
U(1)_{em}$
\\
\hline  II & Identify  $U(1)_X$ as $U(1)_Y$ & $SU(2)_1\times
SU(2)_2\rightarrow SU(2)_L$ & $SU(2)_L\times U(1)_Y\rightarrow
U(1)_{em}$
\\
\hline
\end{tabular}
\end{center}
\end{table}

In addition to specifying the gauge group and the fermion
charge assignments, a complete $\tto$ model should also
include the ingredients of the Higgs sectors and the Yukawa couplings.
While the observed relationships between the masses of $W$ and $Z$
bosons leave little freedom in the Higgs representation used for
electroweak symmetry breaking (EWSB), we have freedoms in the choices of the Higgs representation
used to break the fundamental $\tto$ gauge group to the SM gauge
group.
In  breaking pattern I
we assume the two simplest cases of symmetry breaking: via
a doublet or a triplet Higgs.
In the breaking pattern II
we assume the simplest case of using a bi-doublet Higgs to achieve
this symmetry breaking.
The model-specific Higgs representations and vacuum expectation values (VEV's)
are given in Table \ref{tb:Higgs-stage1}.
For heavy Higgs boson, Wang {\it et al.}~\cite{Wang:2008nk} used a non-linear effective theory
approach to obtain an electroweak chiral Lagrangian for $W^\prime$. In our paper, by assuming a light Higgs,
we analyze the low-energy constraints by using a linearlized effective Lagrangian approach.
%

\begin{table}[h!]
\begin{center}
\vspace{0.125in}
\begin{tabular}
{|l|c|c|}
\hline
\multicolumn{3}{|c|}{First stage breaking}
\\
\hline
 & Rep. & Multiplet and VEV
\\
\hline
\begin{minipage}{1.2in}
LR-D, LP-D \\
HP-D, FP-D
\end{minipage} &
$\Phi \sim (1,2,\tfrac{1}{2})$ &
$\Phi = \begin{pmatrix} \phi^+ \\ \phi^0\end{pmatrix},\
\VEV{\Phi} =
\mfrac{1}{\sqrt{2}}
\begin{pmatrix} 0 \\ \uD \end{pmatrix}$
\\
\hline
\begin{minipage}{1.2in}
LR-T, LP-T \\
HP-T, FP-T
\end{minipage} &
$\Phi \sim (1,3,1)$ &
$\Phi=\frac{1}{\sqrt{2}}
\begin{pmatrix}\phi^{+} &\sqrt{2}\phi^{++} \\ \sqrt{2}\phi^{0} & -\phi^{+} \end{pmatrix},
\
\VEV{\Phi} = \mfrac{1}{\sqrt{2}}
\begin{pmatrix}0 & 0 \\ \uT & 0 \end{pmatrix}$
\\
\hline
\begin{minipage}{1.2in}
UU, NU
\end{minipage}
&
$\Phi \sim (2,\overline{2},0)$ &
$\Phi = \begin{pmatrix} \phi^{0}+\pi^0  & \sqrt{2}\pi^+ \\ \sqrt{2}\pi^- & \phi^0-\pi^0 \end{pmatrix},\
\VEV{\Phi} =
\mfrac{1}{\sqrt{2}}
\begin{pmatrix} \tilde{u} & 0 \\0 & \tilde{u} \end{pmatrix}$
\\
\hline
\end{tabular}
\vspace{0.25in}
\begin{tabular}
{|l|c|c|}
\hline
\multicolumn{3}{|c|}{Second stage breaking}
\\
\hline
 & Rep. & Multiplet and VEV
\\
\hline
\begin{minipage}{1.2in}
LR-D, LP-D \\
HP-D, FP-D
\end{minipage} &
$H \sim (2,\overline{2},0)$ &
$H=\begin{pmatrix}h_1^{0} & h_1^{+} \\ h_2^{-} & h_2^{0} \end{pmatrix}$,
$\VEV{H} = \mfrac{\tilde{v}}{\sqrt{2}}
\begin{pmatrix} c_{\tbeta} & 0 \\ 0 & s_{\tbeta} \end{pmatrix}$
\\
\hline
\begin{minipage}{1.2in}
LR-T, LP-T \\
HP-T, FP-T
\end{minipage} &
$H \sim (2,\overline{2},0)$ &
$H=\begin{pmatrix}h_1^{0} & h_1^{+} \\ h_2^{-} & h_2^{0} \end{pmatrix}$,
$\VEV{H} = \mfrac{\tilde{v}}{\sqrt{2}}
\begin{pmatrix} c_{\tbeta} & 0 \\ 0 & s_{\tbeta} \end{pmatrix}$
\\
\hline
\begin{minipage}{1.2in}
UU, NU
\end{minipage}
&
$H \sim (1,2,\tfrac{1}{2})$ &
$H=\begin{pmatrix} h^{+} \\ h^{0} \end{pmatrix}$,
$\VEV{H} = \mfrac{\tilde{v}}{\sqrt{2}}
\begin{pmatrix} 0  \\ 1 \end{pmatrix}$
\\
\hline
\end{tabular}
\caption{
These tables display the model-specific Higgs representations
and VEVs that achieve the symmetry breaking of $\tto$ models.
}
\label{tb:Higgs-stage1}
\end{center}
\end{table}

The lepto-phobic, hadro-phobic, and un-unified models are, with the current set-up,
incomplete because of gauge anomalies.
It is entirely possible that the additional matter content used to
address the anomalies can alter the low-energy phenomenologies and
the results of our studies.
Nevertheless, for completeness, we include these models with the current set-up in our studies, in which we focus on effects originated from the interactions of $W^\prime$ and $Z^\prime$ bosons to the SM fields.
In the cases of the lepto-phobic and hadro-phobic models, one can view them
as transitions between the left-right models (where both right-handed leptons and
quarks
are charged under $SU(2)_2$) and the fermio-phobic model (where neither are
charged).

There have already been many theoretical and phenomenological studies of various $\tto$ models,
and we focus our brief literature review here mainly to those works that perform a global fitting
in the same spirit as our work.
In the symmetric left-right model (where the couplings of the $W'$ are
of the same strength as those of the $W$), Polak and Zralek obtained the constraints on parameters from
the Z-pole data~\cite{Polak:1991qw} and low energy data~\cite{Polak:1991pc}, separately.
While for the non-symmetric case, Chay, Lee and
Nam~\cite{Chay:1998hd} considered phenomenological constraints on
three parameters: the mass of the $Z'$, the mixing angles $\tilde{\phi}$
(the analog of the Weinberg angle in the breaking of
$SU(2)_R\times U(1)_X\rightarrow U(1)_Y$)
and the $Z$-$Z^{\prime}$ mixing angle $\xi$, by combining the precision electroweak data
from LEP I (through $\epsilon_1, \epsilon_2, \epsilon_3$) and the
low-energy neutral-current experimental data.
For the
non-symmetric case, the combined bounds at the $95\%$ confidence
level are $0.0028 < \xi < 0.0065 $ and $M_{Z'} \geq 400 $ GeV
for all $\tilde{\phi}$, while for the symmetric case, a more
severe bound $M_{Z'} \geq 1.6 $ TeV is obtained.

In the fermio-phobic model, Donini {\it et
al.}~\cite{Donini:1997yu} used the $Z$-pole and low-energy data, and
the flavor physics data from flavor-changing neutral-current (FCNC)
processes and $b\to s\gamma$, to put constraints on the parameter
space ($W$-$W^{\prime}$ mixing angle $\alpha_{\pm}$, and
$Z$-$Z^{\prime}$ mixing angle $\alpha_{0}$) by fixing several sets
of representative values of $M_{W^{\prime}}$ and $x$ (strength of
the coupling of the fermiophobic gauge group, relative to $SU(2)_L$
of the Standard Model).
For the input parameters in the the ranges $100$ GeV $<
M_{W^{\prime}} < 1000$ GeV and $0.6 < x < 15$, and for a low Higgs
mass of $100$ GeV, the best-fit values of $\left|\alpha_{0}\right|$
and $\left|\alpha_{\pm}\right|$ increases with increasing $x$, when
holding $M_{W^{\prime}}$ fixed.
On the other hand, when holding $x$ fixed, increasing $M_{W^{\prime}}$
leads to an increase in the best-fit values of $\left|\alpha_{0}\right|$ and
a decrease in the best-fit values of $\left|\alpha_{\pm}\right|$.
In the entire range of parameter space, the magnitude of
the best-fit values of $\alpha_{0}$ and
$\alpha_{\pm}$ are at the percent level.

In the non-unified model, Malkawi and Yuan~\cite{Malkawi:1996fs}
performed a global fit of the parameter space $(x, \phi)$ using the
Z-pole data, and found that the lower bound is $M_{Z'} = M_{W'}\geq
1.3 \text{ TeV}$ if no flavor physics is included.
Chivukula et.~al~\cite{Chivukula:1994qw} used the
data from precision electroweak measurements to put stringent
bounds on the un-unified Standard Model \cite{Georgi:1989ic}
\cite{Georgi:1989xz}.
They found a lower bound
on the masses of the heavy $W^\prime$ and $Z^\prime$ of approximately $2$ TeV at the
$95\%$ confidence level.

\section{The Effective Lagrangian Approach}
\label{sec:EffLang}
%
To analyze the low-energy constraints, we will
take an effective Lagrangian approach,
and follow the general procedures laid out
by Burgess in Ref.~\cite{Burgess:1993vc}
to extract the effects of new physics.
Although the details of each of these models are different, we first
perform a generic analysis that can be applied to any $\tto$ model
we consider in this work.

Per the convention in Burgess \cite{Burgess:1993vc},
we denote the gauge couplings as $\tgo$, $\tgt$,
and $\tgX$ respectively for the gauge groups $SU(2)_1$,
$SU(2)_2$, and $U(1)_X$.
The tilde ($\tilde{\ }$) on the couplings and VEVs emphasizes the fact that
these are
model parameters, as opposed to quantities that can be directly measured in experiments,
such as the physical mass of the $Z$ boson.
As an extension to the  convention in Burgess \cite{Burgess:1993vc},
we also denote with tilde ($\tilde{\ }$) any combination constructed
from the model parameters.
We also abbreviate the trigonometric functions
\begin{align}
c_{x} \equiv \cos(x),\ s_{x} \equiv \sin(x),\ \mbox{and}\
t_{x}\equiv \tan(x).
\end{align}

\subsection{Mixing Angles and Gauge Couplings}
%
We define the mixing angle $\tphi$ at the first breaking stage as
\begin{align}
t_{\tphi} \left(=\tan\tphi\right)&\equiv
\begin{cases}
\tgX/\tgt & \mbox{(LR, LP, HP, FP models)}
\\
\tgt/\tgo & \mbox{(UU, NU models)},
\end{cases}
\end{align}
and define the couplings
\begin{align}
\tgL &\equiv
\begin{cases}
\tgo, & \mbox{(LR, LP, HP, FP models)}
\\
\left(\mfrac{1}{\tgo^2}+\mfrac{1}{\tgt^2}\right)^{-1/2} &
\mbox{(UU, NU models)},
\end{cases}
\nonumber\\
\tgY &\equiv
\begin{cases}
\left(\mfrac{1}{\tgt^2}+\mfrac{1}{\tgX^2}\right)^{-1/2}
&
\mbox{(LR, LP, HP, FP models)}
\\
\tgX, & \mbox{(UU, NU models)}.
\end{cases}
\label{eq:gYgL-def}
\end{align}
The couplings $\tgL$ and $\tgY$ are respectively the gauge couplings
of the unbroken $SU(2)_L\times U(1)_Y$ gauge groups after the first
stage of symmetry breaking.
 Similarly to the Standard Model, for both breaking patterns we define
the weak mixing angle ($\ttheta$) as
\begin{align}
t_{\ttheta}\left(=\tan\ttheta\right) \equiv \frac{\tgY}{\tgL}.
\end{align}
For both breaking patterns, the electric charge ($\tilde{e}$) is given by
\begin{align}
\frac{1}{\tilde{e}^2}
=\frac{1}{\tgo^2}+\frac{1}{\tgt^2}+\frac{1}{\tgX^2},
\label{eq:electric-charge}
\end{align}
and we also define $\tilde{\alpha}_{e} \equiv \tilde{e}^2/4\pi$.

With the angles $\ttheta$ and $\tphi$, the gauge couplings can be expressed as
\begin{align}
\tgo &= \begin{cases}
\tilde{e}/(s_{\ttheta}), & \mbox{(LR, LP, HP, FP models)}
\\
\tilde{e}/(s_{\ttheta}s_{\tphi}), & \mbox{(UU, NU models)}
\end{cases}
\label{eq:GaugeCouplingExpand0}
\\
\tgt &= \begin{cases}
\tilde{e}/(c_{\ttheta}s_{\tphi}), & \mbox{(LR, LP, HP, FP models)}
\\
\tilde{e}/(s_{\ttheta}c_{\tphi}), & \mbox{(UU, NU models)}
\end{cases}
\label{eq:GaugeCouplingExpand1}
\\
\tgX &= \begin{cases}
\tilde{e}/(c_{\ttheta}c_{\tphi}), & \mbox{(LR, LP, HP, FP models)}
\\
\tilde{e}/(c_{\ttheta}). & \mbox{(UU, NU models)}
\label{eq:GaugeCouplingExpand}
\end{cases}
\end{align}

\subsection{The Effective Lagrangian}

\subsubsection{Gauge Interactions of Fermions}

In this sub-section we parameterize the gauge interactions of the fermions
that is applicable to all the $\tto$ models under considerations  here.
We will obtain both the SM-like effective theory  applicable
at the electroweak scale
as well as the four-fermion effective theory below the electroweak
scale.
We do this by first building up the fundamental Lagrangian in stages, and then
successively integrating out the massive gauge bosons.
The $Z$-pole data measured at the electroweak scale, and measurements of the four-fermion neutral-current interactions
are some of the most precise measurements to-date, and provide
stringent bounds on new physics models.

As discussed earlier, we consider the symmetry breaking to take
two stages:
\begin{align}
SU(2)_1\times SU(2)_2\times U(1)_X \rightarrow
SU(2)_L\times U(1)_Y\rightarrow
U(1)_{\tbox{em}}.
\end{align}
We denote the gauge bosons of the $\tto$ models as:
\begin{align}
SU(2)_1&: W_{1,\mu}^{\pm}, W_{1,\mu}^{3},\nonumber\\
SU(2)_2&: W_{2,\mu}^{\pm}, W_{2,\mu}^{3},\nonumber\\
U(1)_X&: X_{\mu}.
\end{align}
After the first-stage breaking, the neutral gauge eigenstates mix as follows
\begin{align}
\hat{B}_{\mu}
&\equiv
\begin{cases}
s_{\tphi} W^3_{2,\mu}+c_{\tphi} X_{\mu}
&
\mbox{(LR, LP, HP, FP models)}
\\
X_{\mu}
&
\mbox{(UU, NU models)}
\end{cases}
\nonumber
\\
\hat{W}^{3}_{\mu}
&\equiv
\begin{cases}
W_{1,\mu}^{3}
&
\mbox{(LR, LP, HP, FP models)}
\\
s_{\tphi} W^3_{1,\mu}+c_{\tphi}W^3_{2,\mu}
&
\mbox{(UU, NU models)}
\end{cases}
\nonumber
\\
\hat{Z}^{\prime}_{\mu}
&\equiv
\begin{cases}
c_{\tphi} W^3_{2,\mu}-s_{\tphi} X_{\mu}
&
\mbox{(LR, LP, HP, FP models)}
\\
c_{\tphi} W^3_{1,\mu}-s_{\tphi}W^3_{2,\mu},
&
\mbox{(UU, NU models)}
\end{cases}
\label{eq:gauge-boson-basis1}
\end{align}
and for the charged gauge bosons, we have
\begin{align}
\hat{W}_{\mu}^{\pm}
&\equiv
\begin{cases}
W^{\pm}_{1,\mu}
&
\mbox{(LR, LP, HP, FP models)}
\\
s_{\tphi} W^{\pm}_{1,\mu}+ c_{\tphi}W^{\pm}_{2,\mu},
&
\mbox{(UU, NU models)}
\end{cases}
\nonumber
\\
\hat{W}^{\prime\pm}_{\mu}
&\equiv
\begin{cases}
W^{\pm}_{2,\mu},
&
\mbox{(LR, LP, HP, FP models)}
\\
c_{\tphi} W^{\pm}_{1,\mu}- s_{\tphi}W^{\pm}_{2,\mu}.
&
\mbox{(UU, NU models)}
\end{cases}
\label{eq:gauge-boson-basis2}
\end{align}
After the first stage of symmetry breaking, there is still an unbroken $SU(2)_L\times U(1)_Y$,
which may be identified as the Standard Model gauge group.
The gauge bosons $\hat{W}^{\pm,3}$
and $\hat{B}$ are massless, and only $\hat{Z}^{\prime}$ and $\hat{W}^{\prime\pm}$ are massive,
with TeV-scale masses.
The Lagrangian representing
the heavy gauge boson masses
has the form
\begin{align}
\mathcal{L}^{\tbox{stage-1}} &=
 \frac{1}{2}\tM^2_{Z^{\prime}}\hat{Z}^{\prime}_{\mu}\hat{Z}^{\prime\mu}
+ \tM^2_{W^{\prime}} \hat{W}^{\prime +}_{\mu} \hat{W}^{\prime-\mu},
\label{eq:L-stage1}
\end{align}
where $\tM^2_{Z^{\prime}}$ and $\tM^2_{W^{\prime}}$ are given in Table~\ref{tb:MZMW-stage1}.

Before discussing the second stage of symmetry breaking, it
is convenient to define, similarly to the Standard Model, $A_{\mu}$ (which will turn out to be the photon)
and $\hat{Z}_{\mu}$ (approximately the physical $Z$-boson)
in terms of the massless gauge bosons $\hat{W}^3_{\mu}$ and $\hat{B}_{\mu}$
\begin{align}
A_{\mu}
&\equiv
\left(
\frac{\tilde{e}}{\tgo} W^3_{1,\mu} +
\frac{\tilde{e}}{\tgt} W^3_{2,\mu} +
\frac{\tilde{e}}{\tgX} \hat{X}_{\mu}
\right),
\nonumber
\\
&=
s_{\ttheta} \hat{W}^3_{\mu} +
c_{\ttheta} \hat{B}_{\mu},
\nonumber
\\
\hat{Z}_{\mu}
&\equiv
c_{\ttheta} \hat{W}^3_{\mu} -
s_{\ttheta}\hat{B}_{\mu},
\label{eq:gauge-boson-basis6}
\end{align}

At the electroweak scale, the second stage of symmetry breaking
takes place, breaking $SU(2)\times U(1)\rightarrow U(1)_{\tbox{em}}$.
The Higgs vacuum expectation value (VEV) at the second stage not only
gives masses to $\hat{Z}$ and $\hat{W}^{\pm}$, but also induces
further mixing among the gauge bosons $\hat{W}^{\pm}$, $\hat{Z}$, $\hat{W}^{\prime}$ and
$\hat{Z}^{\prime}$.
The masses of the gauge bosons depend not only
on the breaking pattern, but also on the group representations
of the Higgs bosons whose VEV's trigger
the symmetry breaking.
For simplicity, for  breaking pattern I, we consider only
 models with either a
doublet or triplet under $SU(2)_2$, and do not consider models with
\textit{both} doublets and triplets.
Introducing additional Higgses and VEVs would modify the masses
of the $W^{\prime}$ and $Z^{\prime}$ \cite{Polak:1991vf}.
For  breaking pattern II, since the
first stage of symmetry breaking breaks
$SU(2)_1\times SU(2)_2$ to the diagonal subgroup, the masses of $W^{\prime}$ and
$Z^{\prime}$ are degenerate at this stage, and we only consider
the case of an $SU(2)_1\times SU(2)_2$ bi-doublet.
%
%
%
For the convenience of typesetting, we also denote, for example, a
left-right model with first-stage symmetry breaking triggered by an
$SU(2)$-doublet(-triplet) as LR-D (LR-T).

Although different breaking patterns and different group representations of the Higgs bosons will lead to different Lagrangians, we can write down
the Lagrangian involving  the gauge boson masses and fermionic gauge interactions in a general form
\begin{align}
\mathcal{L}_{\tbox{fund}} &=
  \frac{1}{2}\tM^2_{Z}\hat{Z}_{\mu}\hat{Z}^{\mu}
+ \frac{1}{2}(\tM^2_{Z^{\prime}}+\Delta\tM^2_{Z^{\prime}})
  \hat{Z}^{\prime}_{\mu}\hat{Z}^{\prime\mu}
+ \delta \tM^2_{Z} \hat{Z}^{\prime}_{\mu} \hat{Z}^{\mu}
\nonumber\\
&\
+ \tM^2_{W} \hat{W}^{+}_{\mu}\hat{W}^{-\mu}
+ (\tM^2_{W^{\prime}}+\Delta \tM^2_{W^{\prime}})
   \hat{W}^{\prime +}_{\mu} \hat{W}^{\prime-\mu}
+ \delta \tM^2_{W}
(\hat{W}^{\prime+}_{\mu}\hat{W}^{-\mu} + \hat{W}^{\prime-}_{\mu}\hat{W}^{+\mu})
\nonumber\\
&\
+ \hat{W}_{\mu}^{\prime+} K^{+\mu}
+ \hat{W}_{\mu}^{\prime-} K^{-\mu}
+ \hat{Z}_{\mu}^{\prime}  K^{0\mu} \nonumber\\
&\
+ \hat{W}_{\mu}^{+} J^{+\mu}
+ \hat{W}_{\mu}^{-} J^{-\mu}
+ \hat{Z}_{\mu} J^{0\mu} + A_\mu J^\mu,
\label{eq:L-stage2}
\end{align}
where we have denoted the currents that couple
to the primed gauge bosons ($\hat{W}^{\prime}$ and $\hat{Z}^{\prime}$)
as $K^0_{\mu}$ and $K_{\mu}^{\pm}$, and the currents that couple to the SM
gauge bosons as $J_{\mu}$, $J^0_{\mu}$ and $J^{\pm}_{\mu}$.
The SM-like currents have the familiar forms
\begin{align}
J^\mu &= \tilde{e} \sum\limits_{f} Q^f \overline{f} \gamma^\mu f,\\
J^0_{\mu} &= \sqrt{\tgL^2+\tgY^2}\ \sum\limits_{f}
\left(
T_{3L}^{f}\overline{f}_{\tbox{L}}\gamma_{\mu}\PL f_{\tbox{L}}
-
s_{\ttheta}^2 Q^f\  \overline{f}\gamma_{\mu}f
\right),
\\
J^{+}_{\mu}
&=\frac{\tgL}{\sqrt{2}}
\left(
\overline{d}_{\tbox{L}}\gamma_{\mu}\PL u_{\tbox{L}}
+
\overline{e}_{\tbox{L}}\gamma_{\mu}\PL \nu_{\tbox{L}}
\right),
\end{align}
with an implicit sum over the three generations of fermions.
The neutral currents ($K^0_{\mu}$) and charged currents ($K_{\mu}^{\pm}$),
for the various
models are summarized in Tables \ref{tb:Zp-couplings}
and \ref{tb:Wp-couplings}.
\begin{table}
\begin{center}
\vspace{0.125in}
\begin{tabular}
{|c|c|c|c|c|}
\hline
\  &
$\overline{u}  \gamma^{\mu} u$ &
$\overline{d}  \gamma^{\mu} d$ &
$\overline{\nu}\gamma^{\mu} \nu$ &
$\overline{e}  \gamma^{\mu} e$ \\
\hline
LR &
$\begin{matrix}( \tfrac{1}{2}\cphi \tgt - \tfrac{1}{6}\sphi\tgX)\PR \\
- \frac{1}{6}\sphi\tgX\PL \end{matrix}$
&
$\begin{matrix}(-\tfrac{1}{2}\cphi \tgt - \tfrac{1}{6}\sphi\tgX)\PR\\
 - \frac{1}{6}\sphi\tgX\PL\end{matrix}$ &
$\begin{matrix}( \tfrac{1}{2}\cphi \tgt + \tfrac{1}{2}\sphi\tgX)\PR\\
 + \frac{1}{2}\sphi\tgX\PL \end{matrix}$ &
$\begin{matrix}(-\tfrac{1}{2}\cphi \tgt + \tfrac{1}{2}\sphi\tgX)\PR\\
 + \frac{1}{2}\sphi\tgX\PL \end{matrix}$
\\
\hline
LP &
$\begin{matrix}( \tfrac{1}{2}\cphi \tgt - \tfrac{1}{6}\sphi\tgX)\PR \\
- \frac{1}{6}\sphi\tgX\PL \end{matrix}$ &
$\begin{matrix}(-\tfrac{1}{2}\cphi \tgt - \tfrac{1}{6}\sphi\tgX)\PR \\
- \frac{1}{6}\sphi\tgX\PL \end{matrix}$ &
$ \frac{1}{2}\sphi\tgX\PL$ &
$ \sphi\tgX(\frac{1}{2}\PL + \PR)$
\\
\hline
HP &
$-\sphi\tgX( \tfrac{1}{6}\PL + \frac{2}{3}\PR)$ &
$-\sphi\tgX( \tfrac{1}{6}\PL - \frac{1}{3}\PR)$ &
$\begin{matrix} ( \tfrac{1}{2}\cphi \tgt + \tfrac{1}{2}\sphi\tgX)\PR \\
 + \frac{1}{2}\sphi\tgX\PL \end{matrix}$ &
$\begin{matrix}(-\tfrac{1}{2}\cphi \tgt + \tfrac{1}{2}\sphi\tgX)\PR \\
 + \frac{1}{2}\sphi\tgX\PL \end{matrix}$
\\
[2ex]
\hline
FP &
$-\sphi\tgX( \tfrac{1}{6}\PL + \frac{2}{3}\PR)$ &
$-\sphi\tgX( \tfrac{1}{6}\PL - \frac{1}{3}\PR)$ &
$ \frac{1}{2}\sphi\tgX\PL$ &
$ \sphi\tgX(\frac{1}{2}\PL + \PR)$
\\
[2ex]
\hline
\hline
UU &
$ \tfrac{1}{2}\cphi \tgo\PL$ &
$-\tfrac{1}{2}\cphi \tgo\PL$ &
$-\tfrac{1}{2}\sphi \tgt\PL$ &
$ \tfrac{1}{2}\sphi \tgt\PL$
\\
[2ex]
\hline
NU &
$ \frac{1}{2}\begin{pmatrix}\cphi \tgo \\ -\sphi \tgt\end{pmatrix} \PL$ &
$-\frac{1}{2}\begin{pmatrix}\cphi \tgo \\ -\sphi \tgt\end{pmatrix} \PL$ &
$ \frac{1}{2}\begin{pmatrix}\cphi \tgo \\ -\sphi \tgt\end{pmatrix} \PL$ &
$-\frac{1}{2}\begin{pmatrix}\cphi \tgo \\ -\sphi \tgt\end{pmatrix} \PL$
\\
\hline
\end{tabular}
\caption{
This table displays the couplings $\tilde{g}(\overline{f},f,\hat{Z}^{\prime})$ of the
current $K^{0\mu} =
\overline{f}\gamma^{\mu}\tilde{g}(\overline{f},f,\hat{Z}^{\prime})f$.
For the top four models (LR, LP, HP, and FP), $\tan\phi\equiv \tgX/\tgt$.
For the lower two models (UU and NU), $\tan\phi\equiv \tgt/\tgo$.
For the NU model (last row), the top values denote the couplings to
the first two generations of fermions, and the bottom values denote the
couplings to the third generation.
}
\label{tb:Zp-couplings}
\end{center}
\end{table}
\begin{table}
\begin{center}
\vspace{0.125in}
\begin{tabular}
{|c|c|c|}
\hline
\ &
$\overline{d}  \gamma^{\mu} u$&
$\overline{e}\gamma^{\mu} \nu$
\\
\hline
LR &
$\frac{1}{\sqrt{2}}\tgt\PR$ &
$\frac{1}{\sqrt{2}}\tgt\PR$
\\
\hline
LP &
$\frac{1}{\sqrt{2}}\tgt\PR$ &
0
\\
\hline
HP &
0 &
$\frac{1}{\sqrt{2}}\tgt\PR$
\\
\hline
FP &
0 &
0
\\
\hline
\hline
UU &
$ \tfrac{1}{\sqrt{2}}\cphi \tgo\PL$ &
$-\tfrac{1}{\sqrt{2}}\sphi \tgt\PL$
\\
\hline
NU &
$ \frac{1}{\sqrt{2}}\begin{pmatrix}\cphi \tgo \\ -\sphi \tgt\end{pmatrix} \PL$ &
$ \frac{1}{\sqrt{2}}\begin{pmatrix}\cphi \tgo \\ -\sphi \tgt\end{pmatrix} \PL$
\\
\hline
\end{tabular}
\caption{
This table displays the couplings $\tilde{g}(\overline{\psi},\xi,\hat{W}^{\prime +})$ of the
current $K^{+\mu} =  \overline{\psi} \gamma^{\mu}\tilde{g}(\overline{\psi},\xi,\hat{W}^{\prime +}) \xi$.
For the top four models (LR, LP, HP, and FP), $\tan\phi\equiv \tgX/\tgt$.
For the lower two models (UU and NU), $\tan\phi\equiv \tgt/\tgo$.
For the NU model (last row), the top values denote the couplings to
the first two generations of fermions, and the bottom values denote the
couplings to the third generation.
}
\label{tb:Wp-couplings}
\end{center}
\end{table}
We note the following features:
\begin{itemize}
\item
The residual $SU(2)_L\times U(1)_Y$ is broken to the $U(1)_{\tbox{em}}$,
and there are now mass terms for the $\hat{Z}$ and $\hat{W}$
bosons, denoted as $\tM_{Z,W}^2$.
These masses have the familiar form
\begin{align}
\tM_Z^2 &= \frac{1}{4}({\tgL}^2 + {\tgY}^2) \widetilde{v}^2,\\
\tM_W^2 &= \frac{1}{4}{\tgL}^2  \widetilde{v}^2,
\end{align}
where the couplings $\tgL$ and $\tgY$ are defined as in Eq.~\Eref{eq:gYgL-def}
for the two different breaking patterns.

%
\item
There are mass-mixing contributions
$\delta\tM^2_{Z,W}$ that induce $\hat{Z}-\hat{Z}^{\prime}$
and $\hat{W}-\hat{W}^{\prime}$ mixing. They are dependent on the breaking pattern and are given
in Table~\ref{tb:MZMW-stage1}.
%
%
\item
There are additional contributions to  the masses of the
$\hat{Z}^{\prime}$ and $\hat{W}^{\prime}$ after the second stage of
symmetry breaking, which we denote as
$\Delta\tM^2_{Z^{\prime},W^{\prime}}$.
They are also dependent on the breaking pattern and are given
in Table~\ref{tb:MZMW-stage1}.
\end{itemize}
\begin{table}
\begin{center}
\vspace{0.125in}
\begin{tabular}
{|l|c|c||c|c|c|c|}
\hline
\ &
$\tM^2_{Z^{\prime}}$ &
$\tM^2_{W^{\prime}}$ &
$\Delta\tM^2_{Z^{\prime}}$ &
$\Delta\tM^2_{W^{\prime}}$ &
$\delta\tM^2_{Z}$ &
$\delta\tM^2_{W}$
\\
\hline
\begin{minipage}{1.2in}
LR-D, LP-D \\
HP-D, FP-D
\end{minipage}
&
$\mfrac{1}{4}(\tgt^2+\tgX^2)\uD^2$ &
$\mfrac{1}{4}\tgt^2\uD^2$ &
$\mfrac{c^2_{\tphi}}{4}\tgt^2\tilde{v}^2$ &
$\mfrac{1}{4}\tgt^2 \tilde{v}^2$ &
$-\mfrac{c_{\tphi}^2}{4\tilde{e}}\tgo\tgt\tgX\tilde{v}^2$ &
$-\mfrac{1}{4}\tgo\tgt \tilde{v}^2 s_{2\tbeta}$
\\
\hline
\parbox{1.2in}{
LR-T, LP-T \\
HP-T, FP-T
}
&
$(\tgt^2+\tgX^2)\uT^2$ &
$\mfrac{1}{2}\tgt^2\uT^2$ &
$\mfrac{c^2_{\tphi}}{4}\tgt^2\tilde{v}^2$ &
$\mfrac{1}{4}\tgt^2 \tilde{v}^2$ &
$-\mfrac{c_{\tphi}^2}{4\tilde{e}}\tgo\tgt\tgX\tilde{v}^2$ &
$-\mfrac{1}{4}\tgo\tgt \tilde{v}^2 s_{2\tbeta}$
\\
\hline
\begin{minipage}{1.2in}
UU, NU
\end{minipage}
&
$\mfrac{1}{4}(\tgo^2+\tgt^2)\tu^2$ &
$\mfrac{1}{4}(\tgo^2+\tgt^2)\tu^2$ &
$\mfrac{s^2_{\tphi}}{4}\tgt^2 \tilde{v}^2$ &
$\mfrac{s^2_{\tphi}}{4}\tgt^2 \tilde{v}^2$ &
$-\mfrac{s_{\tphi}^2}{4\tilde{e}}\tgo\tgt\tgX\tilde{v}^2$ &
$-\mfrac{1}{4}\tgo\tgt \tilde{v}^2 s^2_{\tphi}$
\\
\hline
\end{tabular}
\caption{
This table displays the model-dependent parameters
$\tM^2_{Z^{\prime},W^{\prime}}$
in Eq.~\Eref{eq:L-stage1},
and
$\Delta\tM^2_{Z^{\prime},W^{\prime}}$ and
$\delta\tM^2_{Z,W}$ in Eq.~\Eref{eq:L-stage2}.
}
\label{tb:MZMW-stage1}
\end{center}
\end{table}
Therefore, the gauge boson mass terms can be written as
\begin{eqnarray}
\mathcal{L}_{\tbox{mass}} &=&
        \left( \begin{array}{cc}\hat{W}^{+}_{\mu} &  \hat{W}^{\prime +}_{\mu} \end{array} \right)
        \left( \begin{array}{cc}
            \tM^2_{W} & \delta \tM^2_{W} \\
            \delta \tM^2_{W} & \tM^2_{W^{\prime}}+\Delta \tM^2_{W^{\prime}}
         \end{array} \right)
        \left( \begin{array}{c}\hat{W}^{-\mu} \\  \hat{W}^{\prime -\mu} \end{array} \right)
    \nonumber\\
        &+&
        \frac{1}{2}
        \left( \begin{array}{ccc} A & \hat{Z}_{\mu} & \hat{Z}^\prime_{\mu} \end{array} \right)
        \left( \begin{array}{ccc}
            0 & 0
            & 0\\
            0 & \tM^2_{Z}
            & \delta \tM^2_{Z} \\
            0 & \delta \tM^2_{Z}
            & \tM^2_{Z^{\prime}}+\Delta\tM^2_{Z^{\prime}}
         \end{array} \right)
        \left( \begin{array}{c} A \\  \hat{Z}^{\mu} \\ \hat{Z}^{\prime\mu} \end{array} \right)
\,.
\end{eqnarray}
In Table~\ref{tb:Higgs-stage1}, we expect that the scale $\tilde{u}^2$
of the first-stage breaking is much larger than the electroweak scale $\tilde{v}^2$.
We work to leading order in $\tilde{v}^2/\tilde{u}^2$, and so if we take the approximation
\begin{align}
\tM^{2}_{Z^{\prime},W^{\prime}}
\gg
\tM^{2}_{Z,W},
\delta\tM^{2}_{Z,W},
\Delta\tM^{2}_{Z,W},
\end{align}
we can expand in large
$\tM^{2}_{Z^{\prime}, W^{\prime}}$.
To order $\mathcal{O}(\tM^{-2}_{W^{\prime},Z^{\prime}})$,
the mass eigenstates, denoted without the hats ($\hat{\ }$), are given by
(similarly for the charged gauge bosons):
\begin{align}
Z_{\mu} &\equiv \hat{Z}_{\mu} - \frac{ \delta \tM^2_{Z} }
{ \tM^2_{Z^{\prime}} - \tM^2_{Z} } \hat{Z}^{\prime}_{\mu}, \\
Z_{\mu}^{\prime} &\equiv  \frac{ \delta \tM^2_{Z} } {
\tM^2_{Z^{\prime}} - \tM^2_{Z} } \hat{Z}_{\mu}
+ \hat{Z}^{\prime}_{\mu}.
\end{align}
Now we can rewrite the fundamental Lagrangian in terms of the mass eigenstates for both neutral and charged gauge bosons
\begin{align}
\mathcal{L}^{\tbox{mass}}_{\tbox{fund}}
&=
\frac{1}{2}\left( \tM_Z^2 - \frac{ \delta \tM^4_{Z} }{ \tM^2_{Z^{\prime}}} \right)
Z_{\mu}Z^{\mu}
+ \left( \tM^2_{W} - \frac{ \delta \tM^4_{W} }{ \tM^2_{W^{\prime}}} \right)
W^{+\mu}W^{-}_{\mu}
\nonumber\\
&\ + \frac{1}{2}\left( \tM_{Z^\prime}^{ 2} + \Delta
\tM^2_{Z^{\prime}} + \frac{ \delta \tM^4_{Z} }{ \tM^2_{Z^{\prime}}}
\right) Z^\prime_{\mu}Z^{\prime\mu} + \left( \tM^2_{W^{\prime}} +
\Delta \tM^2_{W^{\prime}} + \frac{ \delta \tM^4_{W} }{
\tM^2_{W^{\prime}}} \right) W^{\prime+\mu}W^{\prime-}_{\mu}
\nonumber\\
&\
+ Z_{\mu}\left( J^{0\mu}- \frac{\delta \tM^2_{Z} }{\tM^2_{Z^{\prime}}} K^{0\mu} \right)
+ Z^\prime_{\mu}\left( K^{0\mu} + \frac{\delta \tM^2_{Z} }{\tM^2_{Z^{\prime}}} J^{0\mu} \right) +A_\mu J^\mu
\nonumber\\
&\
 + \left[ W_{\mu}^{+} \left( J^{+\mu}- \frac{\delta \tM^2_{W}}{ \tM^2_{W^{\prime}}} K^{+\mu} \right)
+ W_{\mu}^{\prime+} \left( K^{+\mu}+ \frac{\delta \tM^2_{W}}{ \tM^2_{W^{\prime}}} J^{+\mu} \right)
+ (+ \leftrightarrow - )\right].
\label{eq:L-ew-fund}
\end{align}

We can now obtain the effective Lagrangian by successively
integrating out the massive gauge bosons.
In the basis of the mass eigenstates,
integrating out $Z^{\prime}$ and $W^{\prime}$
(whose masses are expected to be at or above the TeV scale)
results in an effective Lagrangian
valid at the electroweak scale:
\begin{align}
\mathcal{L}^{\tbox{EW}}_{\tbox{eff}}
&=
\frac{1}{2}\left( \tM_Z^2 - \frac{ \delta \tM^4_{Z} }{ \tM^2_{Z^{\prime}}} \right)
Z_{\mu}Z^{\mu}
+ \left( \tM^2_{W} - \frac{ \delta \tM^4_{W} }{ \tM^2_{W^{\prime}}} \right)
W^{+\mu}W^{-}_{\mu}
\nonumber\\
&\
+ Z_{\mu}\left( J^{0\mu}- \frac{\delta \tM^2_{Z} }{\tM^2_{Z^{\prime}}} K^{0\mu} \right)
+ \left[ W_{\mu}^{+} \left( J^{+\mu}- \frac{\delta \tM^2_{W}}{ \tM^2_{W^{\prime}}} K^{+\mu} \right)
+ (+ \leftrightarrow - )\right]
\nonumber\\
&\
-\frac{1}{ 2 \tM^2_{Z^{\prime}} } K^{0\mu}K^0_{\mu}
-\frac{1}{ \tM^2_{W^{\prime}} } K^{+\mu}K^{-}_{\mu} +A_\mu J^\mu.
\label{eq:L-eff-stage1}
\end{align}
From Eq.~\Eref{eq:L-eff-stage1}, we see that the low-energy effects
of the heavy gauge bosons are parameterized
by the shifts in the masses of the $W$ and $Z$ gauge bosons, and in the shifts
of their couplings to the fermions, and additional
four-fermion interactions.

We can further integrate out the $Z$ and $W^{\pm}$ gauge bosons
(again to leading order in $\widetilde{M}_{W^{\prime},Z^{\prime}}^{-2}$).
We then have the four-fermion interactions
\begin{align}
\mathcal{L}^{4f}_{\tbox{eff}}
&=
-\frac{1}{2\tM^2_{Z}}
\left[
J^{0\mu}J^0_{\mu}
+
\frac{\tM^2_{Z}}{\tM^2_{Z^{\prime}}}
\left(
  \frac{\delta \tM^4_{Z}}{\tM^4_{Z}} J^0_{\mu}J^{0\mu}
-2\frac{\delta \tM^2_{Z}}{\tM^2_{Z}} J^0_{\mu}K^{0\mu}
+ K^0_{\mu}K^{0\mu}
\right)
\right]
\nonumber\\
&-\frac{1}{\tM^2_{W}}
\left[J^{+\mu}J_{\mu}^-
+
\frac{\tM^2_{W}}{\tM^2_{W^{\prime}}}
\left(
  \frac{\delta \tM^4_{W}}{\tM^4_{W}} J^{+}_{\mu}J^{-\mu}
- \frac{\delta \tM^2_{W}}{\tM^2_{W}}
  (J^{+}_{\mu}K^{-\mu}+J^{-}_{\mu}K^{+\mu})
\right.\right.\nonumber\\&\left.\left.+ K^{+}_{\mu}K^{-\mu}
\right)
\right].
\label{eq:4-fermi}
\end{align}
Before we can compare the predictions of Eq.~\Eref{eq:4-fermi} with
experimental results for the different $\tto$ models,
we first have to properly define some experimental input values
(for example, the Fermi constant $G_F$) for the $\tto$ models under study.
We will discuss this in Section \ref{sec:fit}.

\subsubsection{Triple Gauge Boson Couplings}
In the basis defined through Eqs.~\Eref{eq:gauge-boson-basis1},
\Eref{eq:gauge-boson-basis2} and \Eref{eq:gauge-boson-basis6},
the
triple gauge boson couplings (TGCs) $g(\hat{Z}\hat{W}^{+}\hat{W}^{-})$
and $g(A\hat{W}^{+}\hat{W}^{-})$ have
the standard forms
\begin{align}
g(\hat{Z}\hat{W}^{+}\hat{W}^{-}) &= -\tgL\cos\ttheta,\\
g(      A\hat{W}^{+}\hat{W}^{-}) &= \tilde{e}.
\end{align}
In the basis of mass eigenstates, however, we expect there to be a shift
to these couplings because the mass eigenstate $Z$ ($W$) now is
a mixture of $\hat{Z}$ ($\hat{W}$) and $\hat{Z}^{\prime}$ ($\hat{W}^{\prime}$).
However, because of QED gauge invariance, the  $A W^{+}W^{-}$ coupling does not receive a shift.
On the other hand, the $ZW^+ W^-$ coupling
does shift, and we shall discuss in turn this shift for the two breaking patterns.

In  breaking pattern I (LR, LP, HP, and FP models),
in the hat ($\hat{\ }$) basis of the gauge bosons,
the Lagrangian
contains $\hat{Z}\hat{W}^{\prime}\hat{W}^{\prime}$ and
$\hat{Z}^{\prime}\hat{W}^{\prime}\hat{W}^{\prime}$ vertices in
addition to the typical $\hat{Z}\hat{W}\hat{W}$ vertex.
Since the overlap between $\hat{W}^{\prime}$
and the light mass eigenstate $W$ is of order
$\mathcal{O}(\tM^{-2}_{W^{\prime}})$,
contributions from
$g\left(\hat{Z}\hat{W}^{\prime}\hat{W}^{\prime}\right)$ and
$g\left(\hat{Z}^{\prime}\hat{W}^{\prime}\hat{W}^{\prime}\right)$
to $g\left(ZWW\right)$ are at least of order $\mathcal{O}(\tM^{-4}_{W^{\prime}})$.
As we are only working to leading order in
$\mathcal{O}(\tM^{-2}_{W^{\prime}})$,
there is no shift due to these additional interactions
at this order.

For  breaking pattern II, the story is similar.
There are no $\hat{Z}\hat{W}\hat{W}^{\prime}$
nor $\hat{Z}^{\prime}\hat{W}\hat{W}$ vertices,
only $\hat{Z}\hat{W}^{\prime}\hat{W}^{\prime}$
and $\hat{Z}^{\prime}\hat{W}^{\prime}\hat{W}$ interactions.
The contributions to the $ZWW$ coupling are suppressed by
 fourth powers of the heavy masses
$\tM^{-4}_{W^{\prime},Z^{\prime}}$, and thus of higher order than
those kept in the effective theory.

In both breaking patterns, however, there will be a shift to the
$ZWW$-vertex due to a shift in $\ttheta$ (cf.
Eq.~(\ref{eq:theta-relation})) , the counterpart of the Standard
Model  weak mixing angle $\theta$ , as defined in our fitting
scheme.
The LEP-II experiments, however, do not directly probe the $ZWW$-vertex,
but instead infer the $ZWW$-vertex through the process
$e^{+}e^{-}\rightarrow W^{+}W^{-}$ assuming SM couplings
for all other vertices.
To properly compare the relationship between the experimental measurement
of the $ZWW$-vertex and the theoretical shifts in the $\tto$ models,
we would have to take into account all the other shifts in the
couplings that enter the process $e^{+}e^{-}\rightarrow W^{+}W^{-}$.
We will discuss this in further detail in Section \ref{sec:results}.

\subsubsection{The Yukawa and Higgs Sectors}

We complete our discussion of the effective Lagrangians
of the $\tto$ models with a brief discussion of the Higgs sectors
and the Yukawa interactions.
It is important to stress, however, that despite the complexity of the Higgs
sectors and Yukawa interactions, our results of the global analysis
only depend on the gauge interactions of the fermions, and
not on the details of the Yukawa interactions.
This is because we work only with those observables involving gauge
interactions (which excludes, for example, the branching ratio
$\mbox{Br}(b\rightarrow s\gamma)$), and keep only tree-level
contributions originated from the new physics.

We discuss the Higgs sectors of the two breaking patterns separately.
In  breaking pattern I, we take as an example the left-right model,
where the electroweak symmetry is broken
by a bi-doublet (LR-D).
This is necessary because the VEV's of the bi-doublet should generate
the fermion masses, and the right-handed fermions now are doublets
under the $SU(2)_2$.
With the bi-doublet $H$ in  Table~\ref{tb:Higgs-stage1}, we may have Yukawa couplings
(similarly for leptons) such as:
\begin{align}
-\mathcal{L}\supset
\overline{Q}_{\tbox{R}}\left(\mathcal{Y}_Q H +\tilde{\mathcal{Y}}_Q\tilde{H}\right)Q_{\tbox{L}}
+\mbox{h.c.},
\end{align}
with $\tilde{H}=-i \tau_2 H^{\ast}\tau_2$, and
where $\mathcal{Y}_Q$ and $\tilde{\mathcal{Y}}_Q$ have flavor structures
that may be
related by imposing additional symmetries (for example, left-right parity) on the model.
In any case, unlike the Standard Model where we can solve for Yukawa
couplings in terms of fermion masses and the Higgs VEV, in
 $\tto$ models there are more free parameters in
the Yukawa sectors.
These parameters can lead to interesting flavor phenomena,
particularly in the arena of neutrino physics, and have been studied
in detail in the literature (see, for example, Mohapatra {\it
et.~al.} ~\cite{Mohapatra:2005wg}).
On the other hand, the details of the Yukawa sectors do not affect
the gauge couplings of the fermions at leading order and therefore
do not affect the results of our analysis.

In breaking pattern II, in addition to those Higgs bosons that
are required to break the electroweak symmetry, it may be the case
that the Higgs sector needs to be extended to generate fermion masses.
This is because, with the current set-up, the Higgs boson that generates EWSB can couple
only to leptons (in the case of un-unified model) or
fermions of the third generation (in the case of the non-universal model).
With additional Higgs bosons, the structure of the Higgs
potential may mimic that of the two-Higgs doublet models.
Again, as with breaking pattern I, there are more degrees of
freedom than can be determined from the fermion masses, but
the details of the Yukawa interactions do not affect the results
of our paper, which only depend on the fermionic gauge interactions.

\section{The Global Fit Analysis}
\label{sec:fit} In this section we illustrate our procedure for
performing the global-fit analysis to obtain constraints on new
physics contributions.
From Tables~\ref{tb:Higgs-stage1} and \ref{tb:Zp-couplings},
we see that
the $\tto$ models contains six (five) parameters
for the first (second) breaking pattern:
three (two) VEV's
$\{\tilde{u}_{\tbox{D},\tbox{T}}, \tilde{v}\sin\tbeta, \tilde{v}\cos\tbeta\}$ in Table~\ref{tb:Higgs-stage1}
and three gauge couplings $\{\tgo, \tgt, \tgX\}$ in Table IV.
(For breaking pattern II, there are only two VEV's $\{\tilde{u}, \tilde{v}\}$.)
Compared to the gauge sector of the SM, which contains
only three parameters
(two gauge couplings and one VEV; $g_{\tbox{L}}$, $g_{\tbox{Y}}$ and $v$),
there are three (two) additional
parameters, and our goal is to:
\begin{itemize}
\item find a useful parameterization
of these three additional parameters
so as to parameterize the effects of \textit{new} physics,
and
\item determine
the constraints on these parameters from electroweak precision
measurements through a global-fit analysis.
\end{itemize}
We discuss these two steps in detail in turn.

\subsection{Parameterization}
As stated above, the $\tto$ models contain six (five) parameters
in the gauge sector:
\begin{align}
\{\tgo,\ \tgt,\ \tgX,\ \uD (\uT,\ \mbox{or}\ \tu),\ \tilde{v}^2,\
\tbeta\},
\end{align}
where the parameter $\tilde{\beta}$ only exists in models with
breaking pattern I. Using
Eqs.~\Eref{eq:GaugeCouplingExpand0},\Eref{eq:GaugeCouplingExpand1},
and \Eref{eq:GaugeCouplingExpand}, an equivalent set of parameters
is
\begin{align}
\{\tilde{\alpha}_{e},\ \ttheta,\ \tphi,\ \tilde{x},\ \tilde{v}^2,\
s_{2\tbeta}\}, \label{eq:LR-param-set}
\end{align}
where $\tilde{x}$ is defined as
\begin{align}
\tilde{x}\equiv
\begin{cases}
\uD^2 / \tilde{v}^2\quad \mbox{for LR-D, LP-D, HP-D, and FP-D} \\
\uT^2 / \tilde{v}^2\quad \mbox{for LR-T, LP-T, HP-T, and FP-T} \\
\tu^2 / \tilde{v}^2\quad \mbox{for UU and NU}.
\end{cases}
\end{align}
As we expect $\tilde{x}$ to be large ($\tilde{x}\gtrsim 100$),
we work to leading order in $\tilde{x}^{-1}$.

In addition to these parameters, the loop-level predictions will
require the values of the masses of the top quark ($m_t$) and
the Higgs boson ($M_H$).
For each $\tto$ model, we perform two separate analyses with regard
to these parameters.
In one analysis, we fit these two parameters, $m_t$ and $M_H$,
in addition to the model parameters.
In a second analysis, we fix these two parameters at
the best-fit SM values.

 With regard to the parameters in Eq.~\Eref{eq:LR-param-set}, we will take three
reference observables to constrain three combinations of the
parameters and perform a global-fit over $\{\tilde{x},\ \tphi,\
s_{2\tbeta},\ \overline{m}_t,\ M_H \}$.
The bar ($\bar{\ }$)  over $m_t$ indicates that
we will use the top quark mass as defined in the
$\overline{\mbox{\footnotesize{\textsc{MS}}}}$-scheme.
We take as reference observables the experimental measurements of
\begin{itemize}
\item the mass of the $Z$ boson ($M_Z=91.1876$ GeV), determined from the $Z$-line shape at LEP-I.
\item the Fermi constant ($G_F=1.16637\times 10^{-5}$ GeV$^{-2}$), determined from the lifetime of the muon,
\item the fine structure constant ($\alpha_e^{-1}=127.918$ at the scale $M_Z$).
\end{itemize}
Our task then is to express the model parameters, cf
Eq.~\Eref{eq:LR-param-set}
\begin{align}
\{\tilde{\alpha}_{e},\ \ttheta,\ \tilde{v}^2,\ \tilde{x},\ \tphi,\
s_{2\tbeta},\ \overline{m}_t,\ M_H \}, \nonumber
\end{align}
in terms of the reference and fit parameters
\begin{align}
\{ \alpha_{e},\ M_Z,\ G_F,\ \tilde{x},\ \tphi,\ s_{2\tbeta},\
\overline{m}_t,\ M_H \}. \label{eq:fit-param}
\end{align}
That is, we want the relationships
\begin{align}
\{
\stackrel
{
\mbox{model parameters}
}
{
\overbrace{ \tilde{\alpha}_e,\ \ttheta,\ \tilde{v}^2,\ \tilde{x},\ \tphi,\ s_{2\tbeta}, \bar{m}_t, M_H}
}
\}
\Leftrightarrow
\{
\stackrel
{
\mbox{reference parameters}
}
{
\overbrace{\alpha_e,\ M_Z,\ G_F,\ }
}
\quad
\stackrel
{
\mbox{fit parameters}
}
{
\overbrace{  \tilde{x},\ \tphi,\ s_{2\tbeta},\ \overline{m}_t,\ M_H  }
}
\}
\end{align}

Since $\{\tilde{x},\tphi,\tbeta,\ \overline{m}_t, M_H\}$ appear in both
the model and fit parameters (by construction), we
only have to solve for $\{\tilde{\alpha}_e, \ttheta, \tilde{v}^2\}$
in terms of the reference and fit parameters.
This can be done by analyzing how the reference parameters are
related to the model parameters.

\subsubsection{Electric Charge}
The electric charge in the $\tto$ models is the gauge coupling
of the unbroken $U(1)_{\smbox{em}}$ group, which
we have parameterized as $\tilde{e}$ in Eq.~\Eref{eq:electric-charge}.
There are no tree-level modifications to the wavefunction
renormalization of the photon, so we then simply have the relationship
\begin{align}
\tilde{\alpha}_{e}=\alpha_{e}. \label{eq:alpha-relation}
\end{align}

\subsubsection{The Fermi Constant}
%
The Fermi  constant, $G_F$, is experimentally determined from the muon
lifetime as \cite{Amsler:2008zzb}
\begin{align}
\tau_{\mu}^{-1} = \frac{G_F^2 m_{\mu}^5}{192\pi^3}
\left[1+\mathcal{O}\left(\frac{m_e^2}{m_{\mu}^2}\right)\right]
\left[1+\mathcal{O}\left(\frac{m_{\mu}^2}{M_{W}^2}\right)\right]
\left[1+\mathcal{O}\left(\frac{1}{16\pi^2}\right)\right],
\end{align}
where the precise forms of the higher-order corrections
are given in Ref.~\cite{Amsler:2008zzb}.
Neglecting these higher-order corrections, the SM
 contribution to the muon lifetime
 is
\begin{align}
\tau_{\mu}^{-1} = \frac{g_L^4}{192\cdot 32\pi^3 M_W^4}m_{\mu}^5,
\label{eq:muonlife}
\end{align}
and, using the SM relation $4 M_W^2 = g_L^2 v^2$, we
obtain
\begin{align}
G_F = \frac{1}{\sqrt{2} v^2}.
\label{eq:GF-SM}
\end{align}

In the $\tto$ models, we have extra contributions to the
four-fermion charged-current effective theory below the electroweak
scale,
cf Eq.~\Eref{eq:4-fermi},
\begin{align}
\mathcal{L}_{\tbox{eff}}^{CC,4f}
&=
-\frac{1}{\tM^2_{W}}J^{+}J^{-}
-\frac{1}{\tM^2_{W^{\prime}} }
\left(
K^{+}K^{-}
-\frac{\delta \tM^2_{W} }{\tM^2_{W} }(J^{+}K^{-}+K^{-}J^{+})
+\frac{\delta \tM^4_{W} }{\tM^4_{W} }J^{+}J^{-}
\right),
\nonumber
\end{align}
and these contributions will modify the SM relation in Eq.~\Eref{eq:GF-SM}.
In principle, the fermionic contributions to $K^{+}_{\mu}$ can have
both left- and right-handed components and differ  among the
different generations.
However,
for the $\tto$ models we consider here,
$K^{\pm}_{\mu}$ couples universally to
the first two generations.
Furthermore, $K^{\pm}_{\mu}$ is either purely right-handed (the LR, HP, LP, FP models)
or purely left-handed (the UU and NU models).
We therefore focus on these special cases instead of performing the general analysis.

We first consider the case that $K^{\pm}_{\mu}$ is purely right-handed.
The contributions to the amplitude come from $JJ$, $JK$, and $KK$
operators that do not interfere with one another in the limit of
 neglecting the masses of electrons and
neutrinos.
The squared-amplitudes from the $JK$ and $KK$ operators
are of order $\mathcal{O}(M^{-4}_{W^{\prime}})\sim\mathcal{O}(x^{-2})$ at leading order,
and we do not keep these contributions.
The Fermi  constant is then given by
\begin{align}
\frac{G_F}{\sqrt{2}} = \frac{\tgL^2}{8\tM_W^2}
\left(1+\frac{\delta\tM^4_{W} }{ \tM^2_{W} \tM^2_{W^{\prime}}
}\right) \,,\,\,\mbox{(for breaking pattern I)},
\label{eq:FermiConstant}
\end{align}
independent of the details of $K_{\mu}^{\pm}$.
The expression of $G_F$,  which depends on the details of the
 Higgs representation, is written in terms of
model parameters as
\begin{align}
G_F=
\begin{cases}
\mfrac{1}{\sqrt{2}\tilde{v}^2}
\left(1+\mfrac{ s^2_{2\tbeta} }{ \tilde{x} }\right),
\quad \mbox{(for LR-D, LP-D, HP-D, and FP-D)}\\
\mfrac{1}{\sqrt{2}\tilde{v}^2}
\left(1+\mfrac{ s^2_{2\tbeta} }{ 2\tilde{x} }\right),
\quad \mbox{(for LR-T, LP-T, HP-T, and FP-T)}
\end{cases}
\label{eq:GF-relation01}
\end{align}

Though the left-right and right-right current operators do not
contribute to the total muon decay rate at the order  ${\cal
O}(\tilde{x}^{-1})$ , they do contribute at leading order to the
Michel parameters (for a detailed discussion of the Michel
parameters, see the Muon Decay Parameters article in the Particle
Data Group (PDG) \cite{Amsler:2008zzb}).

In the case that $K^{\pm}_{\mu}$ is purely left-handed,
all the charged-current operators in Eq.~\Eref{eq:4-fermi} contribute, and
$G_F$ is given by
\begin{align}
\frac{G_F}{\sqrt{2}} =
\frac{\tgL^2}{8\tM_W^2}
\left[1+
\frac{\tM^2_{W}}{ M^2_{W^{\prime}} }
\left(
 \frac{\tilde{g}_{W^{\prime}}^{2} }{\tgL^2 } -
2\frac{\delta \tM^2_{W} }{ M^2_{W} }
 \frac{\tilde{g}_{W^{\prime}} }{\tgL} +
\frac{\delta \tM^4_{W} }{ M^4_{W} } \right) \right],
\quad\mbox{(for UU and NU)}
\end{align}
where $\tilde{g}_{W^{\prime}}$ can be looked up in Table~\ref{tb:Wp-couplings}.
For the UU and NU models, these contributions cancel each other, and we
are simply left with
\begin{align}
G_F = \frac{1}{\sqrt{2}\tilde{v}^2}\quad\mbox{(for UU and NU)}.
\label{eq:GF-relation02}
\end{align}

We can rewrite our results in a more suggestive manner
by defining the SM VEV ($v^2$ without tilde $\tilde{\ }$) through the Fermi  constant
\begin{align}
v^2 \equiv \frac{1}{\sqrt{2}G_F}.
\end{align}
We then have
\begin{align}
\tilde{v}^2 =
\begin{cases}
v^2 \left(1+\mfrac{ s^2_{2\tbeta} }{ \tilde{x} }\right),
\quad \mbox{(for LR-D, LP-D, HP-D, and FP-D)}\\
v^2 \left(1+\mfrac{ s^2_{2\tbeta} }{ 2\tilde{x} }\right),
\quad \mbox{(for LR-T, LP-T, HP-T, and FP-T)}\\
v^2. \quad\mbox{(for UU and NU)}
\end{cases}
\label{eq:v-relation}
\end{align}

\subsubsection{$Z$-Mass}
In our effective theory approach, the mass eigenvalue of the
$Z$-boson is given by (using Eq.~\Eref{eq:L-ew-fund},
Table~\ref{tb:MZMW-stage1}, and $\tilde{\alpha}_{e}=\alpha_{e}$)
\begin{align}
M^2_{Z}
&=
\tM_Z^2 - \frac{ \delta \tM^4_{Z} }{ \tM^2_{Z^{\prime}}}
\ \begin{pmatrix}\mbox{general form from the}\\ \mbox{fundamental $\tto$ Lagrangian}\end{pmatrix}
\nonumber
\\
&=
\begin{cases}
\mfrac{\alpha_{e}\pi \tilde{v}^2}{s^2_{\ttheta}c^2_{\ttheta}}
\left(1-\mfrac{c_{\tphi}^4}{\tilde{x}}\right),
\quad \mbox{(for LR-D, LP-D, HP-D, and FP-D)} \\
\mfrac{\alpha_{e}\pi \tilde{v}^2}{s^2_{\ttheta}c^2_{\ttheta}}
\left(1-\mfrac{c_{\tphi}^4}{4\tilde{x}}\right),
\quad \mbox{(for LR-T, LP-T, HP-T, and FP-T)} \\
\mfrac{\alpha_{e}\pi \tilde{v}^2}{s^2_{\ttheta}c^2_{\ttheta}}
\left(1-\mfrac{s_{\tphi}^4}{\tilde{x}}\right), \quad \mbox{(for UU
and NU)}
\end{cases}.
\label{eq:MZ-relation01}
\end{align}
Solving Eq.~\Eref{eq:MZ-relation01} for $c^2_{\ttheta}s^2_{\ttheta}$,
and using Eqs.~\Eref{eq:GF-relation01} and \Eref{eq:GF-relation02},
we can solve for $\ttheta$ in terms
of the reference and fit parameters
\begin{align}
s_{\ttheta}^2c_{\ttheta}^2
=
\begin{cases}
s_{\theta}^2 c_{\theta}^2
\left[
1-
\mfrac{1}{x}\left(c^4_{\tphi}- s^2_{2\tbeta}\right)
\right],
\mbox{(for LR-D, LP-D, HP-D, and FP-D)}
\\
s_{\theta}^2 c_{\theta}^2
\left[1-
\mfrac{1}{x}\left(\tfrac{1}{4}c^4_{\tphi}-\tfrac{1}{2} s^2_{2\tbeta}\right)
\right],
\mbox{(for LR-T, LP-T, HP-T, and FP-T)}
\\
s_{\theta}^2 c_{\theta}^2
\left[1-
\mfrac{s^4_{\tphi}}{x}
\right],
\mbox{(for UU and NU)},
\label{eq:theta-relation}
\end{cases}
\end{align}
where $\theta$ (without a tilde $\tilde{\ }$) is defined in terms
of the reference parameters
\begin{align}
\sin^2\theta \cos^2\theta &\equiv \frac{\pi\alpha_{e}}{\sqrt{2}M_Z^2
G_F}. \label{eq:theta-input-relation}
\end{align}
Eqs.~\Eref{eq:alpha-relation}, \Eref{eq:v-relation},
and \Eref{eq:theta-relation} then enable us
to translate all the model parameters to reference and fit
parameters.

\subsection{Corrections to Observables}
\label{subsec:obs}
%
In this subsection we illustrate the corrections to several example observables
that we include in our global analysis.
These examples elucidate the procedures we had outlined earlier, and
we will refer to these results when we discuss the observables included
in our global analysis.
%

\subsubsection{The $Z$-Partial Widths $\Gamma(Z\rightarrow f\overline{f})$}
%
As a first example, we can then consider the $Z\rightarrow
f\overline{f}$ partial width, which at tree-level has the expression
in the Standard Model
\begin{align}
\Gamma(Z\rightarrow f\overline{f}) = \frac{n_c}{12\pi} M_Z\left(
g_{V}^2 + g_{A}^2 \right), \label{eq:Zdecay-SM}
\end{align}
where $n_c=3$ if $f$ is s quark, and $n_c=1$ for leptons,
and
\begin{align}
g_{V}  &= \frac{e}{2 s_{\theta} c_{\theta} }
\left(T_{3L}^f - 2 Q^f \sin^2\theta\right) , \label{eq:Zdecay-gv}\\
g_{A}  &= \frac{e}{2 s_{\theta} c_{\theta} } T_{3L}^f,\label{eq:Zdecay-ga}
\end{align}
where $T_{3L}^{f}$ and $Q^f$ are respectively
the weak-isospin and electric charge of the fermion $f$.

In the $\tto$ models, the partial decay width can be written in
terms of model parameters as
\begin{align}
\Gamma(Z\rightarrow f\overline{f}) = \frac{n_c}{12\pi}
\tM_Z \left( 1-\frac{\delta\tM_Z^4}{2\tM^2_Z\tM^2_{Z^{\prime}} }
\right) \left( \left[\tilde{g}_{\tmbx{V}}^{\tmbx{Z}}(f)\right]^2 +
\left[\tilde{g}_{\tmbx{A}}^{\tmbx{Z}}(f)\right]^2 \right),
\end{align}
where $\delta\tM_Z^2, \tM^2_{Z^{\prime}}, \tilde{g}_{\tmbx{V}}^{\tmbx{Z}}(f)$,
and $\tilde{g}_{\tmbx{A}}^{\tmbx{Z}}(f)$ depend on the details of the model.
For models that follow  the breaking pattern I (LR-D, LP-D, HP-D,
FP-D), the couplings have the form (to order
$\mathcal{O}(\tilde{x}^{-1})$)
\begin{align}
\tilde{g}_{\tmbx{V}}^{\tmbx{Z}}(f) &= \frac{e}{2 s_{\ttheta} c_{\ttheta} }
\left(
(T_{3L}^f - 2 Q^f s^2_{\ttheta})
+
\frac{c^2_{\tphi}}{2\tilde{x}}
\left[
T_{3R}^f c^2_{\tphi} - (X_L^f+X_R^f) s^2_{\tphi}
\right]
\right)
,
\\
\tilde{g}_{\tmbx{A}}^{\tmbx{Z}}(f) &= \frac{e}{2 s_{\ttheta} c_{\ttheta} }
\left(
T_{3L}^f
-
\frac{c^2_{\tphi}}{2\tilde{x}}
\left[
T_{3R}^f c^2_{\tphi} - (X_R^f-X_L^f) s^2_{\tphi}
\right]
\right),
\label{eq:Z-coup-mod}
\end{align}
where $X_L^f$, and $X_R^f$, and $T_{3R}$ are respectively
the left- and right-handed fermion charges under the $U(1)_X$, and
the $z$-component isospin under the $SU(2)_2$ (which is identified
as $SU(2)_R$ in left-right models).
Expressing $\ttheta$ in terms of the reference and the model parameters
through Eq.~\Eref{eq:theta-relation} and  collecting terms of
$\mathcal{O}(\tilde{x}^{-1})$, we have (in units of GeV)
\begin{align}
\Gamma(Z\rightarrow f\overline{f}) = \Gamma(Z&\rightarrow
f\overline{f})_{\smbox{SM}}
\nonumber\\
+\frac{n_f}{\tilde{x}}
&\left[
s^4_{\ttheta} \left(
   -0.446\ (Q^{f})^2
   +1.773\ Q^{f} T_{3L}^{f}
   -0.310\ Q^{f} T_{3R}^{f}
\right.\right.
\nonumber\\
&\left.\left.
   -0.310\ Q^{f} X_{R}^{f}
   -0.664\ (T_{3L}^{f})^{2}
   \right)
\right.
\nonumber\\
&
+s^2_{\ttheta}
   \left(
   0.582\ (Q^{f})^2
   -1.91\ Q^{f} T_{3L}^{f}
   +0.620\ Q^{f}T_{3R}^{f}
   +0.310\ Q^{f}X_{R}^{f}
   \right)
\nonumber\\
&
+s^2_{2\tbeta}
   \left(
   0.136\ (Q^{f})^2
   -0.136\ Q^f T_{3L}^{f}
   -0.664\ (T_{3L}^{f})^{2}
   \right)
\nonumber\\
&\left.
   -0.136\ (Q^{f})^2
   +0.136\ Q^{f} T_{3L}^{f}
   -0.310\ Q^{f} T_{3R}^{f}
   +0.664\ (T_{3L}^{f})^{2}
\right],
\nonumber\\
&\mbox{(for LR-D, LP-D, HP-D, and FP-D)}
\label{eq:Z-decay-full}
\end{align}
where $\Gamma(Z\rightarrow\overline{f}f)_{\smbox{SM}}$
is given by Eq.~\Eref{eq:Zdecay-SM}, and we have used the numerical
values of the reference parameters.

\subsubsection{The Mass of the $W$-boson}
As a second example, we compute the mass of the $W$-boson in
the $\tto$ models.
The SM expression, for the same set of reference parameters $\{\alpha,\ M_Z,\ G_F\}$, is given by
\begin{align}
M_W = M_Z c_{\theta},
\label{eq:W-mass-SM}
\end{align}
where
$\theta$ is defined in terms of the reference parameters
in Eq.~\Eref{eq:theta-input-relation}.
In the $\tto$ models, the mass of the $W$-boson has the
general form
\begin{align}
M_W = \tM_W
\left(1-
\frac{ \delta\tM^4_{W} }
{ 2 \tM^2_{W} \tM^2_{W^{\prime}} }
\right).
\end{align}
More specifically, in terms of the model parameters for the individual models,
we have
\begin{align}
M_W =
\begin{cases}
\mfrac{\tilde{e}\tilde{v} }{ 2 s_{\ttheta} }
\left(1-\mfrac{s_{\tbeta}^2}{2\tilde{x}}\right)\quad
\mbox{(for LR-D, LP-D, HP-D, FP-D)},
\\
\mfrac{\tilde{e}\tilde{v} }{ 2 s_{\ttheta} }
\left(1-\mfrac{s_{\tbeta}^2}{4\tilde{x}}\right)\quad
\mbox{(for LR-T, LP-T, HP-T, FP-T)},
\\
\mfrac{\tilde{e}\tilde{v} }{ 2 s_{\ttheta} }
\left(1-\mfrac{s_{\tphi}^4}{2\tilde{x}}\right)\quad
\mbox{(for UU, NU)}.
\end{cases}
\end{align}
Using Eqs.~\Eref{eq:alpha-relation}, \Eref{eq:v-relation},
and \Eref{eq:theta-relation}, we can
convert all the model parameters
to reference and fit parameters
\begin{align}
M_W =
\begin{cases}
M_Z\cos\theta
\left[1+\mfrac{1}{2\tilde{x}}\mfrac{c^2_{\theta}}{c^2_{\theta}-s^2_{\theta}}
\left(c^4_{\tphi}-s^2_{2\tbeta}\right)
\right]\quad
\mbox{(for LR-D, LP-D, HP-D, FP-D)},
\\
M_Z\cos\theta
\left[1+\mfrac{1}{2\tilde{x}}\mfrac{c^2_{\theta}}{c^2_{\theta}-2s^2_{\theta}}
\left(\frac{c^4_{\tphi}}{4}-\frac{s^2_{2\tbeta}}{2}\right)
\right]\quad
\mbox{(for LR-T, LP-T, HP-T, FP-T)},
\\
M_Z\cos\theta
\left[1+\mfrac{1}{2\tilde{x}}\mfrac{s^2_{\theta}}{c^2_{\theta}-s^2_{\theta}}
s^4_{\tphi}
\right]\quad
\mbox{(for UU, NU)}.
\end{cases}\label{eq:mwmass}
\end{align}

\subsection{Implementation of the Global Fit and List of Observables}

For a measured observable $O^{\tbox{exp}}$,
the SM prediction can be broken down into
the tree- and loop-level components
\begin{align}
O^{\tbox{th}}_{\tbox{SM}}
=
O^{\tbox{th,tree}}_{\tbox{SM}}+
O^{\tbox{th,loop}}_{\tbox{SM}}(\overline{m}_t,M_H),
\end{align}
where $O^{\tbox{th}}$ is expressed in terms of the reference parameters.
Since the top quark mass ($\overline{m}_t$) and the mass of the
Higgs boson ($M_H$) enter into the loop-calculations in the SM, a
global analysis of precision data and direct detection data
can be used to constrain $M_H$.
In the $\tto$ models, we can express the theoretical prediction as
\begin{align}
O^{\tbox{th}}
=
O^{\tbox{th,tree}}_{\tbox{SM}}+
O^{\tbox{th,loop}}_{\tbox{SM}}(\overline{m}_t,M_H)+
O^{\tbox{th,tree}}_{\tbox{NP}}(\tilde{x},\tphi,\tbeta),
\end{align}
where $O^{\tbox{th,tree}}_{\tbox{NP}}$ is of
the order $\mathcal{O}(1/\tilde{x})$, and
we assume that
\begin{align}
\tilde{x}^{-1}\sim \frac{1}{16\pi^2}\sim
O^{\tbox{th,loop}}_{\tbox{SM}}.
\end{align}
That is, the Born-level new physics contributions from the $\tto$
models are numerically of one-loop order, and loop corrections
involving new physics are of two-loop order
$\mathcal{O}\left(\frac{1}{16\pi^2 \tilde{x}}\right)$, which we
discard in our analysis.

To compare with precision data (from LEP-1 and SLD) and low-energy
observables, we calculate the shifts in observables
$O^{\tbox{th,tree}}_{\tbox{NP}}(\tilde{x},\tphi,\tbeta)$, as in the
previous examples of the partial decay widths of the $Z$-boson and the
mass of the $W$-boson, and we adapt these corrections into a
numerical package GAPP \cite{Erler:1999ug}.
GAPP then computes $O^{\tbox{th,tree}}_{\tbox{SM}}$ and
$O^{\tbox{th,loop}}_{\tbox{SM}}(\bar{m}_t, M_H)$\footnote{The
higher order SM corrections included in our analysis are the same as
those in the default GAPP code used for the PDG analysis.}, together
with the $O^{\tbox{th,tree}}_{\tbox{NP}}(\tilde{x}, \tilde{\phi}, \tilde{\beta})$ to
find
the best-fit values of the fit parameters and the confidence level
contours using the CERN library MINUIT \cite{James:1975dr}.

We perform a global fit over the following classes
of observables
\begin{itemize}
\item LEP-I $Z$-pole observables: the total $Z$-width ($\Gamma_Z$), left-right asymmetries ($A_{LR}$),
and related observables.
\item the mass ($M_W$) and decay width ($\Gamma_W$) of the $W$-boson,
\item the tau  lifetime $\tau_\tau$,
\item the ratios of neutral-to-charged current cross sections measured from
neutrino-hadron deep-inelastic scattering (DIS) experiments
($R_{\nu}\equiv \sigma^{\tbox{NC}}_{\nu N}/ \sigma^{\tbox{CC}}_{\nu
N}$ and similarly defined for $\overline{\nu}$),
\item effective vector and axial-vector neutrino-electron couplings
($g_V^{\nu e}$ and $g_A^{\nu e}$),
\item weak charges ($Q_W$) of atoms and the electron measured from atomic parity experiments.
\end{itemize}
Detailed information  on these observables can be found in
PDG~\cite{Amsler:2008zzb}, and here we only briefly summarize the
observables.
The set of the observables included in our analysis is the same as
that used in the PDG analysis~\cite{Amsler:2008zzb}, with two
exceptions.
\begin{itemize}
\item
First, we do not include the anomalous magnetic moment of the muon
and the decay branching ratio $b\rightarrow s\gamma$.
At leading order, these observables are of one-loop order, and they
depend on the details of the extended flavor structure of the $\tto$
models.
In this work, we assume $W^\prime$ bosons only couple to
fermions in the same generation.
\item
Second, we include the measurements of the decay width of the $W$-boson,
which are not included in the PDG analysis.
However, because of the comparatively low precision of these
measurements, this observable turns out to be insensitive to the new
physics contributions from the $\tto$ models.
\end{itemize}
In total, we include a set of 37 experimental observables in our
global-fit analysis.

Before we give a brief discussion on each of these classes of
observables, we note that for some low-energy observables, such as
the measurements from the atomic parity violation and
neutrino-neucleus DIS experiments, we implement the shifts in the
coefficients of the relevant four-fermion interactions, and rely on
GAPP to compute the theoretical predictions based on these modified
coefficients.
The expressions of the coefficients of the four-fermion interactions
are given in the Appendix.

For the ease of typesetting in the following
subsections, we introduce the abbreviation
for the various forms of the fermionic currents
\begin{align}
\ffLR{f_1}{f_2}{L}^{\mu} &\equiv
\bar{f_1}\gamma^{\mu}\left(1-\gamma_5\right)f_2,
\nonumber\\
\ffLR{f_1}{f_2}{R}^{\mu} &\equiv
\bar{f_1}\gamma^{\mu}\left(1+\gamma_5\right)f_2,
\nonumber\\
\ffLR{f_1}{f_2}{V}^{\mu} &\equiv
\bar{f_1}\gamma^{\mu}f_2,
\nonumber\\
\ffLR{f_1}{f_2}{A}^{\mu} &\equiv
\bar{f_1}\gamma^{\mu}\gamma_5f_2.
\end{align}

\subsubsection{Precision Measurements at the $Z$-Pole}
The precision measurements at the $Z$-pole (including LEP-1 and SLD
experiments) fall into two broad classes: observables that can be
constructed from the partial widths (for example, in
Eq.~\Eref{eq:Z-decay-full}) and the asymmetry (constructed from the
couplings in  Eq.~\Eref{eq:Zdecay-gv} and \Eref{eq:Zdecay-ga}).
We discuss these two classes in turn.

In addition to the total width $\Gamma_Z$, there are also the
following measurements:
\begin{align}
\sigma_{\tbox{had}} &= \frac{12\pi}{M_Z^2
\Gamma_Z^2}\cdot\GZpm{e}\GZhad,
\\
R(\ell)  &= \frac{\GZhad}{\GZ{\ell}} \,,\,\quad\mbox{for}\ \ell =
e,\mu,\tau,
\\
R(q)     &= \frac{\GZ{q}}{\GZhad}\,,\,    \quad\mbox{for}\  q =
u,d,c,s,b,
\\
\mathcal{R}(s) &= \frac{R(s)}{R(u) + R(d) + R(s)},
\end{align}
where  $\Gamma_Z(f\overline{f})$ is the partial decay width
$\Gamma(Z\rightarrow f\overline{f})$, and
\begin{align}
\GZhad  = \sum\limits_{q = u,d,c,s,b}  \Gamma_Z(q\overline{q}).
\end{align}

The left-right asymmetry $A_{LR}(f)$ is defined as
\begin{align}
 A_{LR}(f) \equiv \frac{\gZLt-\gZRt}{\gZLt+\gZRt},
\label{eq:DEFALRf}
\end{align}
where $g_{\tmbx{L}}^{\tmbx{Z}}(f)$ and $g_{\tmbx{R}}^{\tmbx{Z}}(f)$ are the couplings of the fermion $f$ to the $Z$-boson:
\begin{align}
\mathcal{L}\supset Z_{\mu}(
g_{\tbox{L}}^Z(f)\overline{f}_{\tbox{L}}\gamma^{\mu}f_{\tbox{L}}
+
g_{\tbox{R}}^Z(f)\overline{f}_{\tbox{R}}\gamma^{\mu}f_{\tbox{R}}
).
\end{align}
From the quark branching ratios $R(q)$ defined above, the hadronic
left-right asymmetry $Q_{LR}$ can be defined as \cite{Erler:1999ug}
\cite{:2005ema}
\begin{align}
Q_{LR} \equiv \sum_{q = d,s,b} R(q)A_{LR}(q) - \sum_{q = u,c} R(q)A_{LR}(q).
\end{align}
A second class of asymmetries, the forward-backward asymmetries $A_{FB}(f)$,
emerges from the convolution of the $A_{LR}(f)$ asymmetries with the polarization asymmetry $A_{LR}(e)$ of the electron.
The hadronic charge asymmetry $Q_{FB}$ is defined accordingly \cite{Erler:1999ug}
\cite{:2005ema}
\begin{align}
A_{FB}(f) &\equiv \frac{3}{4}A_{LR}(e)A_{LR}(f),
\\
Q_{FB} &\equiv \frac{3}{4}A_{LR}(e)Q_{LR}.
\end{align}

\subsubsection{The Tau Lifetime}
In terms of model parameters, the expression of the tau ($\tau$)
lifetime is similar to the muon ($\mu$) lifetime in the $\tto$
models, cf. Eq.\Eref{eq:muonlife}, with the obvious replacement of
$m_{\mu}$ in the $\mu$ lifetime by $m_{\tau}$ in the $\tau$
lifetime.
This is true even in the non-universal (NU) model, in which third
generation fermions transform under a different gauge group compared
to the  first two generations.
In the four-fermion effective theory of the NU model,
only interactions involving two pairs of third-generation fermions
receive new physics contributions, and the interactions involving
one pair of third-generation fermions with one pair of light-flavor
fermions (those responsible for the decay of the $\tau$) are the
same as those between two pairs of first two generations of fermions
(those responsible for the decay of $\mu$).
This is similar to the case of the un-unified model, where
only interactions involving two pairs of quarks
$(\overline{q}q)(\overline{q}q)$ receive new physics contributions,
while the $(\overline{q}q)(\overline{\ell}\ell)$ interactions are
the same  as the $(\overline{\ell}\ell)(\overline{\ell}\ell)$.
The
lifetime $\tau_\tau$ can be calculated at tree level as
\begin{align}
\tau_{\tau}^{-1} \simeq \frac{G_F^2 m_{\tau}^5}{192\pi^3}\left(1
+3\frac{m_\tau^2}{M_W^2}\right),
\end{align}
in the SM. The dominant new physics contribution from
$\tto$ models can be captured in the shift of $M_W$ as shown in
Eq.~\Eref{eq:mwmass}.
%

\subsubsection{$\nu N$ Deep Inelastic Scattering}
The $\nu N$ deep inelastic scattering experiments probe the
coefficients $\epsL{q}$ and $\epsR{q}$ (for $q$ being $u$ or $d$)
that parameterize the neutral current $\overline{\nu}\nu\overline{q}q$ interactions
below the electroweak scale
\begin{align}
 \Lag\supset
-\frac{G_F}{\sqrt{2}} \ffLRm{\nu}{\nu}{L} \sum_{q=u,d}
\left[\epsL{q}\ffLR{q}{q}{L}^{\mu} + \epsR{q}\ffLR{q}{q}{R}^{\mu}\right].
\label{eq:RESL4fNCnuN}
\end{align}
The DIS experiments measure the ratios of neutral-to-charged current cross sections
\begin{align}
R_{\nu}\equiv \sigma^{\tbox{NC}}_{\nu N}/ \sigma^{\tbox{CC}}_{\nu
N},\,\,\,\,\, R_{\overline{\nu}}\equiv
\sigma^{\tbox{NC}}_{\overline{\nu} N}/
\sigma^{\tbox{CC}}_{\overline{\nu} N},
\end{align}
which can be written in terms of $\epsL{q}$ and $\epsR{q}$ as
\begin{align}
R_{\nu} & = \left(1-\delta\right) \left[a_{\tmbx{L}}(u)\epsLt{u} + a_{\tmbx{L}}(d)\epsLt{d}
+ a_{\tmbx{R}}(u)\epsRt{u} + a_{\tmbx{R}}(u)\epsRt{d}\right], \\
R_{\bar{\nu}} & = \left(1-\bar{\delta}\right) \left[\bar{a}_{\tmbx{L}}(u)
\epsLt{u} + \bar{a}_{\tmbx{L}}(d)\epsLt{d} + \bar{a}_{\tmbx{R}}(u)\epsRt{u}
+ \bar{a}_{\tmbx{R}}(u)\epsRt{d}\right].
\end{align}
The coefficients $\delta$  and $a_{\tmbx{L},\tmbx{R}}$ are related
to the nuclei form factors that are experiment specific.
These coefficients are included in GAPP, and we implement only the
corrections to $\epsL{q}$ and $\epsR{q}$.

\subsubsection{ $\nu e$ Scattering}
The most precise data on neutrino-electron scattering comes from the CHARM II \cite{Vilain:1996yf}
experiment
at CERN that utilized $\nu_{\mu}$ and $\overline{\nu}_{\mu}$.
The relevant parameters $\epsL{e}$ and $\epsR{e}$ are defined similarly
as in the $\nu N$ scattering
\begin{align}
\Lag\supset -\frac{G_F}{\sqrt{2}}
\ffLRm{\nu}{\nu}{L}\left[\epsL{e}\ffLR{e}{e}{L}^{\mu} +
\epsR{e}\ffLR{e}{e}{R}^{\mu}\right]. \label{eq:RESL4fNCnue}
\end{align}
We can further define
\begin{align}
 \geV &\equiv \epsR{e} + \epsL{e},\\
 \geA &\equiv \epsR{e} - \epsL{e},
\end{align}
which are related to the
measured total cross sections $\sigma_{\tmbx{\nu e}}^{\tbox{NC}}$ and $\sigma_{\tmbx{\bar{\nu} e}}^{\tbox{NC}}$
or their ratio $\sigma_{\tmbx{\nu e}}^{\tbox{NC}}/\sigma_{\tmbx{\bar{\nu} e}}^{\tbox{NC}}$.
In the limit of large incident neutrino energies, $E_{\nu} \gg m_e$, the cross sections are given as
\begin{align}
\sigma_{\tmbx{\nu e}}^{\tbox{NC}} \; & = \:
\frac{G_F^2 m_e E_{\nu}}{2\pi}\left[\left(\geV+\geA\right)^2+\frac{1}{3}\left(\geV-\geA\right)^2\right], \\
\sigma_{\tmbx{\bar{\nu} e}}^{\tbox{NC}} \; & = \:
\frac{G_F^2 m_e E_{\nu}}{2\pi}\left[\left(\geV-\geA\right)^2+\frac{1}{3}\left(\geV+\geA\right)^2\right].
\end{align}
We implement corrections to the couplings due to new physics in GAPP
and compute the cross sections that are used in the global-fit
analysis.

\subsubsection{Parity  Violation Experiments}
We consider observables from three different measurements:
atomic parity violation (APV),
M\o ller scattering ($e^{-}e^{-}\rightarrow e^{-}e^{-}$) \cite{Anthony:2005pm},
and $eN$ DIS.
These experiments measure the weak charge ($Q_W$) of
the electron \cite{Anthony:2005pm}, caesium-133 \cite{Wood:1997zq}\cite{Guena:2004sq}
and thallium-205 nuclei \cite{Vetter:1995vf}\cite{Edwards-ThAPV}.
Before defining the weak charge, it is useful to parameterize the
coefficients of the $(\overline{e}e)(\overline{q}q)$ and
$(\overline{e}e)(\overline{e}e)$ interactions in terms of $C_{1q}$, $C_{2q}$, and $C_{1e}$
as
\begin{align}
\Lag\supset -\frac{G_F}{\sqrt{2}}
\sum_q\left[C_{1q}\ffLRm{e}{e}{A}\ffLR{q}{q}{V}^{\mu} +
C_{2q}\ffLRm{e}{e}{V}\ffLR{q}{q}{A}^{\mu}\right]
-\frac{G_F}{\sqrt{2}} C_{1e}\ffLRm{e}{e}{A}\ffLR{e}{e}{V}^{\mu}
\label{eq:RESL4fNCeq}
\end{align}

The weak charges of the quark and electron are defined as
\begin{align}
 Q_W(q) = 2 C_{1q},\,\,\quad Q_W(e)= 2 C_{1e}.
\end{align}
We can express the SM tree-level couplings of quarks to the $Z$-boson as
$\mathcal{L}\supset Z^{\mu}J^Z_{\mu}$, where
\begin{align}
J^Z_{\mu} =
\left|g_{\tmbx{A}}^{\tmbx{Z}}(q)\right|\cdot\left[Q_{W}(q)\ffLRm{q}{q}{V}
\pm \ffLRm{q}{q}{A}\right] ,
\end{align}
 and the $\pm$ on the axial-vector term is the opposite sign of the $T_{L}^{3q}$.
 Hence $Q_W(q)$ can be interpreted as the ratio of the vector current to
 axial-vector current coupling of quark  q to the
$Z$-boson:
\begin{align}
 Q_{W,\tbox{SM}}(q) = \frac{g_{\tmbx{V}}^{\tmbx{Z}}(q)}{\left|g_{\tmbx{A}}^{\tmbx{Z}}(q)\right|}.
\end{align}
The weak charges of the nucleons and nuclei can be built up from
those of the quarks
\begin{align}
 Q_W(p) &= 2 Q_W(u) + Q_W(d), \\
 Q_W(n) &= Q_W(u) + 2 Q_W(d),
\end{align}
and for nucleus $^A\mbox{Z}$ (with atomic number $Z$ and mass number
$A$), which contains $Z$ protons and $N$($=A-Z$) neutrons,
\begin{align}
 Q_W\left(^A\mbox{Z}\right) & =
Z \cdot Q_W(p) + N \cdot Q_W(n)\\
 & = 2  \left[(Z+A) \cdot C_{1u} + (2A-Z) \cdot C_{1d} \right].
\end{align}

There are also measurements of certain linear combinations of the coupling
coefficients $C_{1u}$ and $C_{1d}$ from polarized electron-hadron scattering data \cite{Young:2007zs}.
The particular linear combinations, determined by the experimental data,
\begin{align}
\mathcal{C}_1 &= 9 C_{1u} + 4  C_{1d},
\nonumber\\
\mathcal{C}_2 &= -4 C_{1u} + 9  C_{1d},
\end{align}
are included in our global analysis.

\section{Results}
\label{sec:results}
\subsection{Global Analysis}
\begin{figure}[t!h]
\begin{center}
\includegraphics[width=\picwidth]{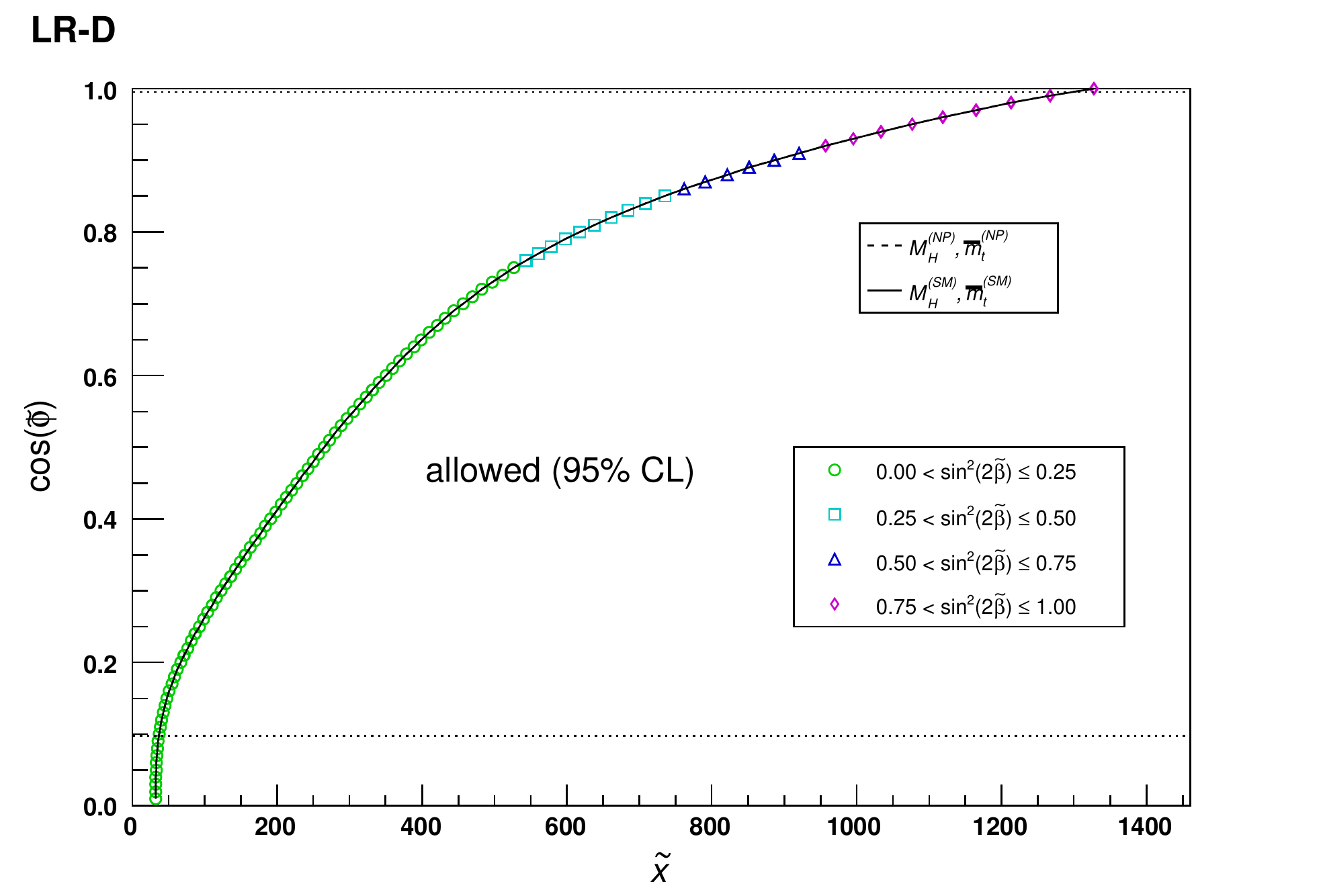}
\includegraphics[width=\picwidth]{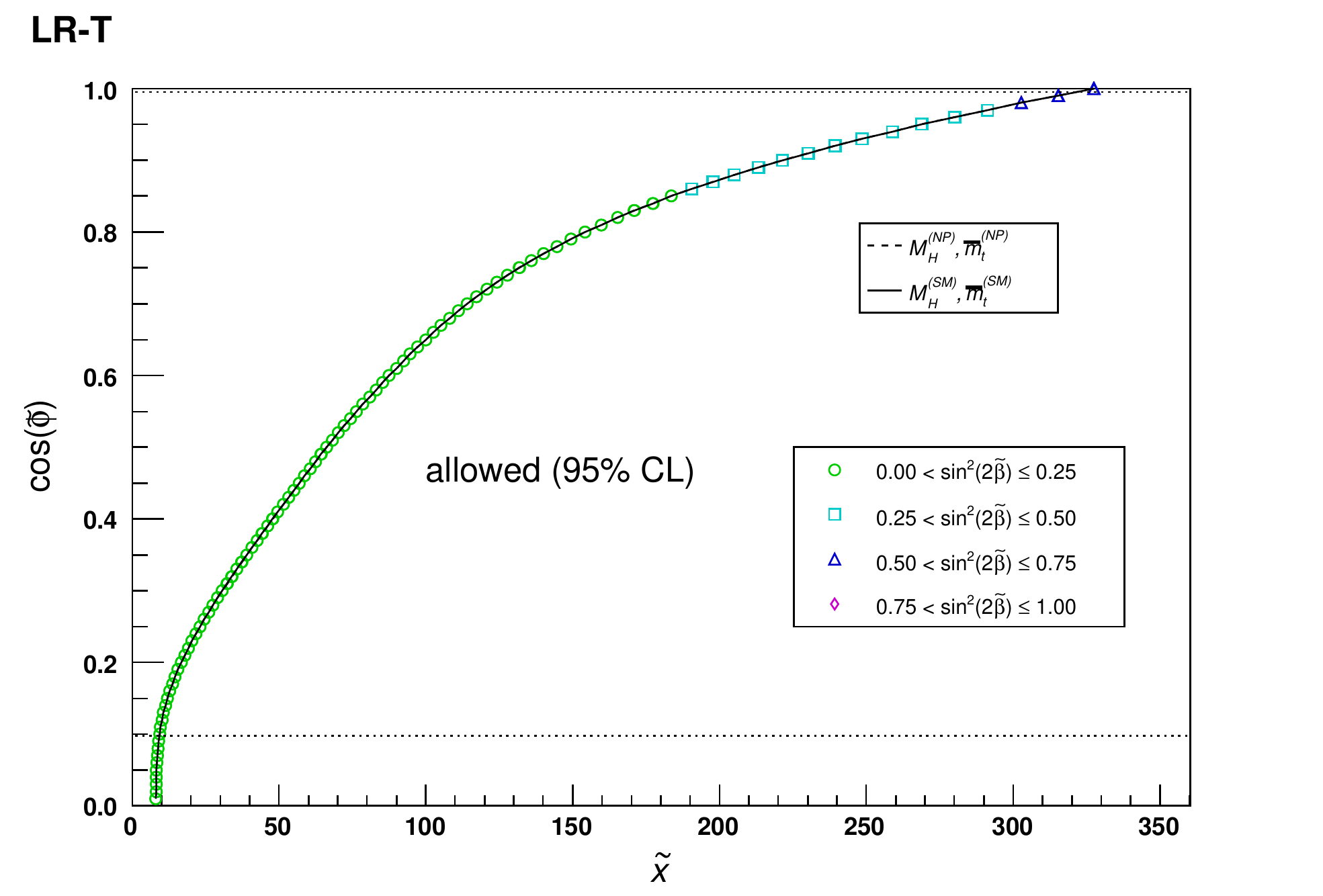}
\includegraphics[width=\picwidth]{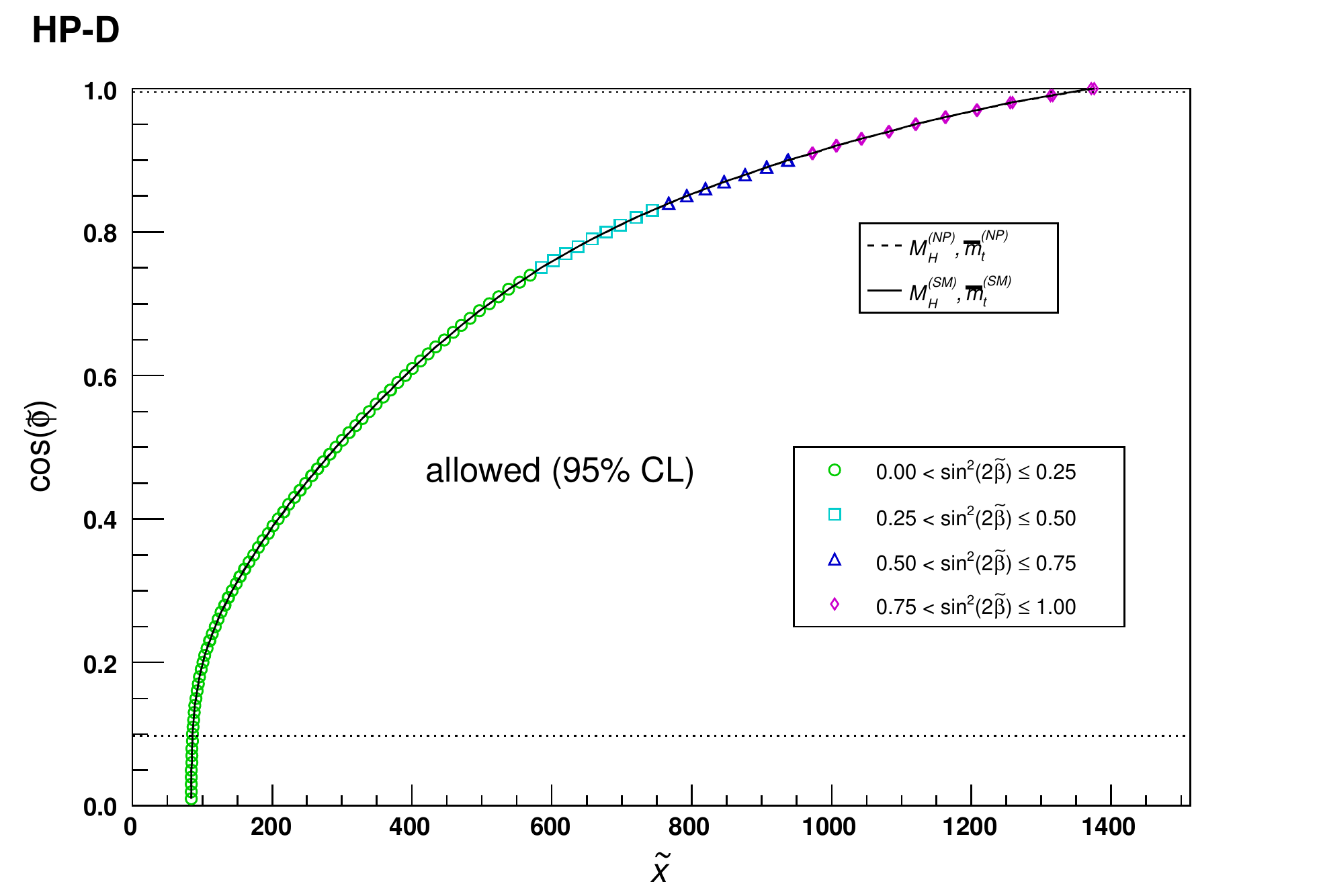}
\includegraphics[width=\picwidth]{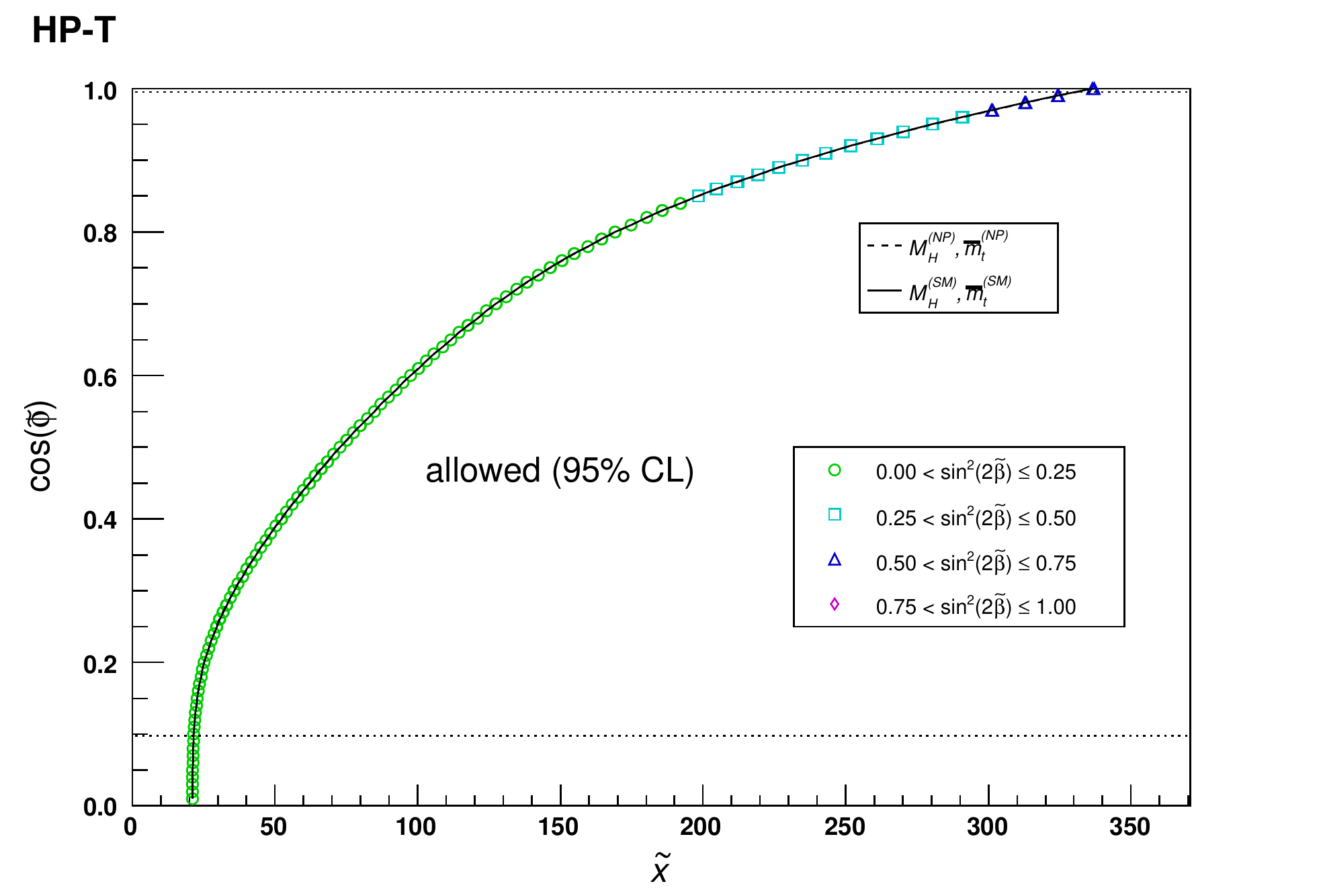}
\includegraphics[width=\picwidth]{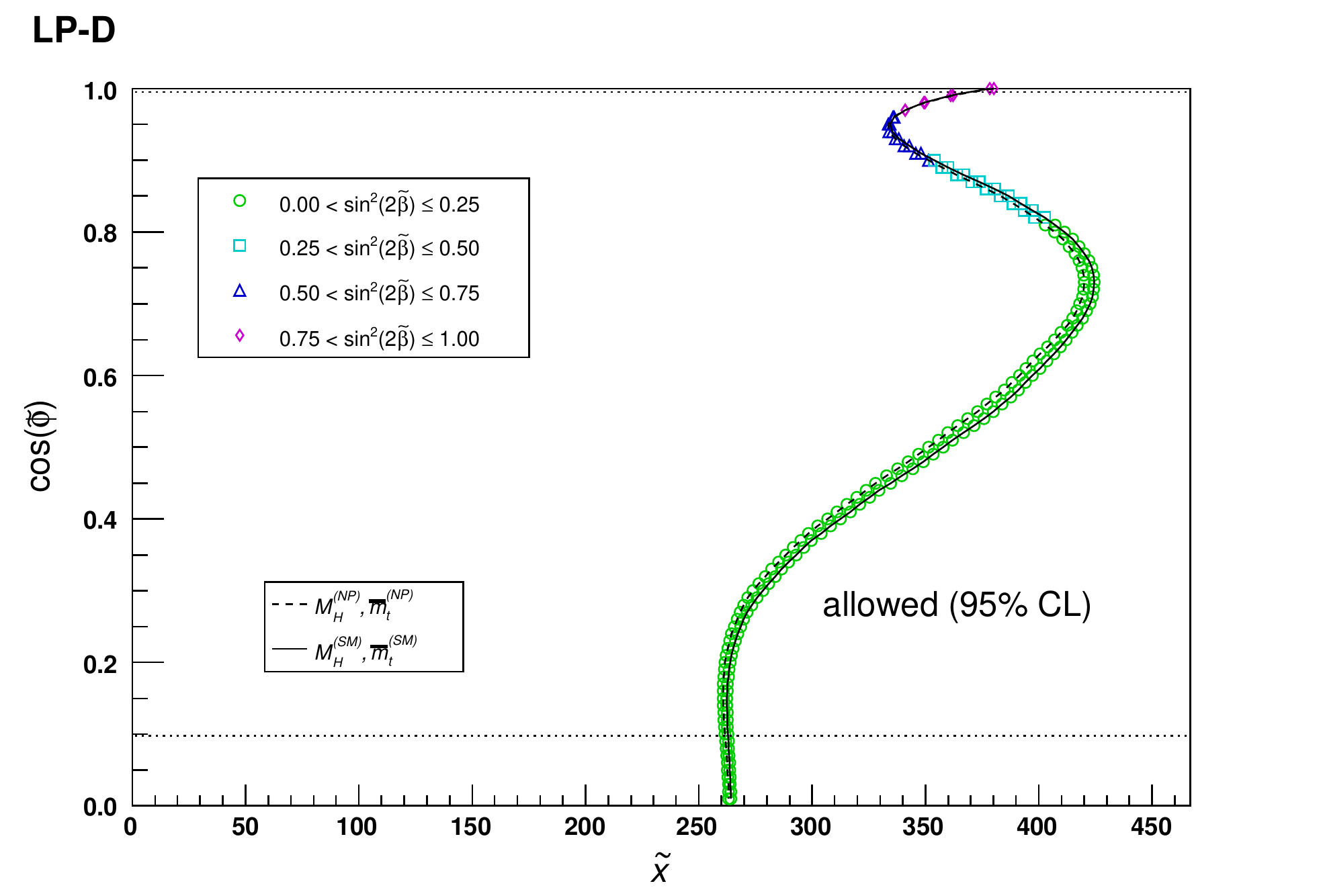}
\includegraphics[width=\picwidth]{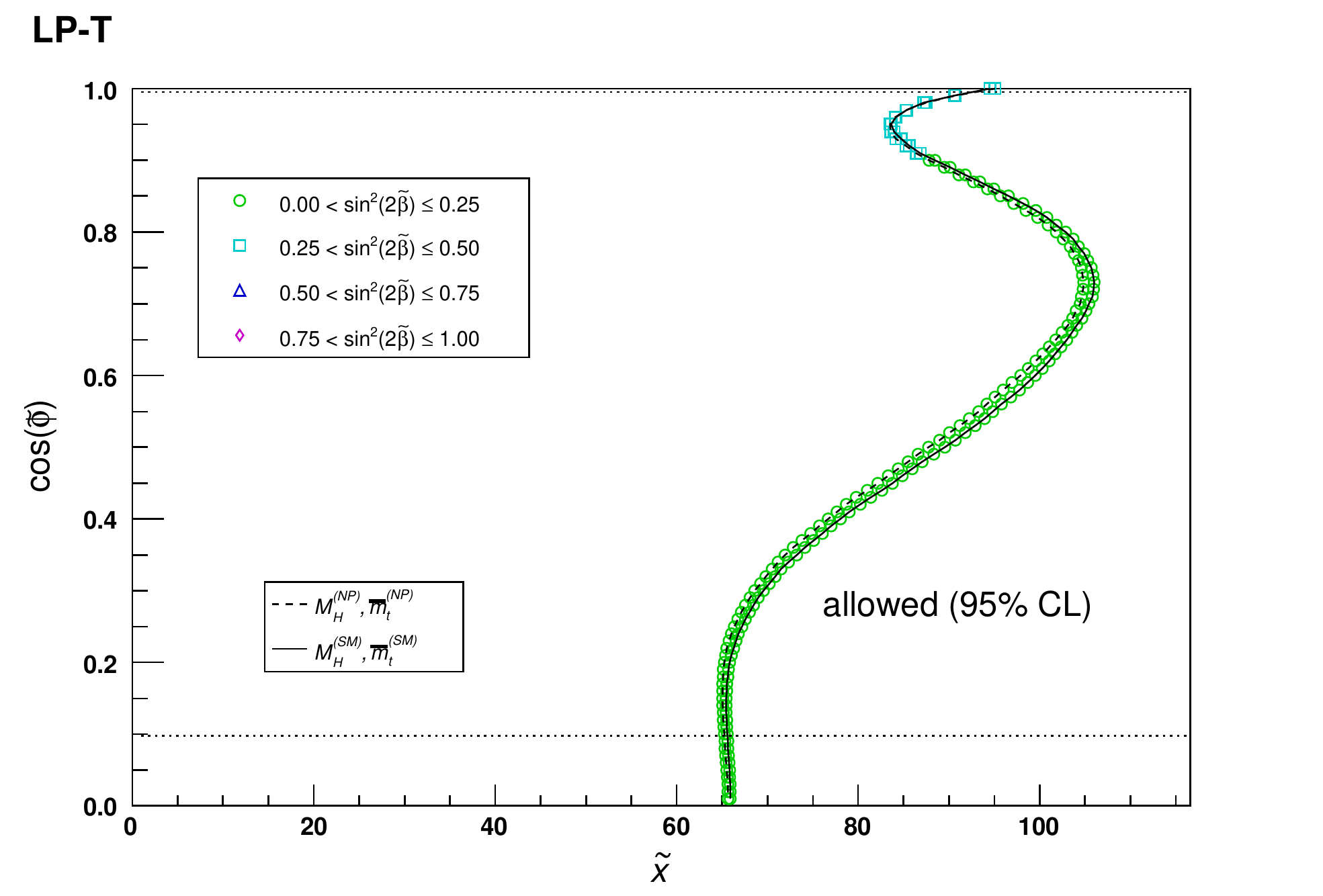}
\includegraphics[width=\picwidth]{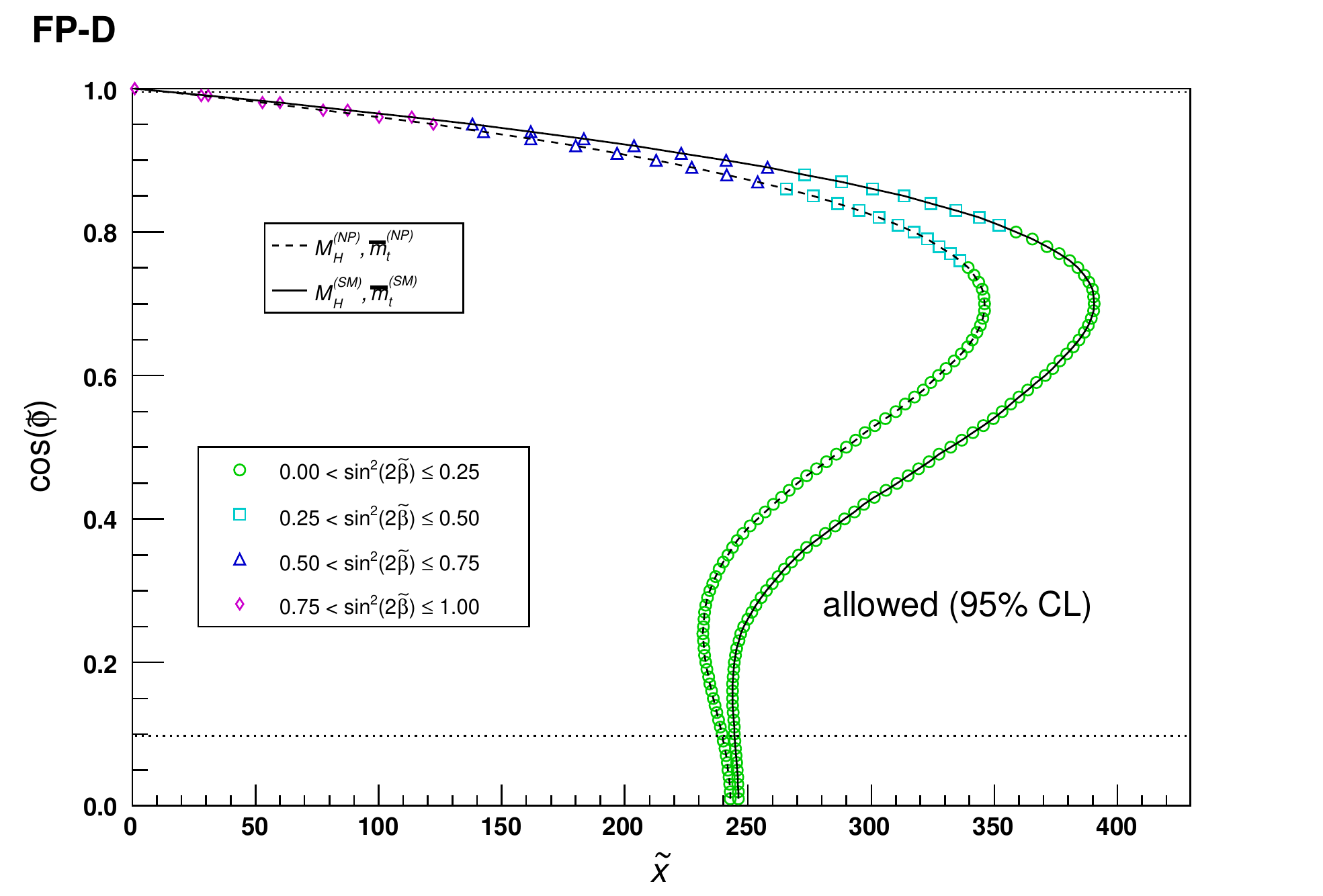}
\includegraphics[width=\picwidth]{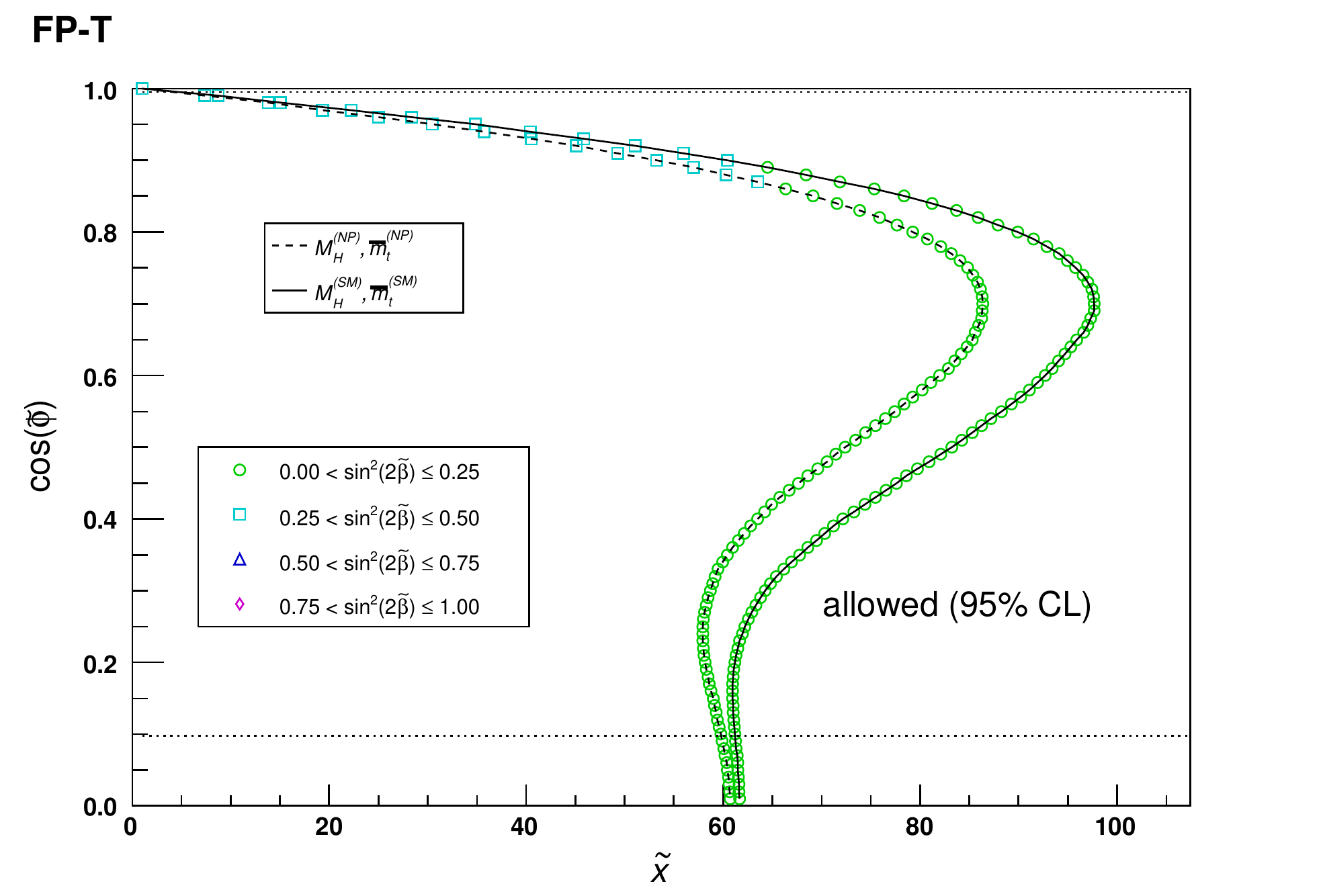}
\caption{
The 95\%  confidence contours of the various models
on the $\tilde{x}$-$\cos\phi$ plane, with $M_H$ and $\overline{m}_t$
either fixed as SM best-fit values (solid) or allowed to
be re-fitted (dashed).
The area to the right of each curve (large $\tilde{x}$ value) is
allowed by the global-fit analysis.
}
\label{fig:x-phi}
\end{center}
\end{figure}

\begin{figure}[h!t]
\begin{center}
\includegraphics[width=\picwidth]{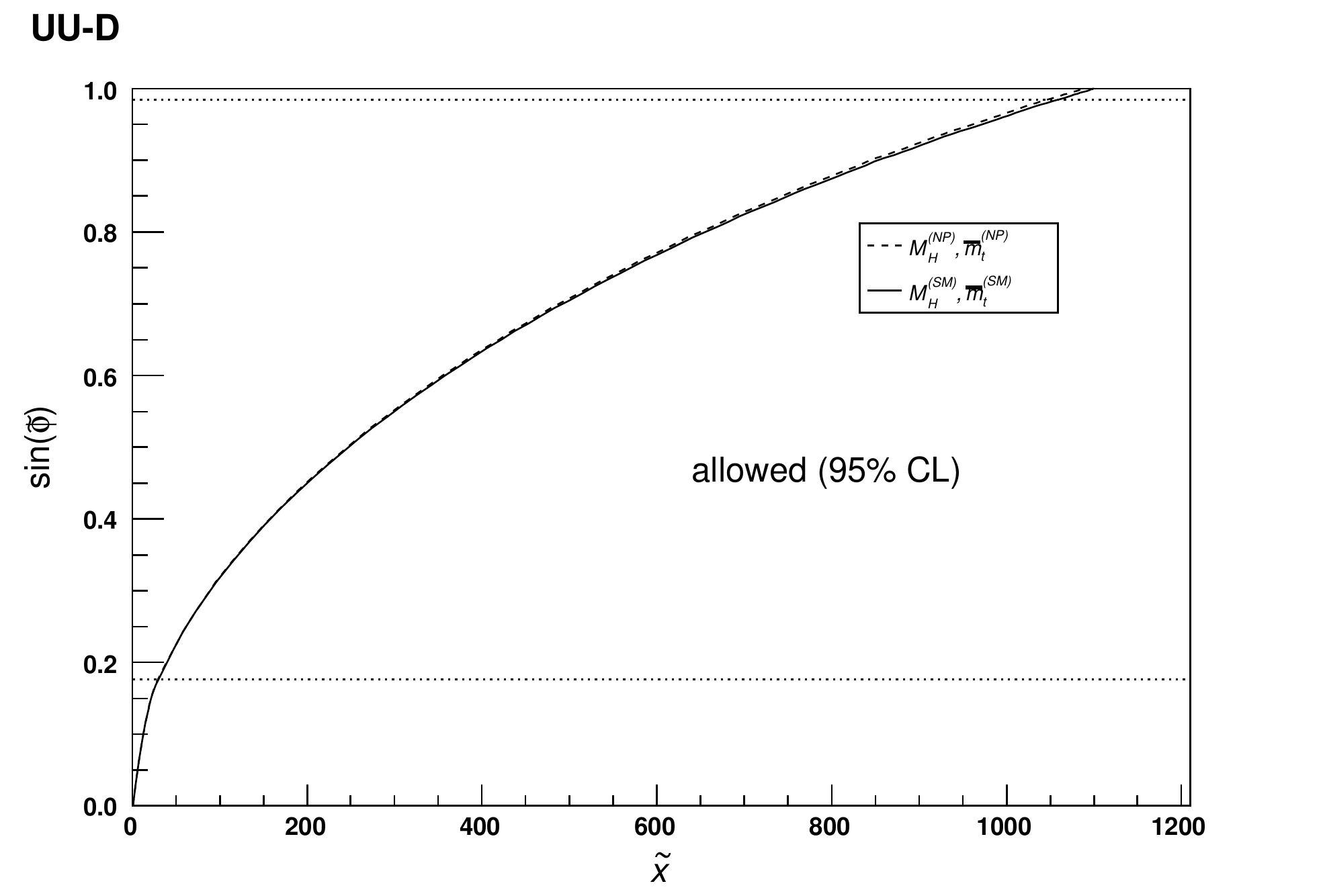}
\includegraphics[width=\picwidth]{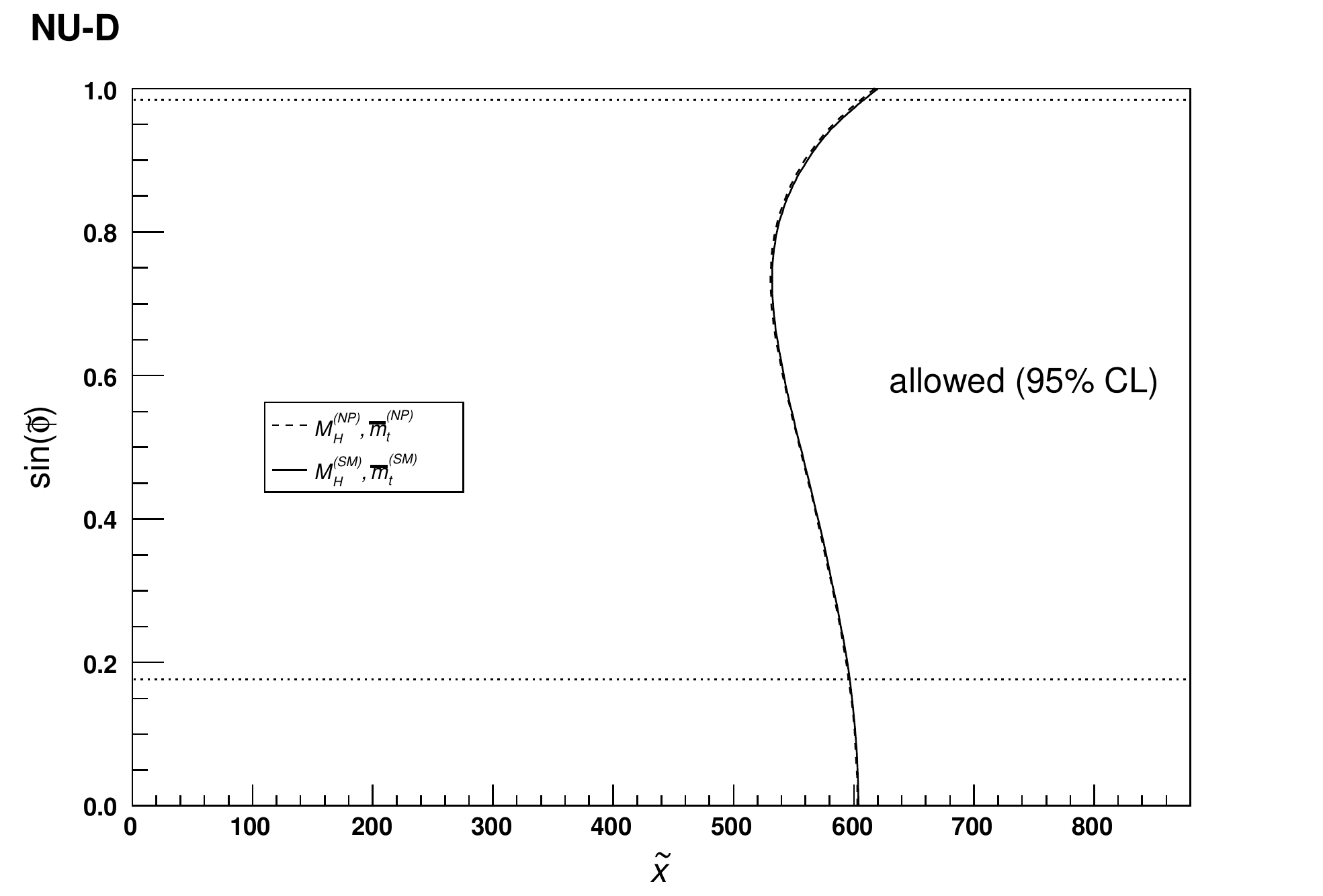}
\caption{
Same as Fig.~\ref{fig:x-phi}, but for the models
that follow the breaking pattern II.
The solid and dashed lines are almost indistinguishable.
}
\label{fig:x-phi2}
\end{center}
\end{figure}

In this section, we present the allowed regions of parameter space
based on the global-fit analysis.
A testament to the success of the SM is that, for all the $\tto$ models,
the global fitting pushes $\tilde{x}$ to large values, decoupling the effects
of the new physics.
This is presented in Figs.~\ref{fig:x-phi} and \ref{fig:x-phi2},
where we show the 95\% confidence level (C.L.) contours on the
$\tilde{x}-c_{\tphi}$ plane.

In addition to the constraints from the precision and low-energy data,
we also require $\cos\phi$ ($\sin\phi$)  to be
greater than 0.1 (0.18) for the first (second) breaking pattern so that all the gauge couplings in
Eq.~\Eref{eq:GaugeCouplingExpand0}, \Eref{eq:GaugeCouplingExpand1} and \Eref{eq:GaugeCouplingExpand} are perturbative and do not exceed
$\sqrt{4\pi}$.
These constraints are shown as horizontal dotted lines in the figures.

\begin{figure}[h!t]
\begin{center}
\includegraphics[width=\picwidth]{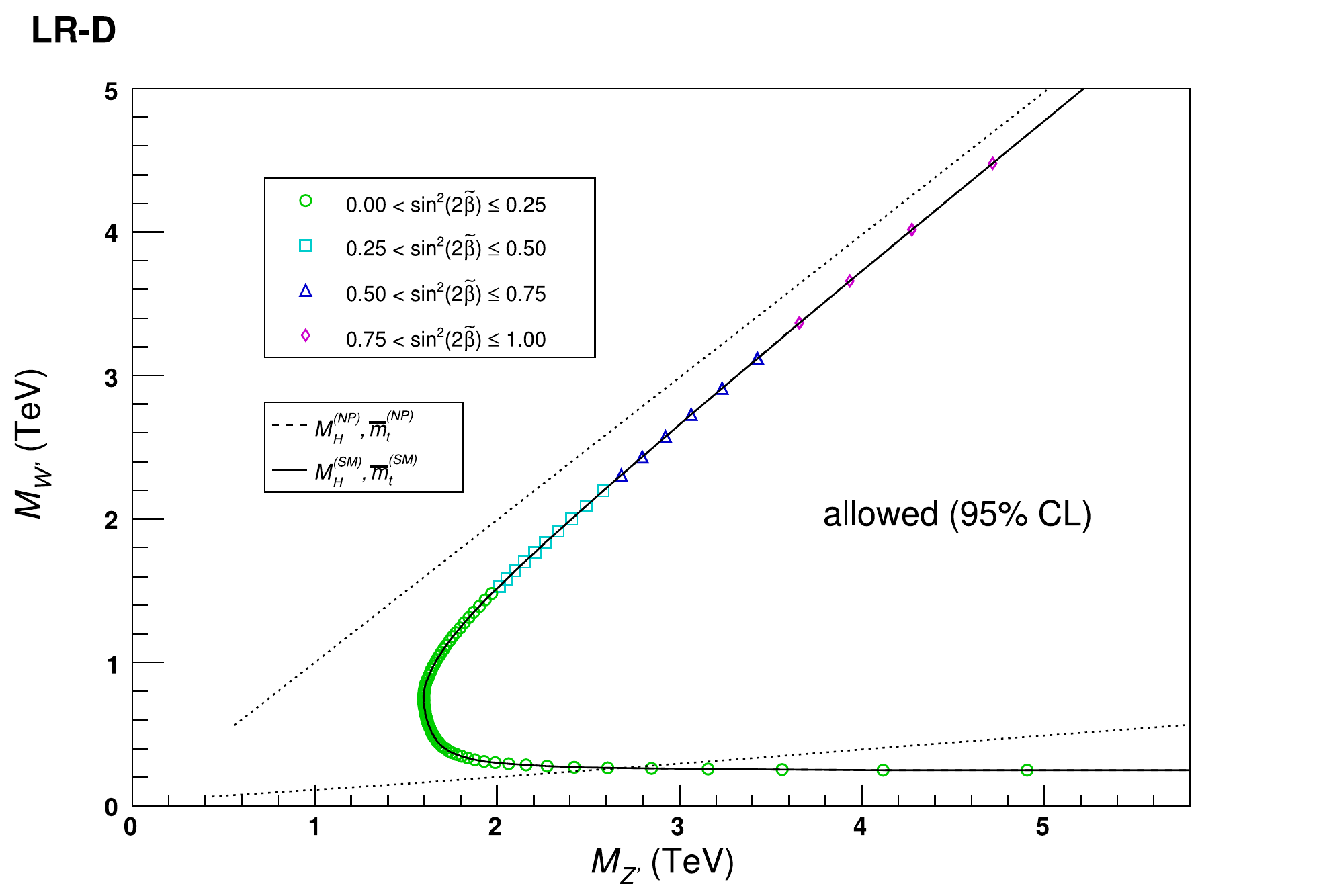}
\includegraphics[width=\picwidth]{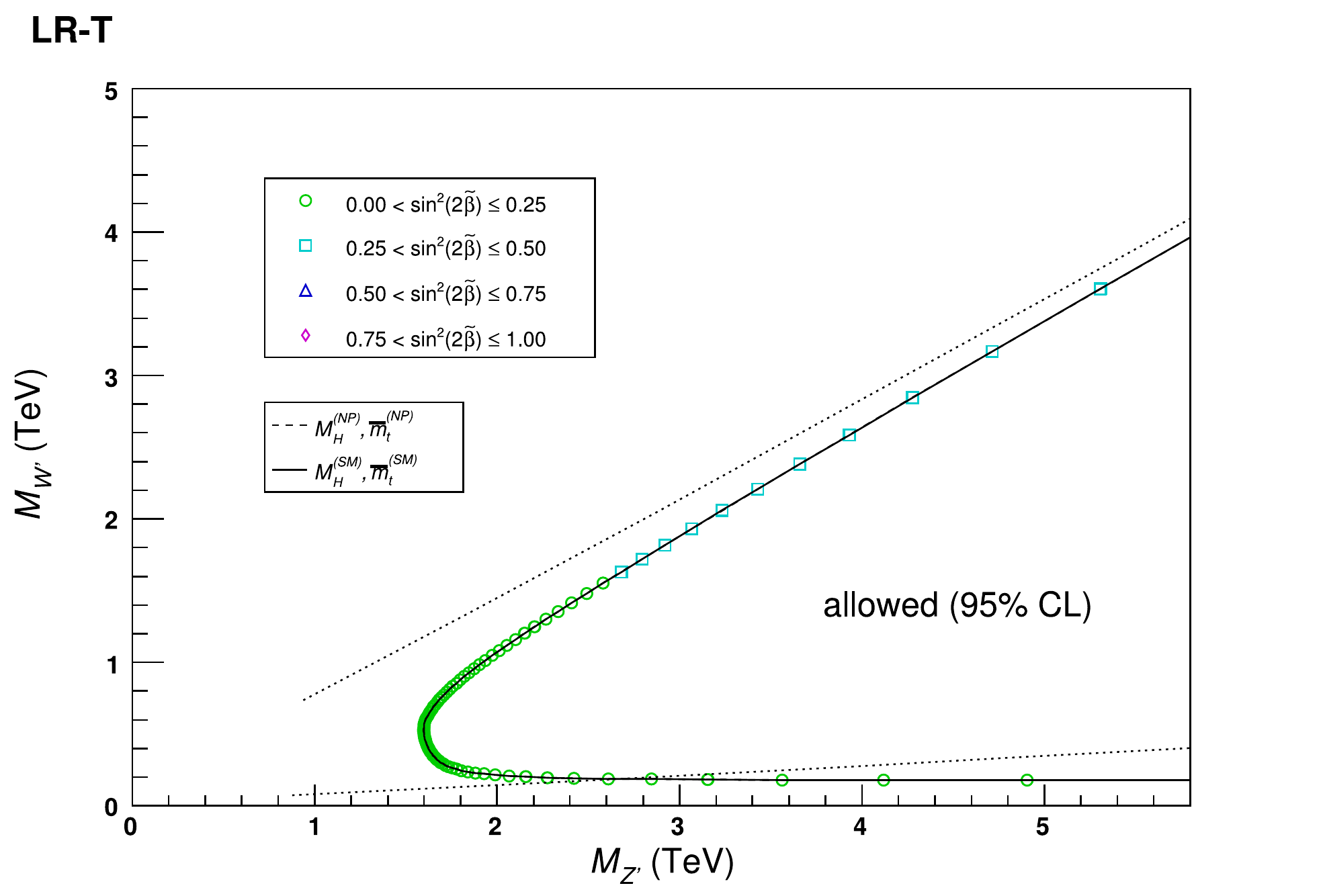}
\includegraphics[width=\picwidth]{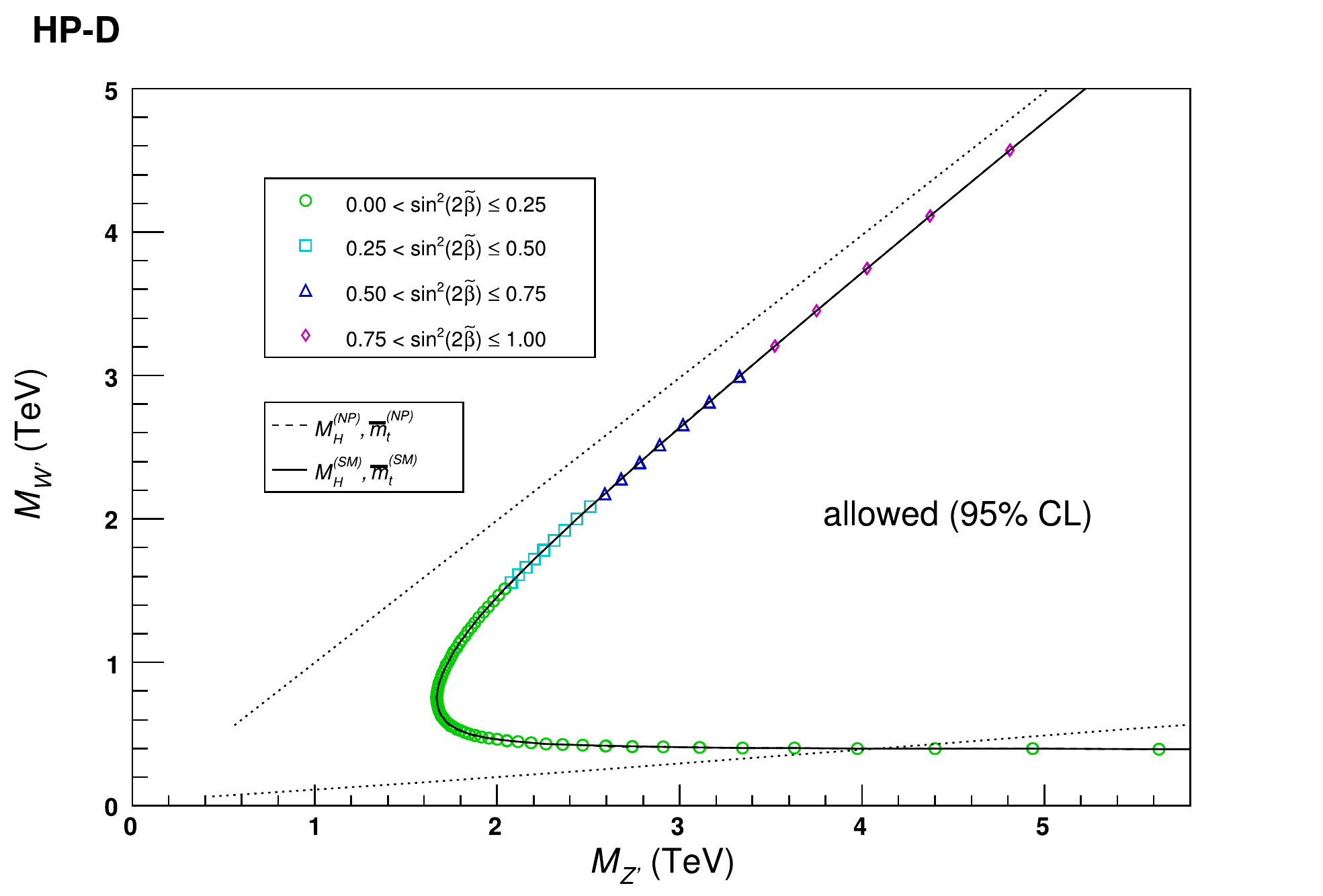}
\includegraphics[width=\picwidth]{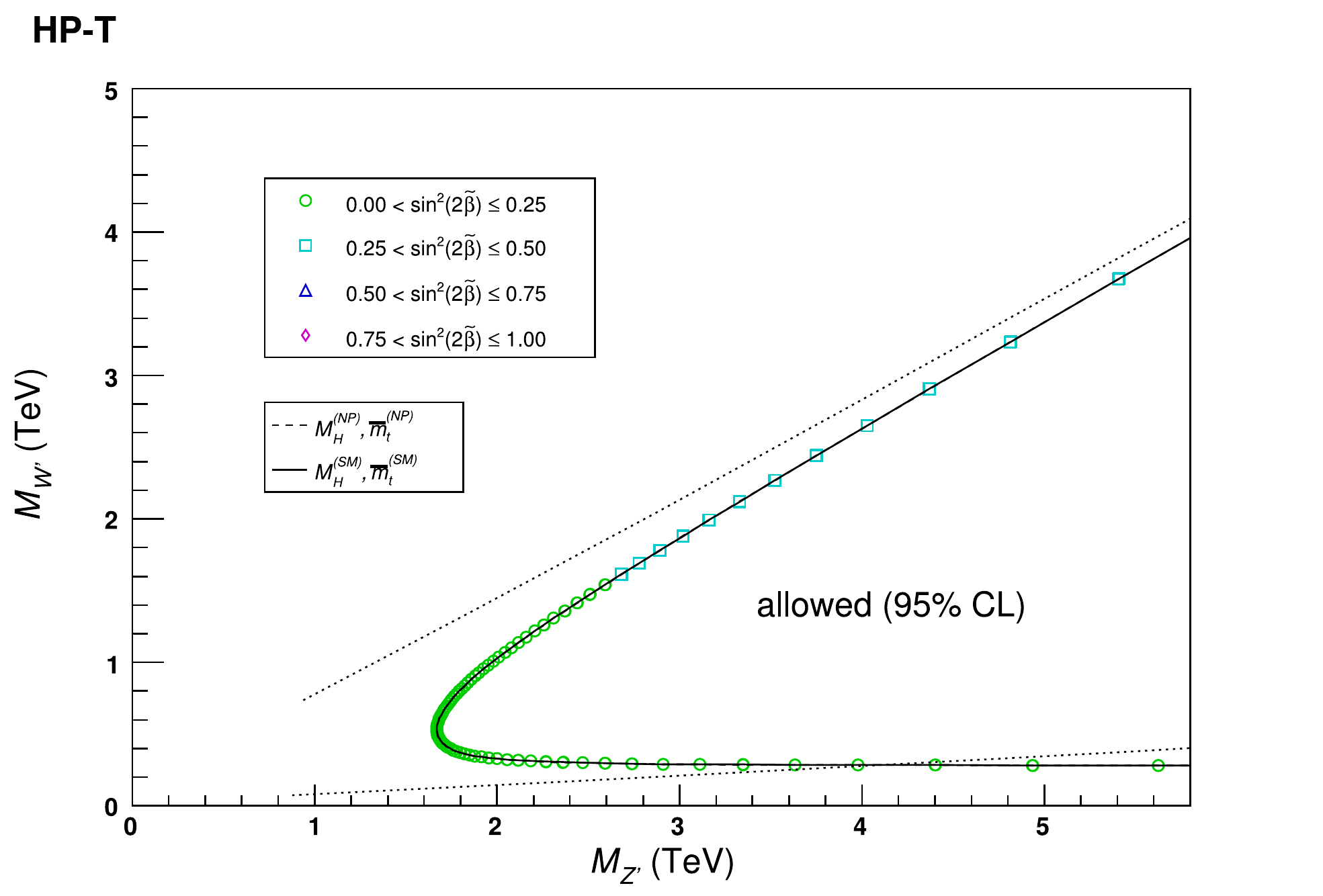}
\includegraphics[width=\picwidth]{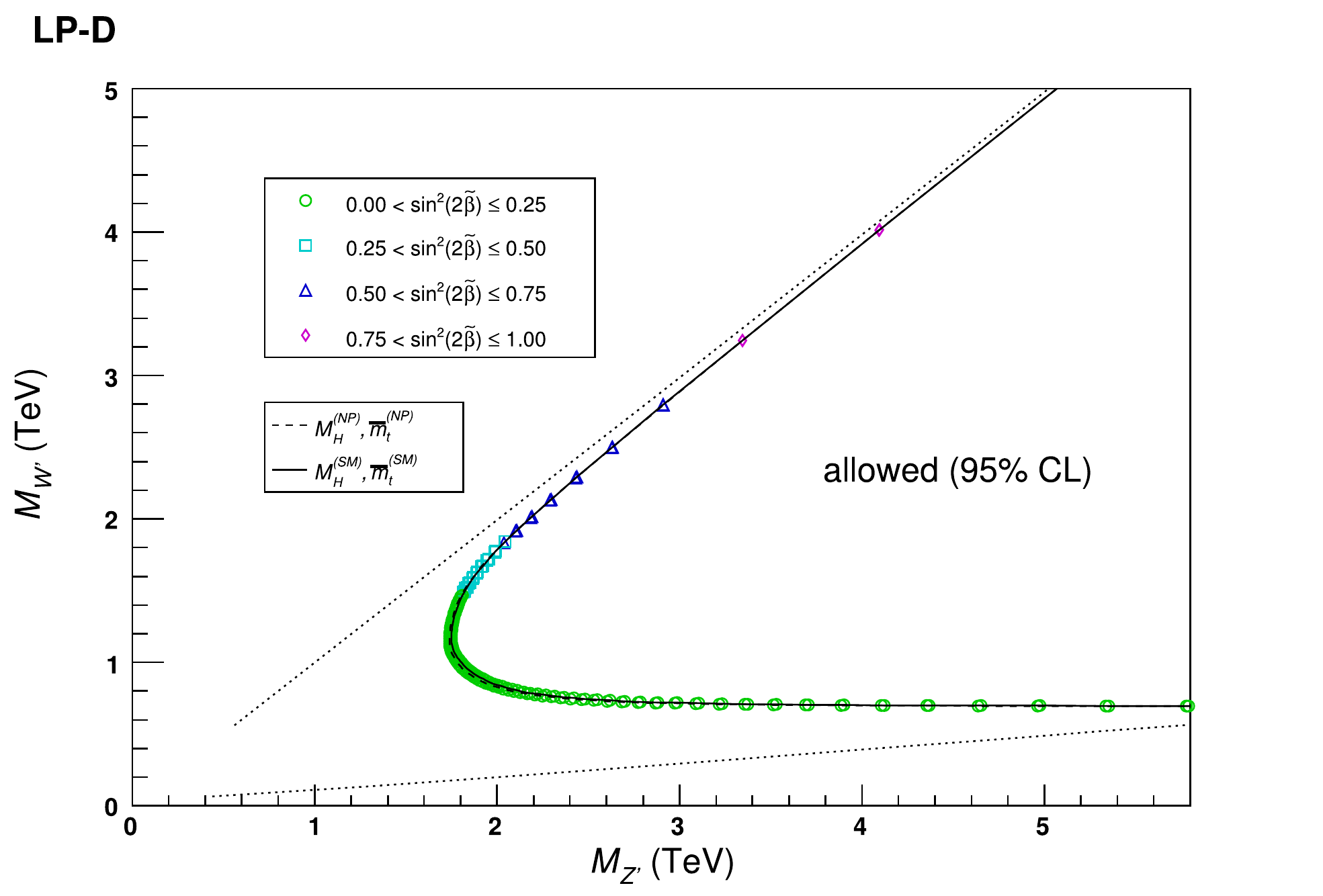}
\includegraphics[width=\picwidth]{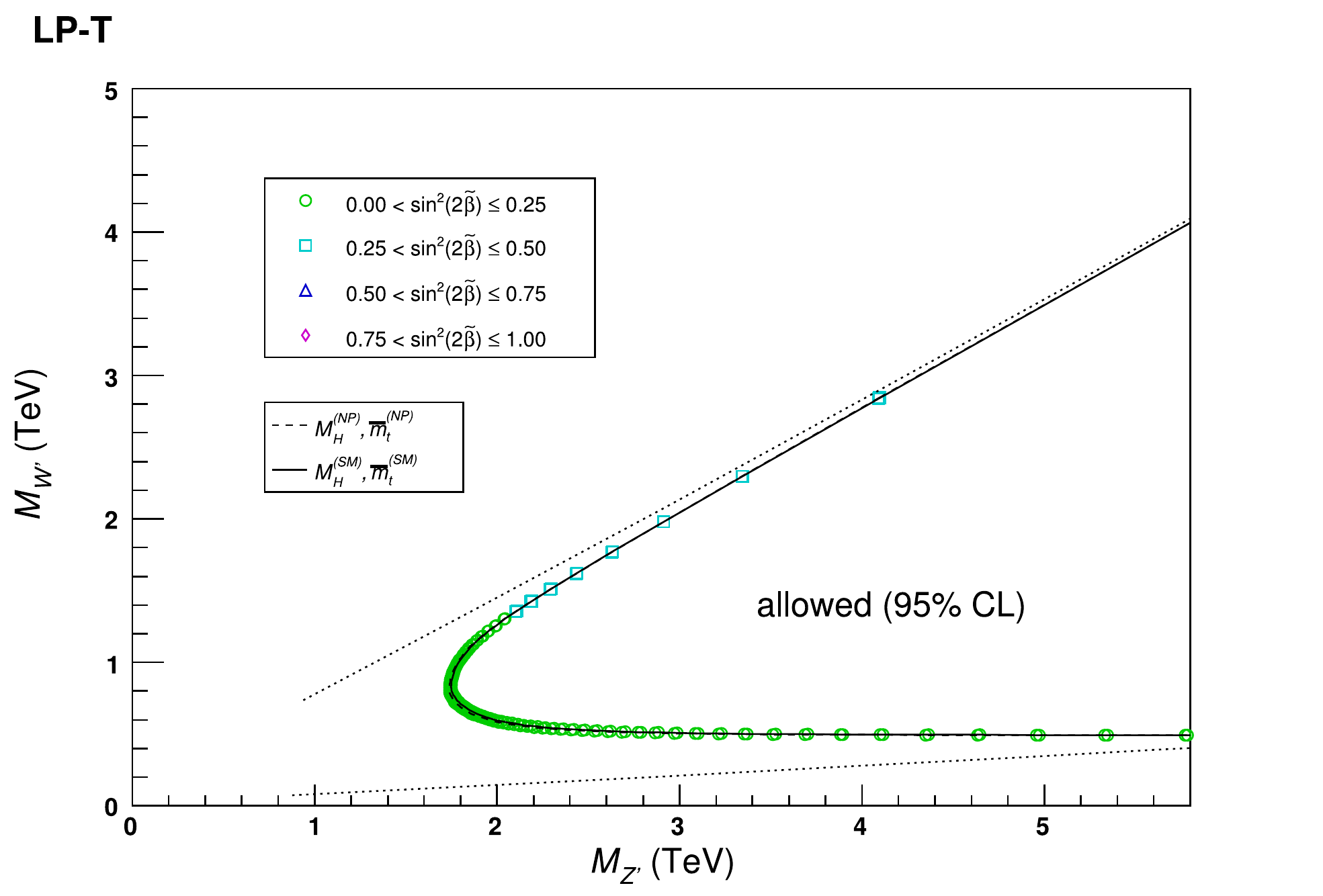}
\includegraphics[width=\picwidth]{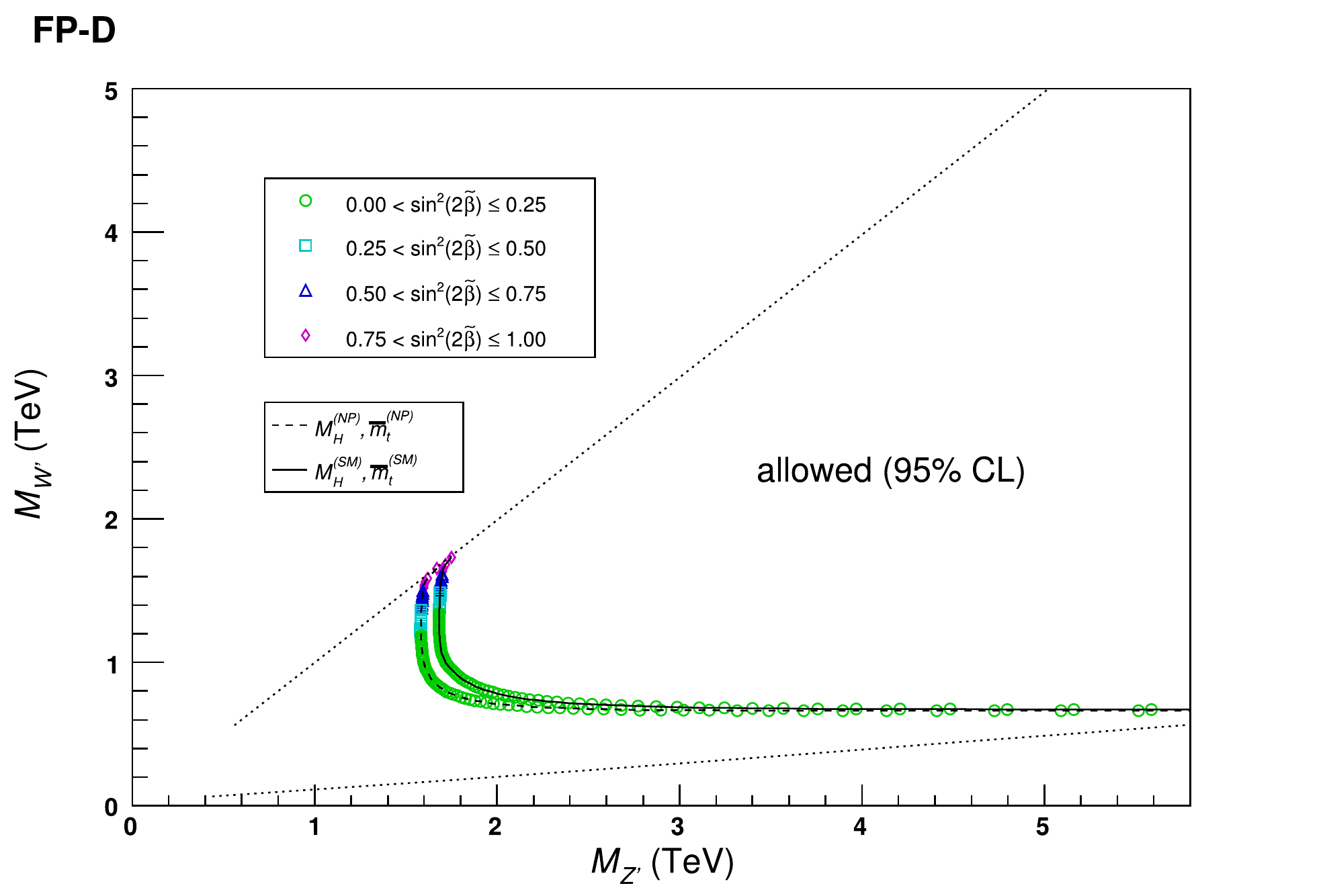}
\includegraphics[width=\picwidth]{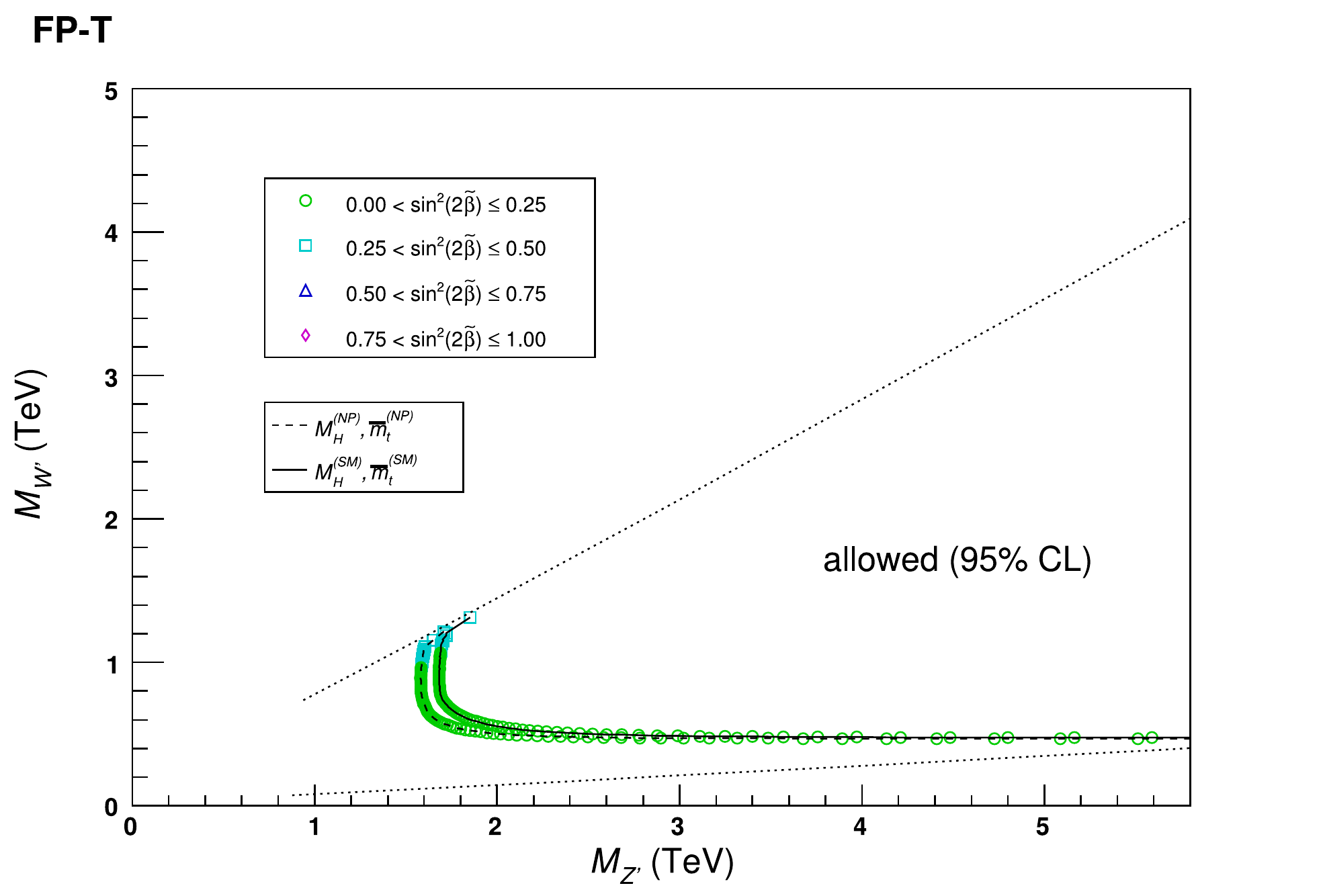}
\caption{
The 95\% confidence  contours of the various models
on the $M_{Z'}$-$M_{W'}$ plane, with $M_H$ and $\overline{m}_t$
either fixed as SM best-fit values (solid) or allowed to
be re-fitted (dashed).
The dotted lines represent lines of constant $\cos\phi$ and $\sin\phi$
at fixed values of 0.1 (0.18) for the first (second) breaking pattern.
Outside the cone surrounded by these regions, one of the gauge
couplings becomes non-perturbative. } \label{fig:MZP-MWP}
\end{center}
\end{figure}

\begin{figure}[h!t]
\begin{center}
\includegraphics[width=\picwidth]{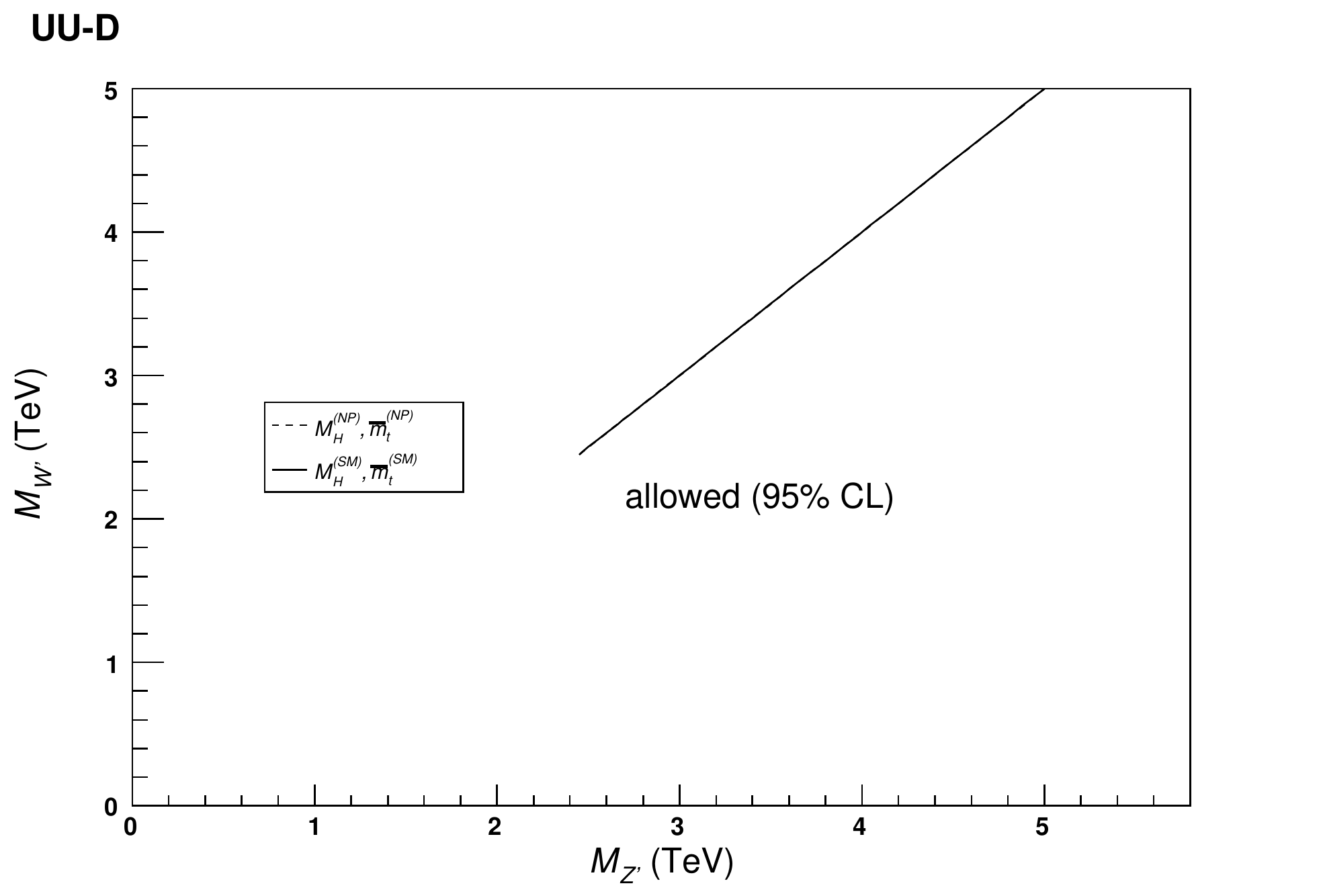}
\includegraphics[width=\picwidth]{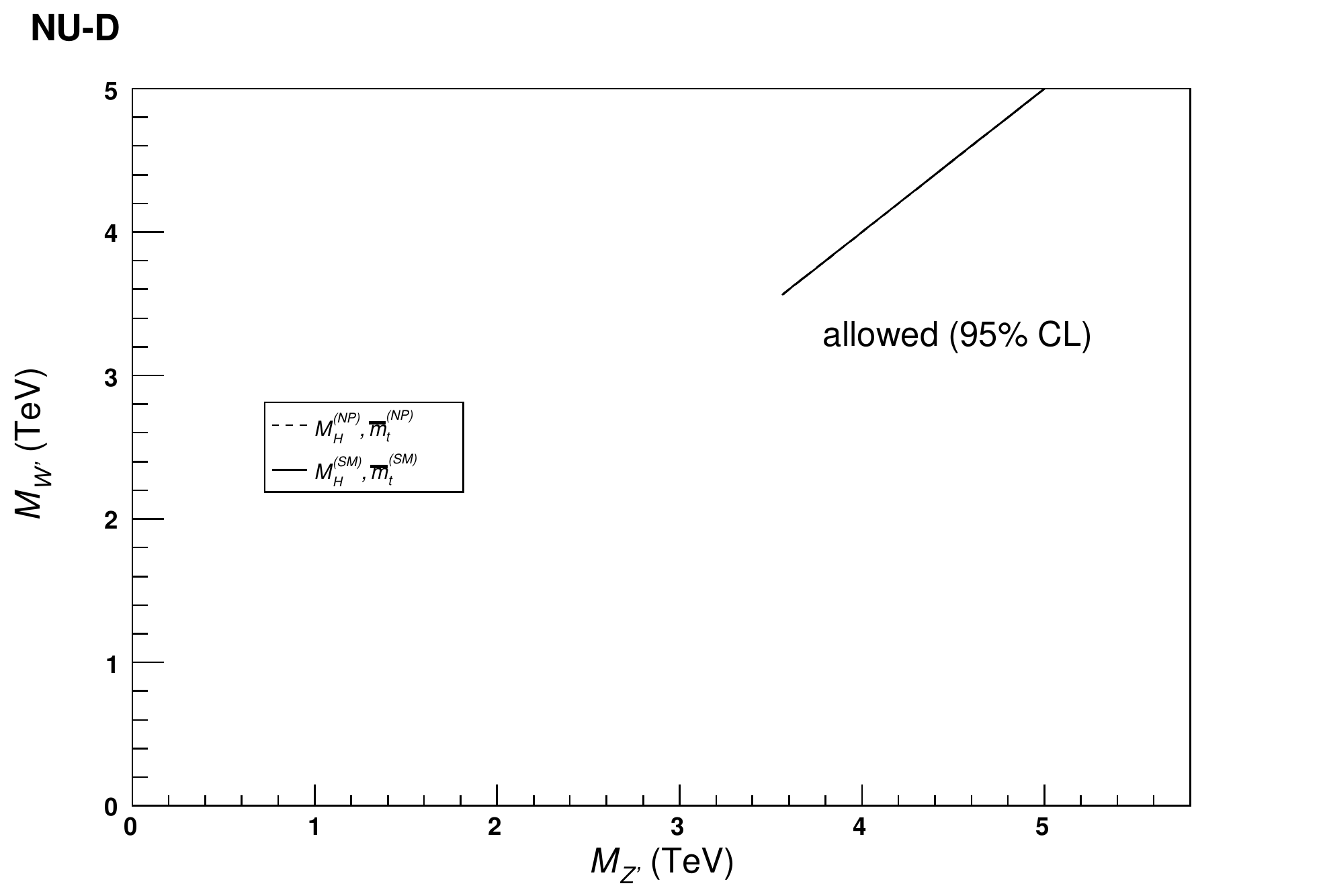}
\caption{
Same as Fig.~\ref{fig:MZP-MWP},
but for the models
that follow the breaking pattern II.
The contours appear as lines because in these models,
$Z^{\prime}$ and $W^{\prime}$ are highly degenerate due
to the pattern of symmetry breaking.
}
\label{fig:MZP-MWP2}
\end{center}
\end{figure}

Since $\tilde{x}$ and $\tphi$ are defined in a model-dependent
manner, it is also useful to show the corresponding contours on the
$M_{Z^{\prime}}$-$M_{W^{\prime}}$ plane to compare different $\tto$
models.
We translate the constraints on the parameter space of $\tilde{x}$
and $\tilde{\phi}$ to constraints on the masses of the new heavy
gauge bosons, and plot these bounds on the $M_{Z'}$-$M_{W'}$ plane
in Figs.~\ref{fig:MZP-MWP}  and \ref{fig:MZP-MWP2}.
From these plots, we can read off the lower bounds on the masses of
the $Z^\prime$ and the $W^\prime$ in these models, which are
presented in Table~\ref{tb:key-results-1}.
In the UU-D and the NU-D models, the masses of the $Z^\prime$ and
the $W^\prime$ bosons are nearly degenerate, as shown in
Fig.~\ref{fig:MZP-MWP2}.
In these two models, the minimum masses of the $Z^\prime$ and the $W^\prime$ consistent with the current experimental
data are respectively $2.48 ~\mbox{TeV}$ and $3.56~\mbox{TeV}$.

\begin{table}[h]
\begin{center}
\caption{
Lower bounds on the masses of the new heavy gauge bosons.
The superscripts SM or NP indicate whether $M_H$ and $\bar{m}_t$
were set to their SM best-fit values or fitted with rest of new
physics parameters.
Compare with the plots of the $M_{\Zp}$--$M_{W^{\prime}}$ plane in Fig.~\ref{fig:MZP-MWP}.
}
\label{tb:key-results-1}
\begin{tabular}{|c|c|c|c|c|}
\hline
 & $M_{\Zp}^{\tbox{(SM)}}$ [TeV] &
   $M_{\Zp}^{\tbox{(NP)}}$ [TeV] &
   $M_{W^{\prime}}^{\tbox{(SM)}}$ [TeV] &
   $M_{W^{\prime}}^{\tbox{(NP)}}$ [TeV] \\
\hline
LR-D & 1.602 & 1.602 & 0.269 & 0.269 \\
LP-D & 1.752 & 1.742 & 0.697 & 0.695 \\
HP-D & 1.674 & 1.673 & 0.403 & 0.403 \\
FP-D & 1.685 & 1.583 & 0.673 & 0.665 \\
\hline
LR-T & 1.607 & 1.607 & 0.197 & 0.197 \\
LP-T & 1.753 & 1.745 & 0.495 & 0.493 \\
HP-T & 1.680 & 1.679 & 0.289 & 0.289 \\
FP-T & 1.687 & 1.587 & 0.478 & 0.472 \\
\hline
UU-D & 2.479 & 2.474 & 2.479 & 2.474 \\
NU-D & 3.562 & 3.558 & 3.562 & 3.558 \\
\hline
\end{tabular}
\end{center}
\end{table}

\subsection{Key Observables and Their Impacts}
For the models that follow the first pattern of symmetry breaking,
we see that models in which Higgs triplets break $SU(2)_2\times
U(1)_X$ lead to smaller bounds on $\tilde{x}$ compared to models
where Higgs doublets break the symmetry.
This is not surprising given
Eqs.~\Eref{eq:GF-relation01},\Eref{eq:MZ-relation01} , where we see
that, in the triplet models, the corrections to the definitions of
the reference parameters are suppressed compared to the doublet
models.
However, the bounds on $M_{W^{\prime}}$ and $M_{Z^{\prime}}$ are comparable.

According to how the contours in parameter space are shaped, the
considered $\tto$ models can be separated into three classes:
\begin{itemize}
\item
In the LR, HP, and UU models, large values of $\ctphi$ ($\stphi$
for UU) are ruled out at small $\tx$.
At small $\ctphi$ ($\stphi$), the parameter contours, however,
extend to relatively low $\tx$ values.

\item
The contours of the LP and FP models are, by contrast, located at
comparatively larger $\tx$ values for small $\ctphi$.
Increasing the values of $\ctphi$ to  about 0.8, the contours of
these models curve to the right (towards higher $\tx$).
However, increasing $\ctphi$ further beyond
 about 0.8, the FP contours bend
towards lower $\tx$, while the LP contours towards higher $\tx$.

\item
The parameter contour of the NU model is unique as it is the only
curve that bends to the left with smaller $\tilde{x}$ value when
going up along the vertical axis with increasing $s_{\tilde{\phi}}$
value.
\end{itemize}

The similarities and differences between the parameter plots can
be traced back to certain key observables.
That is, in the excluded regions of parameter spaces we
consistently observe a pattern that several key observables
contribute with especially large pulls to $\chi^2$, and it is
these observables that drive the overall shapes of the curves in
Fig.~\ref{fig:x-phi}.
\begin{table}[h]
\begin{center}
\caption{Overview of the observables that drive the parameter
plots. The most and second most important observables are
respectively marked with the symbols \ding{192} and \ding{193}. In
the UU-D and NU-D models only one observable significantly
contributes to $\chi^2$.}
\label{tb:Drivers}
\begin{tabular}{|c|c|c|c|c|c|}
\hline
Model  & $\sigma_{\tbox{had}}$ & $A_{FB}(b)$ & $\gNLt$    & $Q_W({}^{133}\mbox{Cs})$ & Set of other obs.\\
\hline\hline
LR, HP & \ding{192}             & \ding{193}  &            &                          & \\
\hline
LP, FP &                        &             & \ding{193} & \ding{192}               & \\
\hline\hline
UU     & \ding{192}             &             &            &                          & \ding{193} \\
\hline
NU     &                        &             & \ding{192} &                          & \ding{193} \\
\hline
\end{tabular}
\end{center}
\end{table}

In Table~\ref{tb:Drivers} we give an overview of these observables
that effectively drive the results in Fig.~\ref{fig:x-phi}.
For each model in the breaking pattern I, we list the two most
important observables.
For the models of the breaking pattern II, we give only one
such observable.
It is important to note that Table~\ref{tb:Drivers} only presents
\textit{qualitative} observations that indicate tendencies, and it
may be the case that some particular points of the parameter
spaces have other observables that contribute with larger pulls
than the ones we indicate here.
Nevertheless, the patterns we give here are useful in indicating
qualitatively which observables are likely to be sensitive to new
physics contributions from the $\tto$ models.

The explicit expressions for the new physics corrections to these
key observables, are listed in Tables \ref{tb:obs-correction-SigAFB}
and \ref{tb:obs-correction-gLQW}. At the $95\%$ C.L.,  the set of observables ($\sigma_{\tbox{had}}$, $A_{FB}(b)$,
$A_{LR}(e)$, $\gNLt$, $Q_W({}^{133}\mbox{Cs})$) is respectively measured with a precision  at the ($0.18, 3.16, 2.80, 0.90, 1.24$) percent level.
\begin{table}[h]
\begin{center}
\caption{ Fractional new physics corrections to the observables
$\sigma_{\tbox{had}}$, $A_{FB}(b)$ and
$Q_W\left({}^{133}\mbox{Cs}\right)$ relative to the corresponding
SM predictions.
To obtain the new physics shifts in the triplet versions of the
LR, LP, HP and FP models, the prefactors of $s_{2\tbet}^2$ need to
be multiplied by $\frac{1}{2}$, all other terms by $\frac{1}{4}$.
}
\label{tb:obs-correction-SigAFB} \vspace{0.125in}
\begin{tabular}{|c|c|c|}
\hline
 & $\tx\ \delta\sigma_{\tbox{had}}/\sigma_{\tbox{had,SM}}$ &
   $\tx\ \delta A_{FB}(b)/A_{FB,\tbox{SM}}(b)$ \\
\hline LR-D & $-1.13 \  \ctphi^2 - 0.142 \  \ctphi^4 + 0.0432 \
s_{2\tbet}^2$ &
       $-30.0 \  \ctphi^2 + 67.6 \  \ctphi^4 - 20.6 \  s_{2\tbet}^2$ \\
LP-D & $+0.346 \  \ctphi^2 - 0.142 \  \ctphi^4 + 0.0432 \
s_{2\tbet}^2$ &
       $-46.1 \  \ctphi^2 + 67.6 \  \ctphi^4 - 20.6 \  s_{2\tbet}^2$   \\
HP-D & $-1.38 \  \ctphi^2 - 0.142 \  \ctphi^4 + 0.0432 \
s_{2\tbet}^2$ &
       $-30.9 \  \ctphi^2 + 67.6 \  \ctphi^4 - 20.6 \  s_{2\tbet}^2$  \\
FP-D & $+0.0985\  \ctphi^2 - 0.142 \  \ctphi^4 + 0.0432 \
s_{2\tbet}^2$ &
       $-47.0 \  \ctphi^2 + 67.6 \  \ctphi^4 - 20.6 \  s_{2\tbet}^2$\\
\hline UU   & $-0.889\  \stphi^2 - 0.0132\  \stphi^4 $ &
       $+0.161 \  \stphi^2 + 6.29 \  \stphi^4$ \\
NU   & $+0.583 \  \stphi^2 - 0.0132\  \stphi^4 $ &
       $+14.2  \  \stphi^2 + 6.29 \  \stphi^4$ \\
\hline
\end{tabular}
\medskip
\begin{tabular}{|c|c|}
\hline
 & $\tx\ \delta Q_W\left({}^{133}\mbox{Cs}\right)/Q_{W,\tbox{SM}}\left({}^{133}\mbox{Cs}\right)$ \\
\hline
LR-D &  $-0.855 \  \ctphi^4 - 0.145 \  s_{2\tbet}^2$ \\
LP-D &  $+3.35 - 1.95 \  \ctphi^2 - 0.855 \  \ctphi^4 - 0.145 \  s_{2\tbet}^2$ \\
HP-D &  $-0.855 \  \ctphi^4 - 0.145 \  s_{2\tbet}^2$ \\
FP-D &  $+2.95- 1.95 \  \ctphi^2 - 0.855 \  \ctphi^4 - 0.145 \  s_{2\tbet}^2$ \\
\hline
UU   & $- 0.855 \  \stphi^4 $ \\
NU   & $+0.406 + 0.594 \  \stphi^2 - 0.855 \  \stphi^4 $ \\
\hline
\end{tabular}
\end{center}
\end{table}

\begin{figure}[ht]
\includegraphics[width=7.5cm]{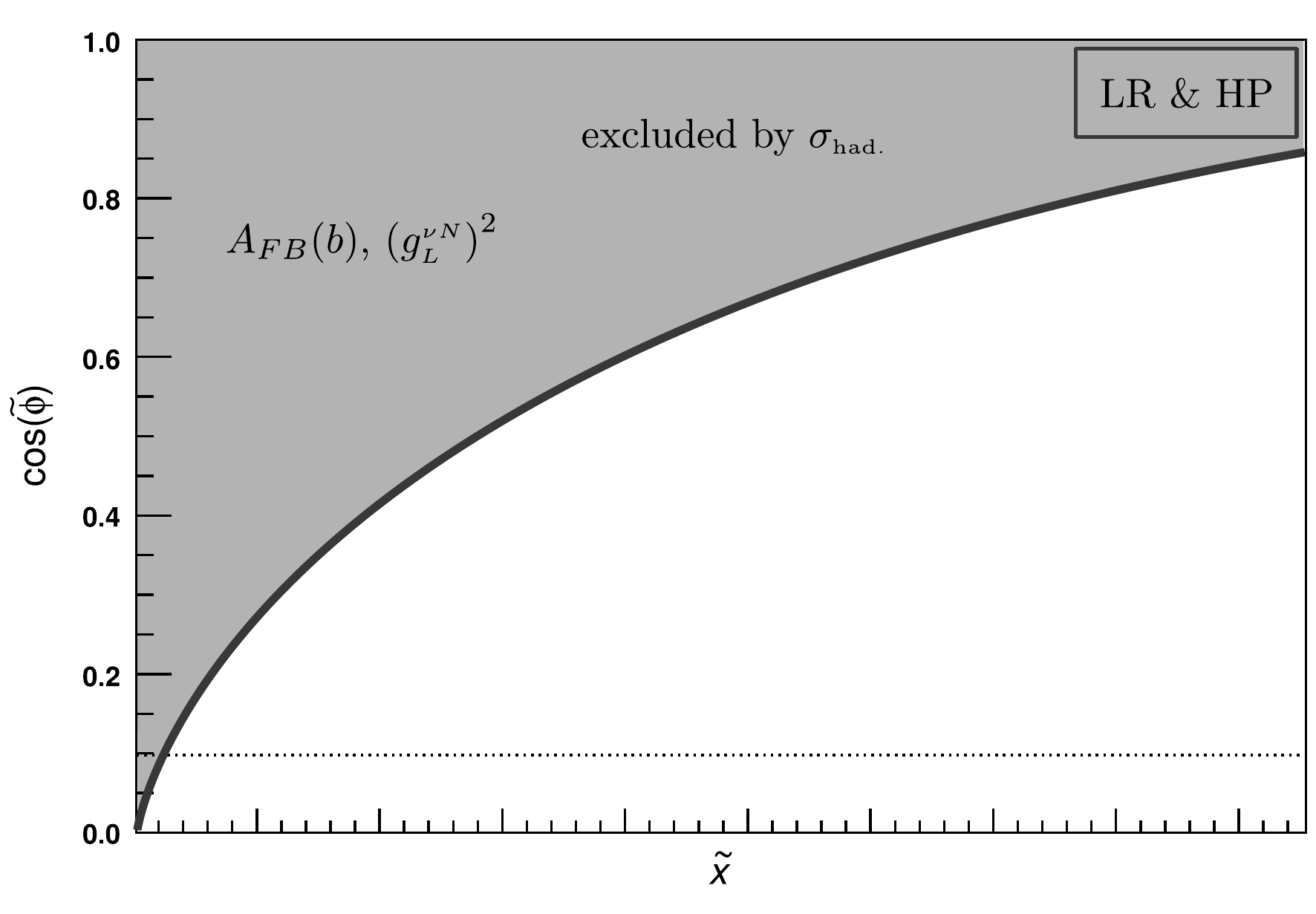}
\includegraphics[width=7.5cm]{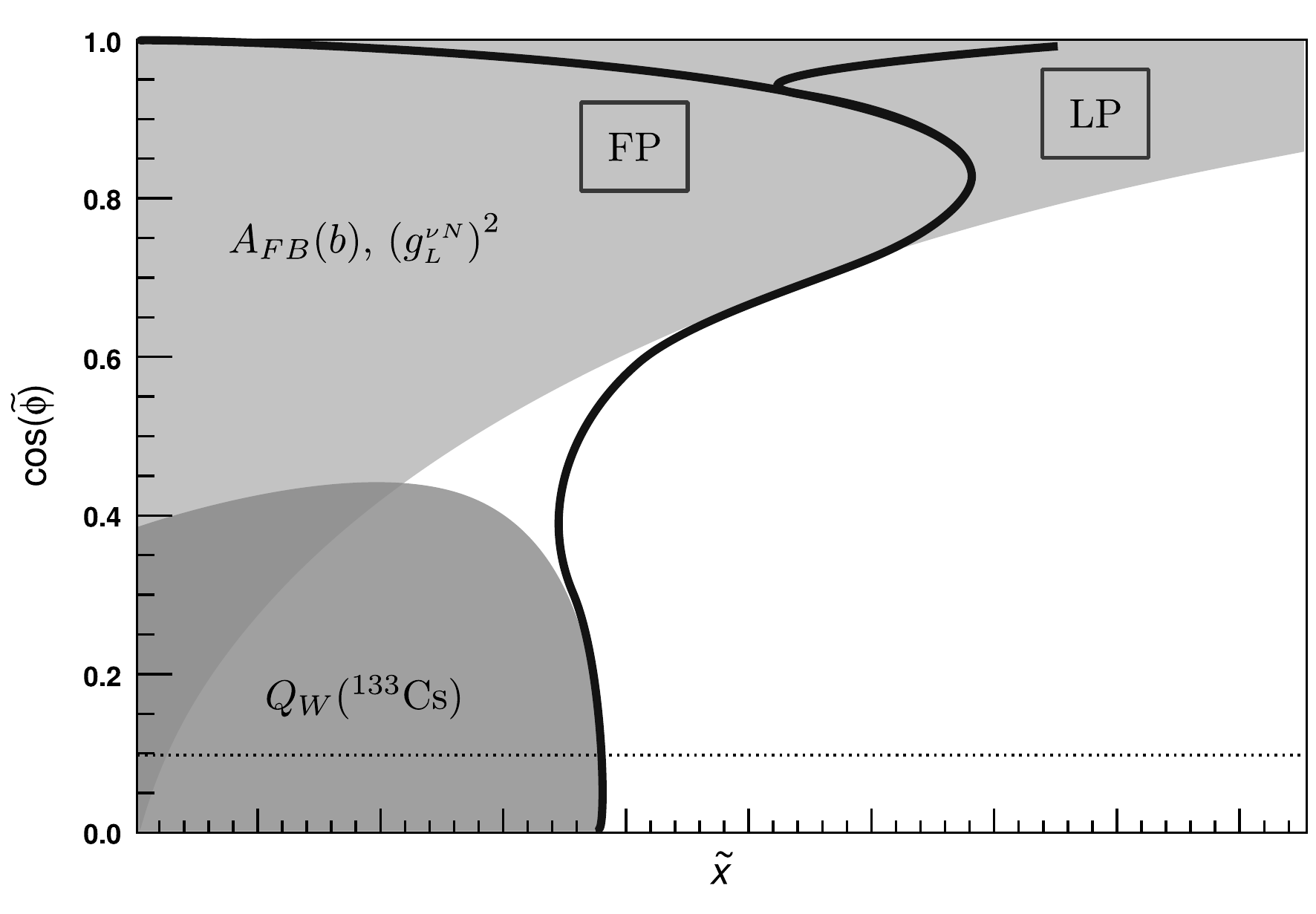}
\caption{Sketches illustrating the influences of some key
observables on the parameter bounds for the models of the first
breaking pattern.
The UU-D parameter contour is driven by
$\sigma_{\tbox{had}}$ as well.
In the NU-D model $\gNLt$ is the
most important observable.
\label{fig:Sketches}}
\end{figure}

\begin{table}[h]
\begin{center}
\caption{New physics corrections to the observable $\gNLt$
relative to the prediction of the SM.}
\label{tb:obs-correction-gLQW} \vspace{0.125in}
\begin{tabular}{|c|c|}
\hline
 & $\tx\ \delta \gNLt/\gNLSMt$ \\
\hline
LR-D, LP-D, HP-D, FP-D & $ 0.0875 + 1.91 \  \ctphi^2 + 0.839 \  \ctphi^4 - 2.84 \  s_{2\tbet}^2 $ \\
LR-T, LP-T, HP-T, FP-T & $ 0.0219 + 0.478 \  \ctphi^2 + 0.210 \  \ctphi^4 - 1.42 \  s_{2\tbet}^2 $ \\
\hline
UU-D & $ 0.839 \  \stphi^4$ \\
NU-D & $ 2.58 - 0.583 \  \stphi^2 + 0.839 \  \stphi^4$ \\
\hline
\end{tabular}
\end{center}
\end{table}
Based on these expressions we can roughly reconstruct the
respective shapes of the contours, and in Fig.~\ref{fig:Sketches}
we illustrate our argumentation.

We find that the shapes of the contours for LR, HP and UU models are
driven by $\sigma_{\tbox{had}}$.
For the LR and HP models, $A_{FB}(b)$ and $A_{LR}(e)$ also play an
important role. Since $A_{FB}(b)$ is defined as $A_{FB}(b)\equiv
\frac{3}{4}A_{LR}(e)A_{FB}(b)$, the large coefficients in $x\delta
A_{FB}^b / A_{FB}^b$ (See Table~\ref{tb:obs-correction-SigAFB})
originate from the smallness of the SM value of $A_{LR}^e$. The combination $x\delta
A_{FB}^b / A_{FB}^b$ imposes about the same constraints as those
derived from $x\delta A_{LR}^e / A_{LR}^e$. Since they are
strongly correlated, we only list the observable $A_{FB}(b)$.
In the LP and FP models, with low $\ctphi$ values,
$Q_W\left({}^{133}\mbox{Cs}\right)$ is the most important
observable, because of the large constant terms independent of
$\tilde{\phi}$ or $\tilde{\beta}$ in $Q_W$, as shown in
Table~\ref{tb:obs-correction-SigAFB}. The constant term for either the LR-D or the
HP-D model vanishes because it is proportional to $(T_L^3-T^3_R)$ of the
electron with the quantum number assignment $T_L^3 = T^3_R =
-1/2$.
 With high $\ctphi$ values, the observable $\gNLt$
determines the shape of the parameter contours.
The NU contour is mainly driven by the pull of $\gNLt$.
( We note that, $A_{FB}(b)$ has  a similar effect on the LP and FP
contours as on the LR and HP contours, though subdominant compared
to the other observables.
The same applies to $\gNLt$ for the LR and HP models.)

 The starting
point of our discussion is that the SM represents the best
description of the present experimental data, and the $\tto$
parameters have to be chosen such that they minimize the new physics
shifts.
We first focus on models in the breaking pattern I, the LR,
LP, HP, and FP models, and note the following points:
\begin{itemize}
\item
The observables $A_{FB}(b)$
 and $\gNLt$ prefer small $\ctphi$. In
$\delta\gNLt$, the $\ctphi^2$ and $\ctphi^4$ terms have the same
sign so that they cannot cancel each other. $A_{FB}(b)$ has the
largest effect on the allowed parameter space in the LR and HP
models as the coefficients of $\ctphi^2$  and $c_{\phi}^4$ are large
in magnitude.

\item
In the case of the LR and HP models the $\ctphi^2$ and $\ctphi^4$
contributions to $\delta\sigma_{\tbox{had}}$ have the same sign and
the $s_{2\tbet}^2$ term is suppressed by a small prefactor.
The pull of $\sigma_{\tbox{had}}$ therefore represents the hindrance
for the LR and HP models to accommodate large values of $\ctphi$ at
smaller $\tx$.

\item
The large impact of $Q_W\left({}^{133}\mbox{Cs}\right)$ on the LP
and FP bounds is due to the  large constant term in $\delta
Q_W\left({}^{133}\mbox{Cs}\right)$.
The only way to make $\delta Q_W\left({}^{133}\mbox{Cs}\right)$
small (other than simply raising $\tx$) is to have large $\ctphi$
so that the negative contributions from the $\ctphi^2$ and
$\ctphi^4$ terms can compete with the positive constant term. 
Consequently, the low-$\ctphi$ region is ruled out in the LP and
FP models and the parameter contours start at higher $\tx$ values
than in the LR and HP models.

\item
In the $\sigma_{had.}$, $A_{FB}(b)$ and $(g^{\nu N}_L)^2$
observables, the $s_{2\tbet}^2$ and $\ctphi^4$ terms always have
opposite sign.
The LP and FP parameter plots suggest that, depending on the exact
interplay between $s_{2\tbet}^2$ and $\ctphi$, the $s_{2\tbet}^2$
terms may be able to overcome the $\ctphi^4$ contributions such
that the contours are pulled back towards lower $\tx$ values.
Note, however, that the expressions given in
Table~\ref{tb:obs-correction-SigAFB} cannot explain the branching
between the LP and FP contours.
To account for that effect we certainly would have to extend our
discussion to a larger set of observables.

\end{itemize}

After these comments on the models of the breaking pattern I,
it is easy to understand the shape of the UU and NU contours.
In the UU model all shifts that we present in Tables
\ref{tb:obs-correction-SigAFB} and \ref{tb:obs-correction-gLQW}
favor small $\stphi$ values.
Especially the fact that the $\stphi^2$ and $\stphi^4$ terms in
$\delta\sigma_{\tbox{had}}$ have the same sign leads to the
exclusion of the high-$\stphi$ region.
For that reason the UU plot looks similar to the plots of the LR
and HP models.
The contour of the NU-D model is mainly influenced by the
correction to $\gNLt$.
Since $\delta\gNLt$ is small if $\stphi$ takes some intermediate
value we observe a bump in the NU-D contour towards lower $\tx$
values for $\stphi$ values around $0.7$.

\begin{figure}[tbh]
\begin{center}
\includegraphics[width=\picwidth]{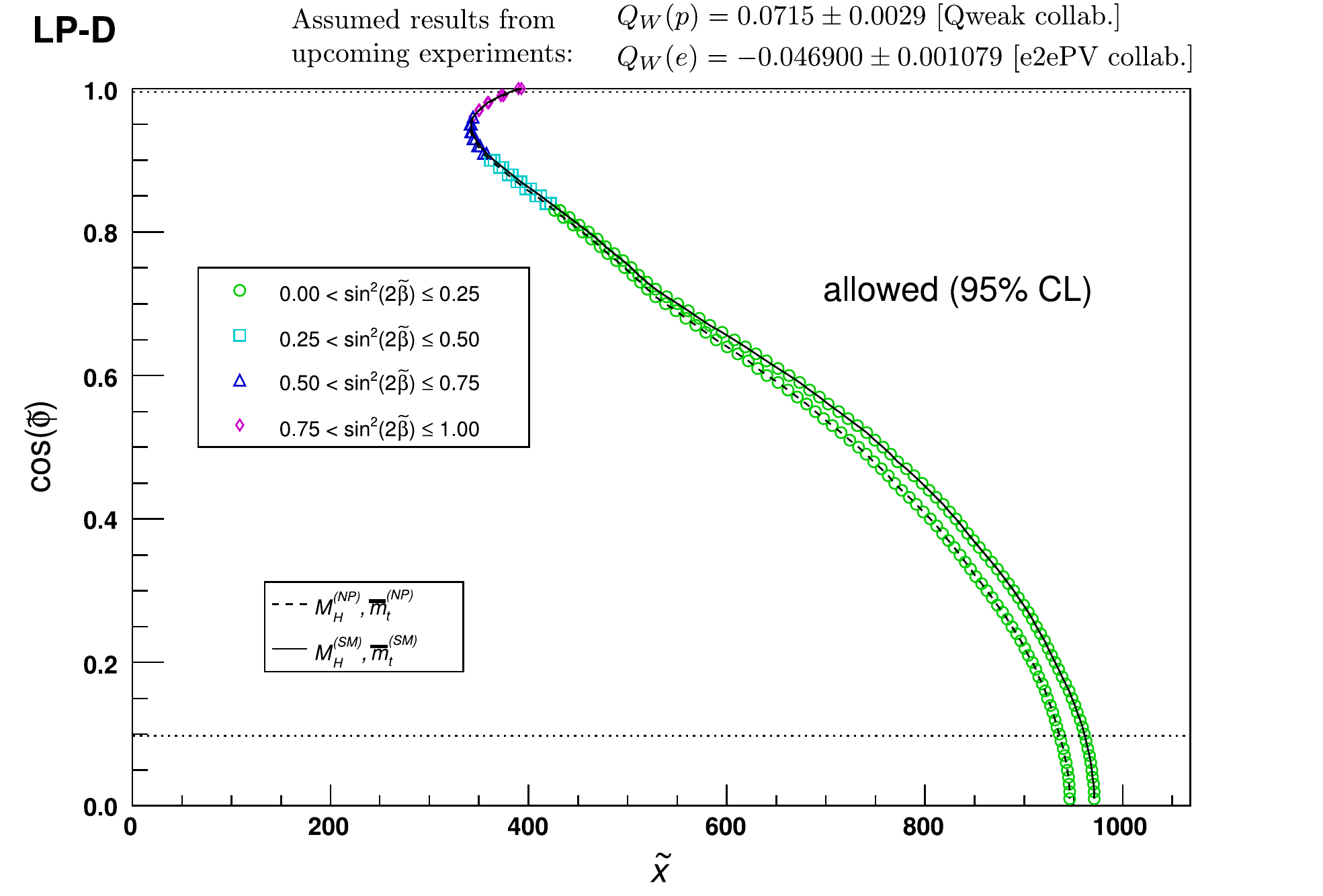}
\includegraphics[width=\picwidth]{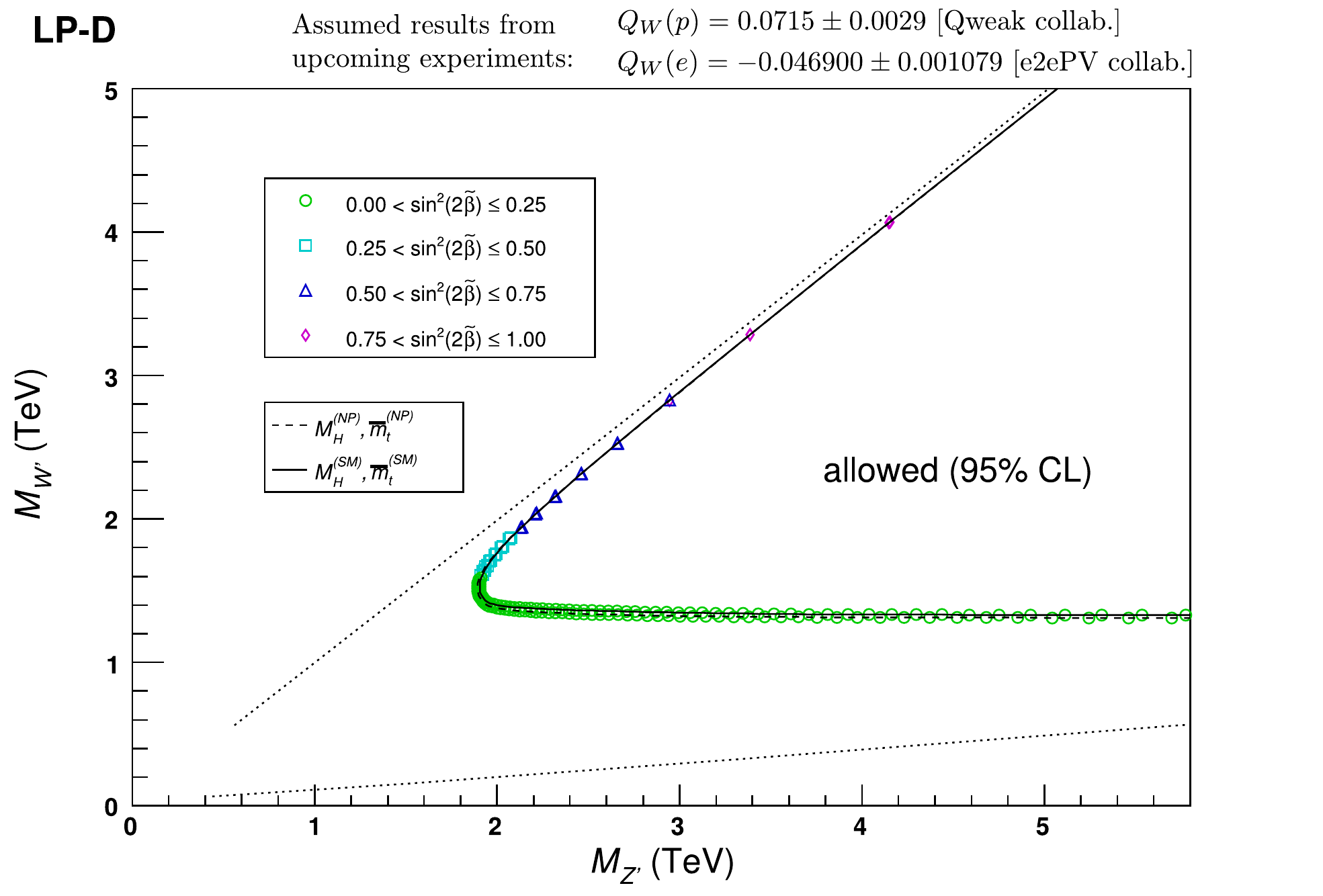}
\caption{
The $\tilde{x}-\cos(\tilde{\phi})$ and $M_{Z^{\prime}}$-$M_{W^{\prime}}$
contours for the LP-D model with an expected uncertainty in the Q-weak
data and the e2ePV data.
 These plots should be compared with the
corresponding plots in Figs.~\ref{fig:x-phi} and \ref{fig:MZP-MWP},and demonstrate that, since $Q_W(e)$ is a key
observable for the LP model, an increase of precision in its
measurement has a large impact on the global analysis.
In particular, at low $c_{\tphi}$, the lower bound on $\tilde{x}$ is substantially increased.
}
\label{fig:Qweak}
\end{center}
\end{figure}

An important consequence of identifying these observables is that we may
anticipate the future impact that the upcoming measurements, with greater precision, may have on
the global analysis.
For example, the Q-weak collaboration~\cite{VanOers:2007zz} and the
e2ePV collaboration~\cite{Mack:2006} at Jefferson Lab
 are expected to have ultra-high precision
measurements of the weak charge of the proton $Q_W(p)$ and the
electron $Q_W(e)$ (with a fractional uncertainties of respectively
4\% and 2.5\%).
As $Q_W(\ ^{133}\mbox{Cs})$ is a  key observable in driving the
results for the lepto-phobic (LP) and fermio-phobic
(FP) models, we would expect that the future measurements of
$Q_W(e)$ and $Q_W(p)$ would have a great impact on the global-fit
analysis.
This is demonstrated in Fig.~\ref{fig:Qweak}, where we perform the
global-fit analysis with the expected future uncertainty of $Q_W(p)$
and $Q_W(e)$. We find that the LP-D contour is drastically different
than those presented in Fig.~\ref{fig:x-phi}.
 As a
result of these further constraints from the $Q_W(p)$ and $Q_W(e)$,
the allowed region in the $M_{Z'}-M_{W'}$ plane shrinks as well. The
lower bounds for the $W^\prime$ mass increase, for instance in LP-D
model, from about $0.7 ~\mbox{TeV}$ to $1.3 \ \mbox{TeV}$.

\subsection{Constraints from Triple Gauge Boson Couplings}
Though we do not include the shifts to the triple gauge boson couplings (TGC)
in our global analysis,
they are nonetheless precise measurements at LEP-II that can be used
to constrain models of new physics.
In particular, the $ZWW$ vertex is measured to a precision of about
2\% and may be used to constrain models of new physics
\cite{Amsler:2008zzb}.
In this subsection, we compute the shift to the $ZWW$ vertex,
and
use it as a complementary constraint when we discuss the
results of our global analysis.

The shift in the $ZWW$ vertex can be parameterized by the Hagiwara's
parameterization \cite{Hagiwara:1986vm}
\begin{align}
g_{\tbox{ZWW}} = g^{\tbox{SM}}_{\tbox{ZWW}}\left(g_{1}^{Z}\right),
\end{align}
where the SM value of $g_{1}^{Z}$ is unity.
At LEP-II, using a partial waves analysis, the measured $ZWW$ vertex is \cite{Schael:2004tq}
\begin{align}
g_{1}^{Z} = 1.001 \pm 0.027 \pm 0.013,
\end{align}
where the uncertainties are respectively the 1$\sigma$ statistical and systematic uncertainties.
Adding these uncertainties in quadrature,
we have
\begin{align}
g_{1}^{Z} = 1.001 \pm 0.030.
\end{align}
It is important to note, however, that the measurement of the $ZWW$
vertex is extracted from the process $e^{+}e^{-}\rightarrow
W^{+}W^{-}$ using an event shape analysis (including angular
distributions) \textit{assuming} SM couplings for all other
vertices.
Thus, to properly use the experimental results, we have to compute the entire
$e^{+}e^{-}\rightarrow W^{+}W^{-}$ amplitude and attribute \textit{all} the shifts
in the amplitude to $\Delta g_{1}^{Z}$.

The full amplitude of the tree-level process $e^{+}e^{-}\rightarrow
W^{+}W^{-}$ is the sum of three diagrams: $s$-channel
$\gamma$-exchange, $s$-channel $Z$-exchange, and $t$-channel
$\nu$-exchange.
We denote these three amplitudes respectively as $\mathcal{A}_{\gamma}$,
$\mathcal{A}_{Z}$, and $\mathcal{A}_{\nu}$.
Even though the LEP experiments utilize unpolarized $e^{\pm}$
beams, it is useful to consider amplitudes with specific
helicities for both $e^{\pm}$ and $W^{\pm}$ and employ the
helicity amplitude method \cite{Xu:1986xb} to analyze the
amplitudes.
The individual amplitudes are computed in Hagiwara et
al.~\cite{Hagiwara:1986vm}, and here we only note the key features
of the dependence in the polar angle $\theta$.
For a given configuration of incoming and outgoing particle
helicities, all three amplitudes $\mathcal{A}_{\gamma,Z,\nu}$ are
proportional to the Wigner's $d$- matrix elements.
For the $s$-channel amplitudes, this is the only dependence in the
scattering angle $\theta$ (not to be confused with the weak mixing angle)
\begin{align}
\mathcal{A}_{\gamma,Z} \propto
d^{\Delta\sigma,\Delta\lambda}_{\Delta J}(\theta).
\end{align}
For the $t$-channel, $\nu$-exchange, we have additional $\theta$
dependence from the $\nu$-propagator
\begin{align}
\mathcal{A}_{\nu} \propto
\left(B-\frac{C}{1+\beta^2-2\beta\cos\theta }\right)
d^{\Delta\sigma,\Delta\lambda}_{\Delta J}(\theta),
\end{align}
where $B$ and $C$ depend on the helicity configuration, but are
independent of $\theta$.
Thus, for a fixed helicity configuration of $e^{+}e^{-}\rightarrow
W^{+}W^{-}$, we can perform a partial wave analysis
to project out the $d^{\Delta\sigma,\Delta\lambda}_{\Delta
J}(\theta)$ component of the amplitudes
\begin{align}
\tilde{\mathcal{A}}_{\gamma,Z,\nu} \equiv \int\limits_{0}^{\pi}
\mathcal{A}_{\gamma,Z,\nu} d^{\Delta\sigma,\Delta\lambda}_{\Delta
J}(\theta) d(\cos\theta)
\end{align}

\begin{figure}[h!t]
\begin{center}
\includegraphics[width=1.5in]{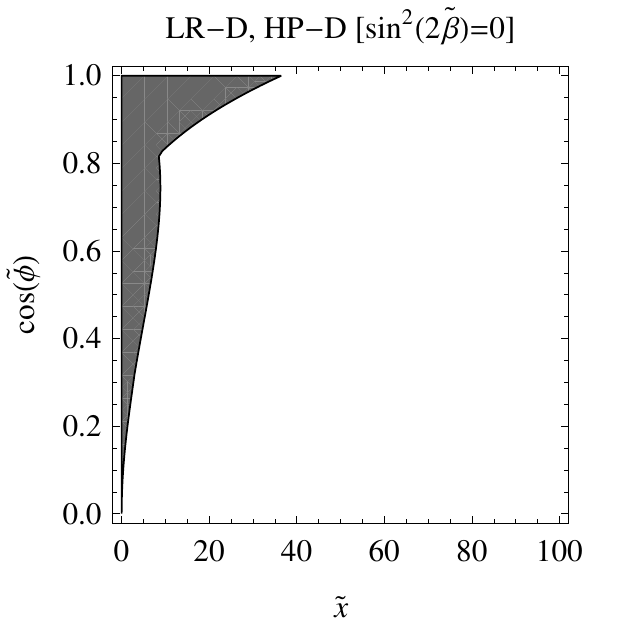}
\includegraphics[width=1.5in]{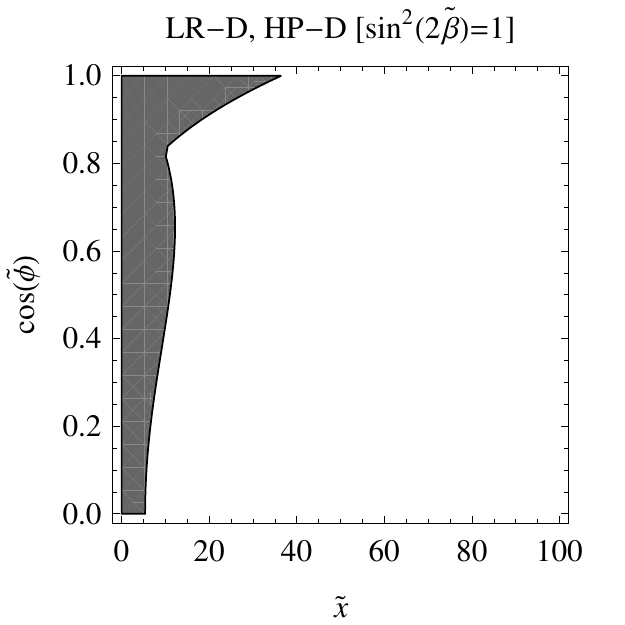}
\includegraphics[width=1.5in]{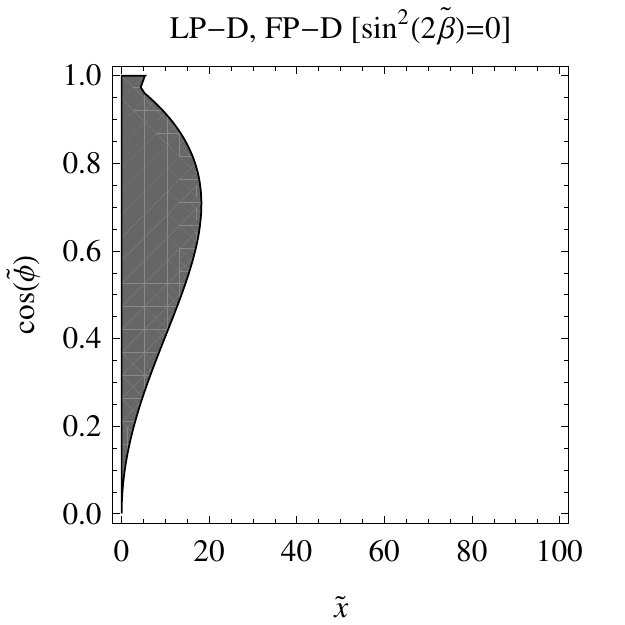}
\includegraphics[width=1.5in]{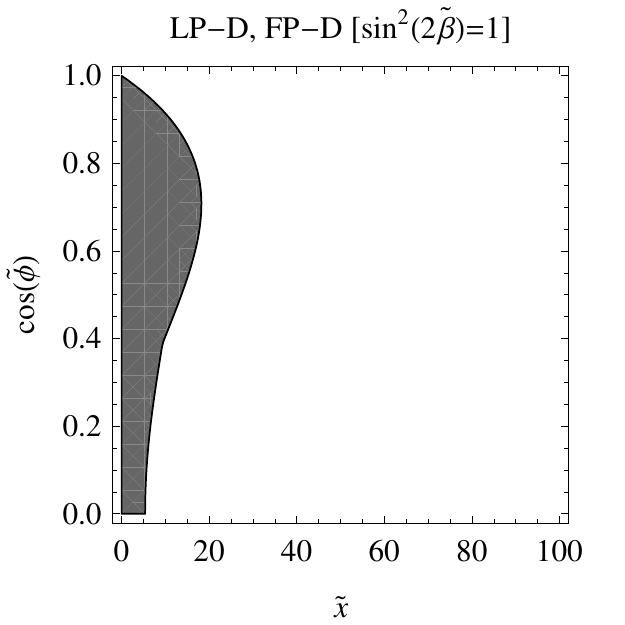}
\includegraphics[width=1.5in]{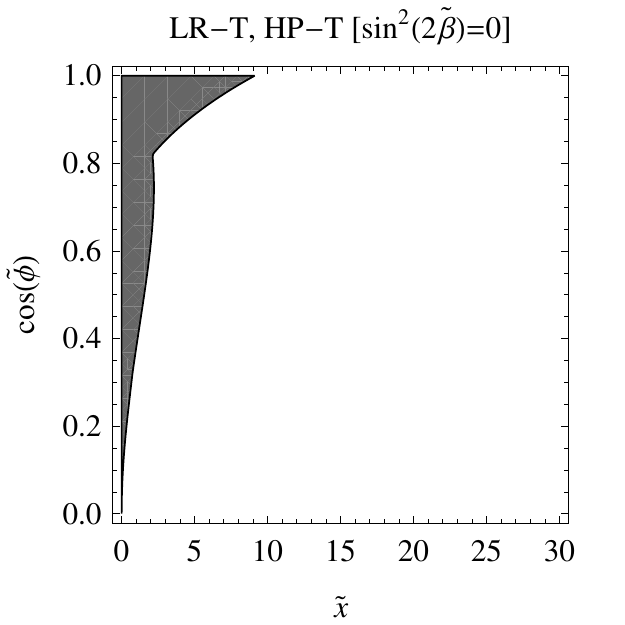}
\includegraphics[width=1.5in]{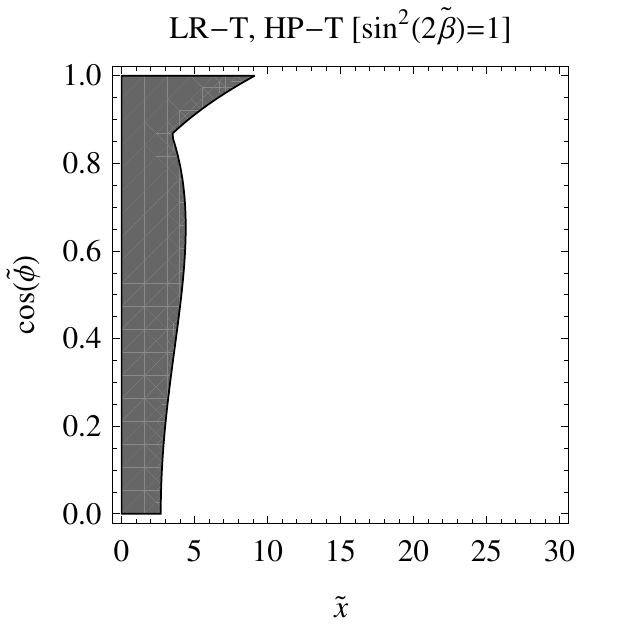}
\includegraphics[width=1.5in]{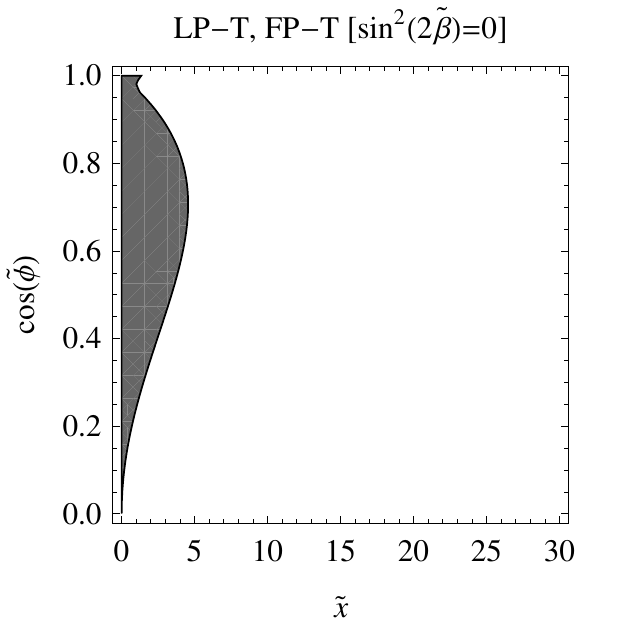}
\includegraphics[width=1.5in]{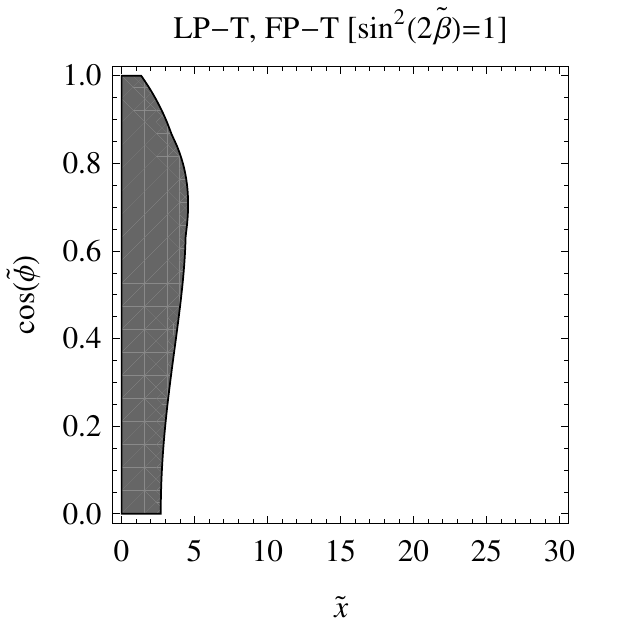}
\caption{
The regions on the $\tilde{x}$-$\cos\phi$ plane excluded by experimental
measurements of $\Delta g_1^{Z}$ (in grey)
for models in the breaking pattern I, for the cases
$\sin^2 2\tbeta=0$ and $1$.
}
\label{fig:ZWW}
\end{center}
\end{figure}

The uncertainties in $g_{1}^{Z}$ (which we denote as $\Delta g_Z^1$)
can be used to constrain the
$\tto$ models by first identifying
\begin{align}
\tilde{\mathcal{A}}_{\gamma}^{\tto}+
\tilde{\mathcal{A}}_{Z}^{\tto}+
\tilde{\mathcal{A}}_{\nu}^{\tto}
=
\tilde{\mathcal{A}}_{\gamma}^{\smbox{SM}}+
\left(1+\Delta g_{1}^{Z}\right)
\tilde{\mathcal{A}}_{Z}^{\smbox{SM}}
+\tilde{\mathcal{A}}_{\nu}^{\smbox{SM}},
\end{align}
and then express each amplitude in the $\tto$ model in terms of the
reference and fit parameters
\begin{align}
\tilde{\mathcal{A}}_{\gamma,Z,\nu}^{\tto} = \tilde{\mathcal{A}}_{\gamma,Z,\nu}^{\smbox{SM}}
\left(1+\frac{1}{\tilde{x}}\delta\tilde{\mathcal{A}}_{\gamma,Z,\nu}\right),
\label{eq:wwz-comp01}
\end{align}
where the fractional shifts in the amplitudes $\delta\tilde{\mathcal{A}}_{\gamma,Z,\nu}$
are functions of $\tphi$ and $\tbeta$.
For a fixed helicity configuration of $e^{+}e^{-}\rightarrow W^{+}W^{-}$,
we can compute $\tilde{\mathcal{A}}_{\gamma,Z,\nu}$ for both the
SM and $\tto$ models and obtain an excluded region on the $\tilde{x}$-$\cos\phi$ plane.
The regions that are allowed by all the helicity combinations are
 shown in Fig.~\ref{fig:ZWW}.
We note that, generally, the bounds given by $\Delta g_1^{Z}$ are
more relaxing than those obtained from the global-fit analysis
earlier.
In fact, for small values of $\tbeta$, the regions excluded by
$\Delta g_1^{Z}$ are already excluded by the global-fit analysis
presented in the previous subsection.

\section{Conclusions}
\label{sec:conc} 

In this work we analyze the constraints
 on the masses of the heavy gauge bosons of the
$\tto$ (including the left-right (LR), leptophobic (LP), hadrophobic
(HP) and fermiophobic (FP) as well as the ununified (UU) and
non-universal (NU)) models in a unified view based on the
classification of the $\tto$ models in terms of the patterns of
symmetry breaking and the gauge couplings of fermions.
Adapting the framework of effective field theory, we give the
effective Lagrangians at the electroweak scale and low energy scale
, applicable to any $G(221)$ model, and perform a global-fit
analysis about a set of $37$ electroweak observables, including $Z$
pole data, the mass and the width of the $W^{\pm}$ boson, and
various low-energy observables.
The experimental precision with which these observables have been
measured allows us to put strong bounds on the parameter space of
the $\tto$ models and to constrain the masses of the $Z^\prime$ and
 $W^{\prime \pm}$ bosons.
At the same time, we show that the bounds from the  triple gauge
boson couplings on the parameters do not affect the result of the
global-fit analysis.
We present our key results in terms of 95\% C.L. contours of the
allowed regions both on the $\tilde{x}$-$\cos\tphi$ plane, as well
as on the $M_{Z^{\prime}}$-$M_{W^{\prime}}$ plane, from which we can
readily give the lower bounds on the masses of the $W^{\prime}$ and
$Z^{\prime}$, which are presented in Table~\ref{tb:key-results-1},
which can be used as a guide for future collider search.
We show that, in the first breaking pattern, although the mass of
$Z^\prime$ is about $1.7$ TeV, the mass of the $W^\prime$ in some
models can be relatively light (of a few hundreds of GeV),
particularly in the left-right (LR), hadrophobic (HP)
models.
In the case of the second breaking pattern, due the near-degeneracy between the masses of the $Z^\prime$ and the $W^{\prime \pm}$,
the $W^\prime$ is necessarily heavy.

In addition to the constraints on the parameters and bounds on the
extra gauge boson masses, we also find associations between certain
key observables and the $\tto$ models discussed in this paper.
As these observables are responsible for `driving' the shape of the 95\% contour plots,
future measurements on these particular observables would have a tremendous impact
on our results.
We demonstrate such an example in Fig.~\ref{fig:Qweak}, showing that
an anticipated precision on the measurement of $Q_W(e)$ could
largely increase the lower bound on the $W^{\prime}$ mass from the
current value of  $0.7 ~\mbox{TeV}$ to $1.3 \ \mbox{TeV}$ in the LP-D
model.

In this work, we focus on the interactions of the heavy $W^\prime$
and $Z^\prime$ bosons to fermions. To extend our results to include
flavor-dependent observables, such as the branching ratio
$\mbox{Br}(b\rightarrow s\gamma)$ and the anomalous magnetic moment
of the muon, requires a detailed specification of the flavor sectors
of the $\tto$ models. Though it is difficult to enumerate the many
models in the literature, in the advent of the Large Hadron Collider
(LHC), it would be useful to extend to the flavor sector the
insights provided in this work.
Moreover, the  direct search of the  $W^\prime$ and $Z^\prime$ bosons at
the Fermilab Tevatron could further constrain the $\tto$ model
parameter space.
The potential of the Tevatron and LHC to observe the $W^\prime$ and
$Z^\prime$ bosons will be presented in a separate work.

\appendix

\section{Effective Lagrangians}
In this appendix we show how to obtain the coulping coefficients in
the effective Lagrangian.
There are two effective Lagrangians in our framework.
The first effective Lagrangian is SM-like, and is applicable at the
electroweak scale after integrating out $Z^{\prime}$ and
$W^{\prime}$, we parameterize this Lagrangian as
\begin{align}
\mathcal{L}_{\tbox{eff}}^{\tbox{ew}}
&=
 g^{\tbox{L}}_{Z\overline{f}f} Z_{\mu}\overline{f}\gamma^{\mu}\PL f
+g^{\tbox{R}}_{Z\overline{f}f} Z_{\mu}\overline{f}\gamma^{\mu}\PR f
+\left(
 g^{\tbox{L}}_{W\overline{f_1}f_2} W^{+}_{\mu}\overline{f}_1\gamma^{\mu}\PL f_2
+g^{\tbox{R}}_{W\overline{f_1}f_2} W^{+}_{\mu}\overline{f}_1\gamma^{\mu}\PR f_2
+\mbox{h.c.}
  \right)
\nonumber\\
&-
B\left(\left(\overline{f}_1 f_2\right)_{\tbox{L,R}}, \left(\overline{f}_3f_4\right)_{\tbox{L,R}}\right)
\frac{G_F}{\sqrt{2}}
\left(\overline{f}_1 f_2\right)_{\tbox{L,R}} \left(\overline{f}_3f_4\right)_{\tbox{L,R}}.
\label{eq:app-EW-eff}
\end{align}
%
%
Except for the non-universal (NU) model, the couplings
$g^{\tbox{L,R}}_{Z\overline{f}f}$, $g^{\tbox{L,R}}_{Z\overline{f}f}$
and $B\left(\left(\overline{f}_1 f_2\right)_{\tbox{L,R}},
\left(\overline{f}_3f_4\right)_{\tbox{L,R}}\right)$ are all
flavor-universal.
The second effective Lagrangian is Fermi's theory of four-fermion
interactions. It is obtained upon further integrating
out $Z$ and $W$ bosons, and is applicable below the
electroweak scale.
We parameterize this Lagrangian as
\begin{align}
\mathcal{L}_{\tbox{eff}}^{\tbox{4-fermion}}
&=
-C\left(\left(\overline{f}_1 f_2\right)_{\tbox{L,R}}, \left(\overline{f}_3f_4\right)_{\tbox{L,R}}\right)
\frac{G_F}{\sqrt{2}}
\left(\overline{f}_1 f_2\right)_{\tbox{L,R}} \left(\overline{f}_3f_4\right)_{\tbox{L,R}}.
\label{eq:app-4F-eff}
\end{align}

The above coefficient functions $g^{\tbox{L,R}}_{Z\overline{f}f}$,
$g^{\tbox{L,R}}_{W\overline{f_1}f_2}$ and
$C\left(\left(\overline{f}_1 f_2\right)_{\tbox{L,R}},
\left(\overline{f}_3f_4\right)_{\tbox{L,R}}\right)$ can be obtained
by the following steps:
\begin{itemize}
\item
write down the specific form of the effective Lagrangians at the electroweak scale
and low energy scale by plugging the formulae in Tables~\ref{tb:Zp-couplings}
and \ref{tb:Wp-couplings} into the Eqs.~\Eref{eq:L-eff-stage1} and \Eref{eq:4-fermi};
\item
extract the coulpings $g^{\tbox{L,R}}_{Z\overline{f}f}$,
$g^{\tbox{L,R}}_{W\overline{f_1}f_2}$ and $C$ by comparing the
parametrized Lagrangians in Eqs.~\Eref{eq:app-EW-eff} and
\Eref{eq:app-4F-eff} with the specific form of the effective
Lagrangians;
\item replace all the model parameters $\tilde{e}$, $\tilde{v}$,
and $\ttheta$ in the expressions of
$g^{\tbox{L,R}}_{Z\overline{f}f}$,
$g^{\tbox{L,R}}_{W\overline{f_1}f_2}$, and $C$ by reference
parameters $\alpha$, $G_F$ and $M_Z$, using the relations in
Eqs.~\Eref{eq:MZ-relation01}, \Eref{eq:v-relation} and
\Eref{eq:theta-relation}.
\end{itemize}
For future reference, we list below the coefficient functions
$g^{\tbox{L,R}}_{Z\overline{f}f}$,
$g^{\tbox{L,R}}_{W\overline{f_1}f_2}$ and
$C\left(\left(\overline{f}_1 f_2\right)_{\tbox{L,R}},
\left(\overline{f}_3f_4\right)_{\tbox{L,R}}\right)$ in terms of the
model parameters.
We also give some examples of the final form of the coefficients in
terms of the fit parameters.

\subsection{The LR-D, LP-D, HP-D, and FP-D Models}
For the four models that follow the first breaking pattern with a
doublet (LR-D, LP-D, HP-D, and FP-D models), the difference in the
 coefficient functions is originated from
the quantum numbers of the fermions.
In Table \ref{tb:QN1}, we give the quantum numbers
of the fermions, and present the coefficients of the effective
Lagrangian in terms of these quantum numbers.

\begin{table}[h]
\begin{center}
\caption{
The charge assignments of the SM fermions for the first
breaking patter.
$T_{3L}^f$ and $T_{3R}^f$ are respectively
the third component of the isospin for the $SU(2)_1$ and $SU(2)_2$ gauge groups
(which are conventionally called $SU(2)_L$ and $SU(2)_R$ in left-right models).
These charge assignments
apply to all three generations.
}
\label{tb:QN1}
\vspace{0.125in}
\begin{tabular}{|c|c|c|c|c||c|c|c|c|c||c|c|c|c|c||c|c|c|c|c|}
\hline
\multicolumn{2}{|c|}{ } & $T_{L}^3$ & $T_{R}^3$ & $X^{f}$ &
\multicolumn{2}{|c|}{ } & $T_{L}^3$ & $T_{R}^3$ & $X^{f}$ &
\multicolumn{2}{|c|}{ } & $T_{L}^3$ & $T_{R}^3$ & $X^{f}$ &
\multicolumn{2}{|c|}{ } & $T_{L}^3$ & $T_{R}^3$ & $X^{f}$
\\ \hline
\multirow{8}{*}{LR}
& $\nu_{\tbox{L}}$ & $+\tfrac{1}{2}$ & 0 & $-\tfrac{1}{2}$
& \multirow{8}{*}{LP}
& $\nu_{\tbox{L}}$ & $+\tfrac{1}{2}$ & 0 & $-\tfrac{1}{2}$
& \multirow{8}{*}{HP}
& $\nu_{\tbox{L}}$ & $+\tfrac{1}{2}$ & 0 & $-\tfrac{1}{2}$
& \multirow{8}{*}{FP}
& $\nu_{\tbox{L}}$ & $+\tfrac{1}{2}$ & 0 & $-\tfrac{1}{2}$
\\
& $e_{\tbox{L}}$ & $-\tfrac{1}{2}$ & 0 & $-\tfrac{1}{2}$ &
& $e_{\tbox{L}}$ & $-\tfrac{1}{2}$ & 0 & $-\tfrac{1}{2}$ &
& $e_{\tbox{L}}$ & $-\tfrac{1}{2}$ & 0 & $-\tfrac{1}{2}$ &
& $e_{\tbox{L}}$ & $-\tfrac{1}{2}$ & 0 & $-\tfrac{1}{2}$
\\
& $u_{\tbox{L}}$ & $+\tfrac{1}{2}$ & 0 & $+\tfrac{1}{6}$ &
& $u_{\tbox{L}}$ & $+\tfrac{1}{2}$ & 0 & $+\tfrac{1}{6}$ &
& $u_{\tbox{L}}$ & $+\tfrac{1}{2}$ & 0 & $+\tfrac{1}{6}$ &
& $u_{\tbox{L}}$ & $+\tfrac{1}{2}$ & 0 & $+\tfrac{1}{6}$
\\
& $d_{\tbox{L}}$ & $-\tfrac{1}{2}$ & 0 & $+\tfrac{1}{6}$ &
& $d_{\tbox{L}}$ & $-\tfrac{1}{2}$ & 0 & $+\tfrac{1}{6}$ &
& $d_{\tbox{L}}$ & $-\tfrac{1}{2}$ & 0 & $+\tfrac{1}{6}$ &
& $d_{\tbox{L}}$ & $-\tfrac{1}{2}$ & 0 & $+\tfrac{1}{6}$
\\
& $\nu_{\tbox{R}}$ & $0$ & $+\tfrac{1}{2}$ & $-\tfrac{1}{2}$ &
& $\nu_{\tbox{R}}$ & $0$ & $0$             & $0$ &
& $\nu_{\tbox{R}}$ & $0$ & $+\tfrac{1}{2}$ & $-\tfrac{1}{2}$ &
& $\nu_{\tbox{R}}$ & $0$ & $0$             & $0$
\\
& $e_{\tbox{R}}$ & $0$ & $-\tfrac{1}{2}$ & $-\tfrac{1}{2}$ &
& $e_{\tbox{R}}$ & $0$ & $0$             & $-1$ &
& $e_{\tbox{R}}$ & $0$ & $-\tfrac{1}{2}$ & $-\tfrac{1}{2}$ &
& $e_{\tbox{R}}$ & $0$ & $0$             & $-1$
\\
& $u_{\tbox{R}}$ & $0$ & $-\tfrac{1}{2}$ & $+\tfrac{1}{6}$ &
& $u_{\tbox{R}}$ & $0$ & $-\tfrac{1}{2}$ & $+\tfrac{1}{6}$ &
& $u_{\tbox{R}}$ & $0$ & $0$             & $+\tfrac{2}{3}$ &
& $u_{\tbox{R}}$ & $0$ & $0$             & $+\tfrac{2}{3}$
\\
& $d_{\tbox{R}}$ & $0$ & $-\tfrac{1}{2}$ & $+\tfrac{1}{6}$ &
& $d_{\tbox{R}}$ & $0$ & $-\tfrac{1}{2}$ & $+\tfrac{1}{6}$ &
& $d_{\tbox{R}}$ & $0$ & $0$             & $-\tfrac{1}{3}$ &
& $d_{\tbox{R}}$ & $0$ & $0$             & $-\tfrac{1}{3}$
\\
\hline
\end{tabular}
\end{center}
\end{table}

Performing the above procedure, we obtain
\begin{align}
    g^{\tbox{L}}_{W\overline{f_1}f_2} &= \mfrac{\tilde{e}}{\sqrt{2}s_{\ttheta}}, \\
    g^{\tbox{R}}_{W\overline{f_1}f_2} &= \begin{cases}
    \mfrac{\tilde{e}}{\sqrt{2}s_{\ttheta}} \mfrac{s_{2\tbeta}}{\tilde{x}}, &
\mbox{(for $f_{1,2}$ as quarks in LR and LP, and leptons in LR and HP)}\\
0, & \mbox{(for $f_{1,2}$ as quarks in HP and FP, and leptons in LP and FP)}
    \end{cases}\\
    g^{\tbox{L}}_{Z\overline{f}f} &= \mfrac{\tilde{e}}{s_{\ttheta}c_{\ttheta}}\left( T^3_L - s_{\ttheta}^2 Q + \mfrac{s_{\tphi}^2 c_{\tphi}^2}{\tilde{x}}(T^3_L - Q)\right),\\
    g^{\tbox{R}}_{Z\overline{f}f} &= \mfrac{\tilde{e}}{s_{\ttheta}c_{\ttheta}}\left(  - s_{\ttheta}^2 Q + \mfrac{c_{\tphi}^2}{\tilde{x}}(T^3_R - s_{\tphi}^2 Q)\right),
\end{align}
and the neutral-current four-fermion coupling coefficients
\begin{align}
    \mfrac{1}{2}C\left(\left(\overline{f} f\right)_{\tbox{L}}, \left(\overline{f}^\prime f^\prime \right)_{\tbox{L}}\right)
    &= (1+\mfrac{c_{\tphi}^4}{\tilde{x}}-\mfrac{s_{2\tbeta}^2}{\tilde{x}})(T^3_L - s_{\tth}^2 Q)(T^{\prime 3}_L - s_{\ttheta}^2 Q^\prime)
     \nonumber\\
     & +\mfrac{s_{\tphi}^2 c_{\tphi}^2}{\tilde{x}}\left[(T^3_L - s_{\ttheta}^2 Q)(T^{\prime 3}_L - Q^\prime)+ (T^3_L - Q)(T^{\prime 3}_L - s_{\ttheta}^2 Q^\prime)\right] \nonumber\\
     &+\mfrac{s_{\tphi}^4}{\tilde{x}}(T^3_L -  Q)(T^{\prime 3}_L - Q^\prime),\\
    \mfrac{1}{2}C\left(\left(\overline{f} f\right)_{\tbox{R}}, \left(\overline{f}^\prime f^\prime \right)_{\tbox{R}}\right)
    &= (1+\mfrac{c_{\tphi}^4}{\tilde{x}}-\mfrac{s_{\tbeta}^2}{\tilde{x}})( - s_{\tth}^2 Q)( - s_{\ttheta}^2 Q^\prime)
     \nonumber\\
     & +\mfrac{c_{\tphi}^2}{\tilde{x}}\left[(T^3_R -s_{\tphi}^2  Q)(-s_{\ttheta}^2 Q^\prime)+ (- s_{\ttheta}^2 Q)(T^{\prime 3}_R - s_{\tphi}^2 Q^\prime)\right]\nonumber\\
     & +\mfrac{1}{\tilde{x}}(T^3_R - s_{\tphi}^2 Q)(T^{\prime 3}_R -s_{\tphi}^2 Q^\prime),\\
     \mfrac{1}{2}C\left(\left(\overline{f} f\right)_{\tbox{L}}, \left(\overline{f}^\prime f^\prime \right)_{\tbox{R}}\right)
    &= (1+\mfrac{c_{\tphi}^4}{\tilde{x}}-\mfrac{s_{\tbeta}^2}{\tilde{x}})( T^3_L- s_{\tth}^2 Q)( - s_{\ttheta}^2 Q^\prime)
     \nonumber\\
     & +\mfrac{s_{\tphi}^2 c_{\tphi}^2}{\tilde{x}}(T^3_L -  Q)( - s_{\ttheta}^2 Q^\prime) + \mfrac{c_{\tphi}^2}{\tilde{x}}( T^3_L- s_{\tth}^2 Q)(T^{\prime 3}_R -s_{\tphi}^2 Q^\prime) \nonumber\\
     &+\mfrac{s_{\tphi}^2}{\tilde{x}}(T^3_L -  Q)(T^{\prime 3}_R -s_{\tphi}^2 Q^\prime),
\end{align}
where $T^3_{L,R}$ and $Q$ are the isospin charge and electric charge
of the fermion $f$, and $T^{\prime 3}_{L,R}$ and $Q^\prime$ are the
isospin charge and electric charge for the fermion $f^\prime$, respectively. To
obtain $C\left(\left(\overline{f} f\right)_{\tbox{R}},
\left(\overline{f}^\prime f^\prime \right)_{\tbox{L}}\right)$, we
need to exchange $Q$ with $Q^\prime$ and $T^3_L $ with $T^{\prime
3}_R$ in the coefficient function $C\left(\left(\overline{f}
f\right)_{\tbox{L}}, \left(\overline{f}^\prime f^\prime
\right)_{\tbox{R}}\right)$. For the charged-current four-fermion
coupling coefficients, we only list the results for LR-D models as follows:
\begin{align}
    C\left(\left(\overline{f}_1 f_2\right)_{\tbox{L}}, \left(\overline{f}^\prime_3 f^\prime_4 \right)_{\tbox{L}}\right)
    &= 1,\\
    C\left(\left(\overline{f}_1 f_2\right)_{\tbox{L}}, \left(\overline{f}^\prime_3 f^\prime_4 \right)_{\tbox{R}}\right)
    &= C\left(\left(\overline{f}_1 f_2\right)_{\tbox{R}}, \left(\overline{f}^\prime_3 f^\prime_4 \right)_{\tbox{L}}\right) = \mfrac{s_{2\tbeta}}{x},\\
    C\left(\left(\overline{f}_1 f_2\right)_{\tbox{R}}, \left(\overline{f}^\prime_3 f^\prime_4 \right)_{\tbox{R}}\right)
    &= \mfrac{1}{x},
\end{align}
The final form of $g^{\tbox{L,R}}_{Z\overline{f}f}$ and $C$ can be
obtained by replacing the model parameters $\ttheta$ by the reference
parameters $\alpha$, $G_F$ and $M_Z$. Below we only list the results
for $C\left(\left(\overline{u} u\right)_{\tbox{L}},
\left(\overline{e} e \right)_{\tbox{L}}\right)$ in the LR-D model:
\begin{align}
    g^{\tbox{L}}_{Z\overline{e}e} &= g^{\tbox{L,SM}}_{Z\overline{e}e} + \delta g^{\tbox{L}}_{Z\overline{e}e}
    =-0.197 +\mfrac{1}{x}\left(-0.348 + 0.348 s_{2\tbeta}^2 +1.07 s_{\tphi}^2 -0.718 s_{\tphi}^4\right),
\end{align}
\begin{align}
C\left(\left(\overline{u} u\right)_{\tbox{L}}, \left(\overline{e} e
\right)_{\tbox{L}}\right) &= C^{\tbox{SM}}\left(\left(\overline{u}
u\right)_{\tbox{L}}, \left(\overline{e} e \right)_{\tbox{L}}\right)
+\delta C\left(\left(\overline{u} u\right)_{\tbox{L}},
\left(\overline{e} e \right)_{\tbox{L}}\right)
\nonumber\\
&= -0.183 +\mfrac{1}{x}\left(-0.534 + 0.534 s_{2\tbeta}^2 +1.50 s_{\tphi}^2 -1.13 s_{\tphi}^4 \right).
\end{align}

\subsection{The LR-T, LP-T, HP-T, and FP-T Models}
The coefficients of the effective operators in the LR-T, LP-T, HP-T,
and FP-T models take a similar form as those presented above for the
LR-D, LP-D, HP-D, and FP-D models respectively, with the following
replacements after applying the identity $c_{\tilde{\phi}}^2 = 1-
s_{\tilde{\phi}}^2 $:
\begin{align}
s_{\tilde{\phi }}^2 &\rightarrow \frac{1}{4}s_{\tilde{\phi }}^2,
\\
s_{\tilde{\phi }}^4 &\rightarrow \frac{1}{4}s_{\tilde{\phi }}^4,
\\
s_{2 \tilde{\beta }}^2 &\rightarrow \frac{1}{2}s_{2 \tilde{\beta }}^2,
\end{align}
and terms that are suppressed by $1/\tilde{x}$, but without the factors listed above, are further divided by
a factor of 4. For example, for the LR-T model, we have
\begin{align}
    g^{\tbox{L}}_{Z\overline{e}e} &
    =-0.197 +\mfrac{1}{x}\left(-0.087 + 0.174 s_{2\tbeta}^2 +0.268 s_{\tphi}^2 -0.180 s_{\tphi}^4\right).
\end{align}

\subsection{The UU and NU Models}

For the un-unified and non-universal models,
we classify the fermions as
\begin{align}
\mbox{UU}  &: \begin{cases}
f\equiv\ \mbox{leptons}\\
F\equiv\ \mbox{quarks},
\end{cases},\\
\mbox{NU}  &: \begin{cases}
f\equiv\ \mbox{fermions of the third generation},\\
F\equiv\ \mbox{fermions of the first two generation}.
\end{cases}
\end{align}
We denote the coefficients of the effective Lagrangians with
the notations in Eqs.~\Eref{eq:app-EW-eff} and \Eref{eq:app-4F-eff}.
Compared to the first breaking pattern presented earlier, there are
considerably less coefficients because there are no right-handed
charged currents in the NU and UU models.

Similar to the LR-D model, we obtain the electroweak couplings:
\begin{align}
    g^{\tbox{L}}_{W\overline{f}_1 f_2} & = \mfrac{\tilde{e}}{\sqrt{2}s_{\ttheta}}\left(1-\mfrac{s_{\tphi}^4}{\tilde{x}}\right),\\
    g^{\tbox{L}}_{W\overline{F}_1 F_2} & = \mfrac{\tilde{e}}{\sqrt{2}s_{\ttheta}}\left(1+\mfrac{s_{\tphi}^2c_{\tphi}^2}{\tilde{x}}\right),\\
    g^{\tbox{L}}_{Z\overline{f}f} &= \mfrac{\tilde{e}}{s_{\ttheta}c_{\ttheta}}\left( T^3 - s_{\ttheta}^2 Q - \mfrac{s_{\tphi}^4 }{\tilde{x}}T^3 \right),\\
    g^{\tbox{L}}_{Z\overline{F}F} &= \mfrac{\tilde{e}}{s_{\ttheta}c_{\ttheta}}\left( T^3 - s_{\ttheta}^2 Q + \mfrac{s_{\tphi}^2 c_{\tphi}^2}{\tilde{x}}T^3\right),\\
    g^{\tbox{R}}_{Z\overline{f}f} &= g^{\tbox{R}}_{Z\overline{F}F} =\mfrac{\tilde{e}}{s_{\ttheta}c_{\ttheta}}\left(  - s_{\ttheta}^2 Q \right),
\end{align}
and the neutral-current four-fermion coupling coefficients
\begin{align}
    \mfrac{1}{2}C\left(\left(\overline{f} f\right)_{\tbox{L}}, \left(\overline{f}^\prime f^\prime \right)_{\tbox{L}}\right)
    &= (1+\mfrac{s_{\tphi}^4}{\tilde{x}})(T^3 - s_{\tth}^2 Q)(T^{\prime 3} - s_{\ttheta}^2 Q^\prime)
    \nonumber\\
    & -\mfrac{ s_{\tphi}^4 }{\tilde{x}}
    \left(  (T^3 - s_{\ttheta}^2 Q) T^{\prime 3} +T^{3} (T^{\prime 3} - s_{\ttheta}^2 Q^\prime)  -
    T^3 T^{\prime 3} \right),\\
    \mfrac{1}{2}C\left(\left(\overline{f} f\right)_{\tbox{R}}, \left(\overline{f}^\prime f^\prime \right)_{\tbox{R}}\right)
    &= (1+\mfrac{s_{\tphi}^4}{\tilde{x}})( - s_{\tth}^2 Q)( - s_{\ttheta}^2 Q^\prime)
     ,\\
     \mfrac{1}{2}C\left(\left(\overline{f} f\right)_{\tbox{L}}, \left(\overline{f}^\prime f^\prime \right)_{\tbox{R}}\right)
    &= (1+\mfrac{s_{\tphi}^4}{\tilde{x}})( T^3- s_{\tth}^2 Q)( - s_{\ttheta}^2 Q^\prime)
     - \mfrac{s_{\tphi}^4}{\tilde{x}} T^3 ( - s_{\ttheta}^2 Q^\prime) ,\\
       \mfrac{1}{2}C\left(\left(\overline{F} F\right)_{\tbox{L}}, \left(\overline{F}^\prime F^\prime \right)_{\tbox{L}}\right)
    &= (1+\mfrac{s_{\tphi}^4}{\tilde{x}})(T^3 - s_{\tth}^2 Q)(T^{\prime 3} - s_{\ttheta}^2 Q^\prime)
    \nonumber\\&+\mfrac{ s_{\tphi}^2 c_{\tphi}^2 }{\tilde{x}}
    \left[T^3 (T^{\prime 3}-s_{\ttheta}^2 Q^\prime) +  (T^{3}-s_{\ttheta}^2 Q)T^{\prime 3}\right]+\mfrac{c_{\tphi}^4 }{\tilde{x}}T^3 T^{\prime 3},\\
    \mfrac{1}{2}C\left(\left(\overline{F} F\right)_{\tbox{R}}, \left(\overline{F}^\prime F^\prime \right)_{\tbox{R}}\right)
    &= (1+\mfrac{s_{\tphi}^4}{\tilde{x}})( - s_{\tth}^2 Q)( - s_{\ttheta}^2 Q^\prime)
     ,\\
     \mfrac{1}{2}C\left(\left(\overline{F} F\right)_{\tbox{L}}, \left(\overline{F}^\prime F^\prime \right)_{\tbox{R}}\right)
    &= (1+\mfrac{s_{\tphi}^4}{\tilde{x}})( T^3- s_{\tth}^2 Q)( - s_{\ttheta}^2 Q^\prime)
     + \mfrac{s_{\tphi}^2 c_{\tphi}^2}{\tilde{x}} T^3 ( - s_{\ttheta}^2 Q^\prime) ,
\end{align}
and
\begin{align}
       \mfrac{1}{2}C\left(\left(\overline{F} F\right)_{\tbox{L}}, \left(\overline{f}^\prime f^\prime \right)_{\tbox{L}}\right)
    &= (1+\mfrac{s_{\tphi}^4}{\tilde{x}})(T^3 - s_{\tth}^2 Q)(T^{\prime 3} - s_{\ttheta}^2 Q^\prime)
    \nonumber\\
    &-\mfrac{ s_{\tphi}^4 }{\tilde{x}}  (T^{ 3}-s_{\ttheta}^2 Q) T^{\prime 3}
    + \mfrac{ s_{\tphi}^2 c_{\tphi}^2 }{\tilde{x}} T^3 (T^{\prime 3}-s_{\ttheta}^2 Q^\prime) - \mfrac{ s_{\tphi}^2 c_{\tphi}^2 }{\tilde{x}} T^3 T^{\prime 3},\\
    \mfrac{1}{2}C\left(\left(\overline{F} F\right)_{\tbox{R}}, \left(\overline{f}^\prime f^\prime \right)_{\tbox{R}}\right)
    &= (1+\mfrac{s_{\tphi}^4}{\tilde{x}})( - s_{\tth}^2 Q)( - s_{\ttheta}^2 Q^\prime)
     ,\\
     \mfrac{1}{2}C\left(\left(\overline{F} F\right)_{\tbox{L}}, \left(\overline{f}^\prime f^\prime \right)_{\tbox{R}}\right)
    &= (1+\mfrac{s_{\tphi}^4}{\tilde{x}})( T^3- s_{\tth}^2 Q)( - s_{\ttheta}^2 Q^\prime)
     + \mfrac{s_{\tphi}^2 c_{\tphi}^2}{\tilde{x}} T^3 ( - s_{\ttheta}^2 Q^\prime) ,\\
     \mfrac{1}{2}C\left(\left(\overline{F} F\right)_{\tbox{L}}, \left(\overline{f}^\prime f^\prime \right)_{\tbox{R}}\right)
    &= (1+\mfrac{s_{\tphi}^4}{\tilde{x}})( T^3- s_{\tth}^2 Q)( - s_{\ttheta}^2 Q^\prime)
     - \mfrac{s_{\tphi}^4}{\tilde{x}}  ( - s_{\ttheta}^2 Q) T^{\prime 2}.
\end{align}
We can get $C\left(\left(\overline{f} f\right),
\left(\overline{F}^\prime F^\prime \right)\right)$ by exchanging
$Q\leftrightarrow Q^\prime$ and $T^3 \leftrightarrow T^{\prime 3}$
in the expresion $C\left(\left(\overline{F} F\right),
\left(\overline{f}^\prime f^\prime \right)\right)$. For the
charged-current four-fermion coupling coefficients, we only list the
results for UU-D:
\begin{align}
    C\left(\left(\overline{f}_1 f_2\right)_{\tbox{L}}, \left(\overline{f}_3 f_4 \right)_{\tbox{L}}\right)
    &= 1,\\
    C\left(\left(\overline{F}_1 F_2\right)_{\tbox{L}}, \left(\overline{f}_3 f_4 \right)_{\tbox{L}}\right)
    &= C\left(\left(\overline{f}_1 f_2\right)_{\tbox{L}}, \left(\overline{F}_3 F_4 \right)_{\tbox{L}}\right) = 1,\\
    C\left(\left(\overline{F}_1 F_2\right)_{\tbox{L}}, \left(\overline{F}_3 F_4 \right)_{\tbox{L}}\right)
    &= 1+\mfrac{1}{x},
\end{align}
For the final forms in terms of fit parameters, we only list the
results for $g^{\tbox{L}}_{Z\overline{e}e}$ and
$C\left(\left(\overline{u} u\right)_{\tbox{L}}, \left(\overline{e} e
\right)_{\tbox{L}}\right)$ in the UU-D model:
\begin{align}
    g^{\tbox{L}}_{Z\overline{e}e} &= g^{\tbox{L,SM}}_{Z\overline{e}e} + \delta g^{\tbox{L}}_{Z\overline{e}e} =  -0.197 + 0.0227 \mfrac{ s_{\tphi}^4}{x},\\
    C\left(\left(\overline{u} u\right)_{\tbox{L}}, \left(\overline{e} e \right)_{\tbox{L}}\right)
&= C^{\tbox{SM}}\left(\left(\overline{u} u\right)_{\tbox{L}}, \left(\overline{e} e \right)_{\tbox{L}}\right)+
    \delta C\left(\left(\overline{u} u\right)_{\tbox{L}}, \left(\overline{e} e \right)_{\tbox{L}}\right)    \nonumber\\& = -0.183 +\mfrac{1}{x}\left(0.234 s_{\tphi}^2 -0.424 s_{\tphi}^4 \right).
\end{align}

\begin{acknowledgments}
We are very grateful to Jens Erler for providing us with the latest version
of GAPP, and patiently guiding us through using it correctly.
We are also grateful to Sekhar Chivukula, Elizabeth Simmons, and Jim Linnemann for
helpful discussions.
We
acknowledge the support of the U.S. National Science Foundation
under grant PHY-0555545 and PHY-0855561. C.P.Y. would also like to
thank the hospitality of National Center for Theoretical Sciences in
Taiwan and Center for High Energy Physics, Peking University, in
China, where part of this work was done.
\end{acknowledgments}



\end{document}